# Magnetic Fields in the Milky Way and in Galaxies [1]

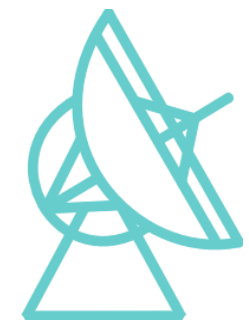

Rainer Beck & Richard Wielebinski

Max-Planck-Institut für Radioastronomie, Bonn, Germany

**Abstract.** Most of the visible matter in the Universe is ionized, so that cosmic magnetic fields are quite easy to generate and due to the lack of magnetic monopoles hard to destroy. Magnetic fields have been measured in or around practically all celestial objects, either by in-situ measurements of spacecrafts or by the electromagnetic radiation of embedded cosmic rays, gas, or dust. The Earth, the Sun, solar planets, stars, pulsars, the Milky Way, nearby star-forming galaxies, distant radio galaxies, quasars, and even the intergalactic space in clusters of galaxies have significant magnetic fields. Even larger volumes of the Universe may be permeated by so-far invisible magnetic fields. Information on cosmic magnetic fields has increased enormously as the result of the rapid development of observational methods, especially in radio astronomy. In the Milky Way, a wealth of magnetic phenomena was discovered that are only partly related to objects visible in other spectral ranges. The large-scale structure of the Milky Way's magnetic field is still under debate. The available data for external spiral galaxies can well be explained by field amplification and ordering via the dynamo mechanism. The measured field strengths and the similarity of field patterns and flow patterns of the diffuse ionized gas give strong indication that galactic magnetic fields are dynamically important. They may affect the formation of spiral arms, outflows, and the general evolution of galaxies. In spite of our increasing knowledge on magnetic fields, many important questions on the origin and evolution of magnetic fields, like their first occurrence in young galaxies or the existence of large-scale intergalactic fields, remain unanswered. Progress can be expected from new-generation radio telescopes like LOFAR and the SKA, for which "cosmic magnetism" is a key science project.

## 1. Introduction

The first report of a cosmic magnetic field outside the Earth was the result of a direct measurement of the Zeeman effect in the magnetic fields in sunspots in 1908. In 1950 it was suggested that the observed cosmic rays would require magnetic fields for their creation and their containment within the Galaxy. Optical polarization observations were first successful in 1949. Polarization of optical and infrared emission can also be caused by elongated dust grains that are aligned in magnetic fields first proposed in 1951. This interpretation was not accepted for a long time in the optical astronomy community. With the advent of radio astronomy this controversy was resolved and an active epoch of magnetic field studies could begin.

Radio astronomy started in 1932 with the detection of continuum radio emission from the Milky Way. It became quickly clear that the observed radio waves were of non-thermal nature and an interpretation of this phenomenon was actively sought. This was given in 1950 – the radio emission is due to relativistic cosmic-ray electrons gyrating in magnetic fields, emitting radio waves by the synchrotron process – when the theory of synchrotron emission theory was developed. In particular, it was soon pointed out that synchrotron emission should be highly polarized. In fact, in homogenous magnetic fields, up to 75% linear polarization of the continuum emission is expected. This suggestion was taken up by observers of optical radiation, who found in 1954 that the Crab Nebula was highly polarized and hence emitting light through the synchrotron process. The radio confirmation of the polarization of the

[1] *Chapter 13 in: Planets, Stars and Stellar Systems, Vol. 5, ed. G. Gilmore, Springer, Berlin 2013.*

***Revised version of Sept. 2023***

Crab Nebula followed in 1957. The first definite detection of linear polarization of Galactic radio waves was published in 1962. At the same time the polarization of the bright radio galaxy Cygnus A and the Faraday rotation of the polarization angles of the linearly polarized radio emission in Centaurus A were detected. Observations at two frequencies of a section of the Milky Way showed that the interstellar medium of the Milky Way can also cause the Faraday effect. During this exciting time of definite detections of interstellar and extragalactic magnetic fields by observations of linear polarization, the Zeeman effect of radio spectral lines proved to be more elusive. Several groups attempted to measure magnetic fields by this direct method. It was in 1968 that finally the Zeeman effect at radio wavelengths was successfully observed in the absorption profile of the HI line in the direction towards Cassiopeia A. From that time onward considerable data was collected on the distribution of magnetic fields in the Milky Way. The all-sky maps of synchrotron polarization obtained with ground-based radio telescopes and the WMAP and PLANCK space observatories shed new light on the large-scale structure of magnetic fields in the Milky Way.

In the optical and near-infrared ranges, polarization is produced by the different extinction along the minor and major axis of dust grains, while at far-infrared and sub-millimeter wavelengths the elongated dust grains themselves emit polarized emission that was first detected in the 1980s. Progress has been slow, until recently an increase in reliable data became possible with the advent of sub-mm telescopes on excellent sites and sensitive polarimeters. The data on polarized dust emission from PLANCK and the HAWC+ polarimeter on board of the SOFIA airborne telescope opened a new era in studying magnetic fields in dense clouds of the Milky Way and of external galaxies.

The first suggestions about the presence of magnetic fields in nearby galaxies were made in 1958, based on observations of the polarization of stars in the Andromeda galaxy, M31. In 1967 observations of the linear polarization of diffuse starlight started in bright nearby galaxies. In 1970 the polarization of stars in the Magellanic Clouds implied the presence of magnetic fields in these neighboring galaxies. Low-frequency radio observations of galaxies showed non-thermal spectra and hence indicated the presence of magnetic fields. The first detection of the linear polarization of the radio emission from nearby galaxies in 1972 led the way to massive improvement on our knowledge of the morphology of magnetic fields in galaxies. These early radio observations were in good agreement with the early optical polarization studies of galaxies. A history of radio polarization measurements was compiled by Wielebinski (2012).

In this review, the status of our knowledge about the magnetic fields in our Milky Way and in nearby star-forming galaxies is summarized. Magnetic fields are a major agent in the interstellar and intra-cluster medium and affect the physical processes in various ways. They contribute significantly to the total pressure, which balances the gas disk of a galaxy against gravitation. Magnetic reconnection is a possible heating source for the interstellar medium (ISM) and the halo gas. Magnetic fields affect the dynamics of the turbulent ISM and the gas flows around spiral arms. The shock strength in spiral density waves is decreased and structure formation is reduced in the presence of a strong field. The interstellar fields are closely connected to gas clouds. Magnetic fields stabilize gas clouds and reduce the star-formation efficiency to the observed low values. On the other hand, magnetic fields are essential for the onset of star formation as they enable the removal of angular momentum from protostellar clouds via ambipolar diffusion. MHD turbulence distributes energy from supernova explosions within the ISM and drives field amplification and ordering via the dynamo mechanism. In galaxies with low star-formation activity or in the outer disks, the magneto-rotational instability can generate turbulence and heat the gas. Magnetic fields control the density and distribution of cosmic rays in the ISM. Cosmic rays accelerated in supernova remnants can provide the pressure to drive a galactic outflow and generate buoyant loops of magnetic fields through the Parker instability. Understanding the interaction between the gas and the magnetic field is a key to understand the physics of galaxy disks and halos and the evolution of galaxies.

The magnetic field of the Milky Way is of particular importance for experiments to detect *ultrahigh-energy cosmic rays* (UHECRs). Results from the first years of AUGER indicate that the arrival directions of detected UHECRs with energies of more than $10^{19}$ eV show some coincidence with the positions of known nearby active galaxies. This interpretation only holds if the deflections in the magnetic fields of the intergalactic medium and the Milky Way halo are not too large. However, little is known about the structure and strength of the magnetic field in the halo of our Milky Way and beyond.

There is one class of galaxies, where magnetic fields play a crucial role: “active” galaxies that are governed by a central Black Hole. The formation of jets and radio lobes can only be understood with the presence of magnetic fields. The physics of these phenomena is quite different from that in "normal" star-forming galaxies and will not be discussed in this review.

Magnetic fields have also been detected in the intergalactic medium surrounding the galaxies in a cluster through observations of non-thermal diffuse radio halos and the Faraday effect of background radio sources seen through the cluster medium. These intracluster magnetic fields are probably generated by turbulent gas motions as the result of massive interactions between galaxies and the intracluster gas. Magnetic fields affect thermal conduction in galaxy clusters and hence their evolution. Intracluster magnetic fields are reviewed elsewhere (e.g. Feretti et al. 2012; Han 2017). Outflows from galaxies may have magnetized the intergalactic medium, so that the general intergalactic space may be pervaded with magnetic fields. Unfortunately, cosmic rays and dust grains are rare outside of galaxies and galaxy clusters, so that magnetic fields remain invisible.

Cosmological models of structure formation indicate that the intergalactic space is probably permeated by magnetic filaments. Galactic winds, jets from active galaxies, and interactions between galaxies can magnetize the intergalactic medium. The detection of magnetic fields in intergalactic filaments and observations of the interaction between galaxies and the intergalactic space is one of the important tasks for future radio telescopes. Until now the arguments for the presence of magnetic fields in the distant Universe is based on observations of the non-thermal radio emission and Faraday rotation in galaxies at high redshift. Magnetic fields existed already in Quasi Stellar Objects (QSOs) at epochs with redshifts of at least $z \approx 8$ and in starburst galaxies at redshifts of at least $z \approx 4$, but the earliest magnetic fields are yet to be discovered (section 5).

For further reading, the books by Ruzmaikin et al. (1988), Wielebinski & Beck (2005), Rüdiger et al. (2013), Klein & Fletcher (2015), Lazarian et al. (2015), Kronberg (2016), and Shukurov & Subramanian (2022) are recommended.

## 2. Observational methods

As the methods of measuring of magnetic fields have been discussed widely in the literature, a short summary of the methods clarifies the present limitations.

### 2.1 Optical and infrared polarization

Elongated dust grains can be oriented with their major axis perpendicular to the field lines by paramagnetic alignment (Davis & Greenstein 1951) or, more efficiently, by radiative torque alignment (Hoang & Lazarian 2008; 2014). When particles on the line of sight to a star are oriented with their major axis perpendicular to the line of sight (and the field is oriented in the same plane), the different levels of extinction along the major and the minor axis leads to polarization of the starlight, with the polarization orientation pointing parallel to the field *orientation*, while the field *direction* remains unknown. This is the basis to measure magnetic fields with optical and near-infrared polarimetry by observing individual stars or diffuse starlight. Extinction is most efficient for grains of sizes similar to the wavelength. These small particles are well aligned in the medium between molecular clouds, but less efficiently in the clouds themselves (Cho & Lazarian 2005).

The detailed physics of the alignment is complicated and depends on the magnetic properties of the particles. The degree of polarization p (in optical magnitudes) due to a volume element along the line of sight $\delta L$ is given by Ellis & Axon (1978):

$$p = \frac{K\, B_{\perp}^{2}\, \zeta\, \delta L}{N_H\, T_g\, T^{1/2}}$$

where $T$ is the gas temperature,
$T_g$ the grain temperature,
$N_H$ the gas density,
$\zeta$ the space density of grains,
$B_{\perp}$ the magnetic field strength perpendicular to the line of sight.

Optical light can also be polarized by scattering, a process unrelated to magnetic fields. This contamination is small when observing stars, but needs to be subtracted from diffuse light, requiring multi-colour measurements, which has not been attempted so far.

In the far-infrared (FIR) and sub-mm wavelength ranges, the emission of elongated dust grains is intrinsically polarized and scattered light is negligible. As the grains are aligned perpendicular to the magnetic field lines, the polarization orientation points perpendicular to the field orientation. FIR polarimetry probes dust particles in the warmer parts of molecular clouds (section 4.1), while sub-mm polarimetry probes grains with large sizes that are aligned in the densest and coldest regions. The field strength can be crudely estimated from the velocity dispersion of the molecular gas along the line of sight and the dispersion of the polarization angles in the sky plane, the *Davis-Chandrasekhar-Fermi method* (Davis 1951, Chandrasekhar & Fermi 1953), further developed for the case of a mixture of large-scale and turbulent fields (Hildebrand et al. 2009; Houde et al. 2009), and for compressible ISM gas by Skalidis & Tassis (2021). An uncertainty by a factor of at least two remains (Crutcher 2012).

## 2.2 **Synchrotron emission**

Charged particles moving at relativistic speeds (cosmic rays) around magnetic fields lines on spiral trajectories generate electromagnetic waves. The diffuse radio emission from the Milky Way emerges mostly from cosmic-ray electrons in interstellar magnetic fields (Fermi 1949; Kiepenheuer 1950). For particles with a continuous power spectrum of energies, the effective contribution at a given frequency comes from electrons with energy E (in GeV) emitting in a magnetic field with a component perpendicular to the line of sight of strength $B_\perp$ (in μG) (Webber 1980):

$$\nu \approx 16\ \mathrm{MHz}\ E^2\ B_\perp$$

where $B_\perp$ is the strength of the magnetic field component perpendicular to the line of sight.
The half-power lifetime of the observable synchrotron-emitting cosmic-ray electrons (CREs) is:

$$t_{syn} \approx 8.35\ 10^9\ \mathrm{yrs}\ \ B_\perp^{-2}\ E^{-1}\ \approx 1.06\ 10^9\ \mathrm{yrs}\ \ B_\perp^{-1.5}\ \nu^{-0.5}$$

where $B_\perp$ is measured in μG, E in GeV, and $\nu$ in GHz.

The emissivity $\varepsilon$ from cosmic-ray electrons with a power-law energy spectrum in a volume with a magnetic field strength $B_\perp$ is given by:

$$\varepsilon \sim N_0\ \nu^{(\gamma+1)/2}\ B_\perp^{(1-\gamma)/2}$$

where $\nu$ is the frequency,
$N_0$ the density of cosmic-ray electrons per energy interval,
$\gamma$ the spectral index of the power-law energy spectrum of the cosmic-ray electrons ($\gamma \approx -2.8$ for typical spectra in the interstellar medium of galaxies).

A source of size L along the pathlength has the intensity:

$$I_\nu \sim N_0\ B_\perp^{(1-\gamma)/2}\ L$$

A power-law energy spectrum of the cosmic-ray electrons with the spectral index $\gamma$ leads to a power-law synchrotron spectrum $I \sim \nu^\alpha$ with the spectral index $\alpha = (\gamma + 1)/2$. The initial spectrum of young particles injected by supernova remnants with $\gamma_0 \approx -2.2$ (Caprioli 2011) leads to an initial synchrotron spectrum with $\alpha_0 \approx -0.6$. These particles are released into the interstellar medium. A stationary energy spectrum with continuous injection and dominating synchrotron loss has $\gamma = (\gamma_0 - 1) \approx -3.2$ and $\alpha = (\alpha_0 - 0.5) \approx -1.1$. If the cosmic-ray electrons escape from the galaxy faster than within the synchrotron loss time, the initial spectrum is observed. The diffusion coefficient D of high-energy cosmic-ray electrons with energies E larger than about 5 GeV depends on energy ($D = D_0\ (E/E_0)^\delta$, where $\delta$ is the exponent of the energy dependence of the electron diffusion coefficient, $\delta \approx 0.3 - 0.6$, Blasi & Amato 2012), so that the spectral slope becomes $\gamma = (\gamma_0 - \delta) \approx -2.8$ and $\alpha = (\alpha_0 - \delta/2) \approx -0.9$. The synchrotron spectrum can be flattened at low radio frequencies by absorption in ionized gas clouds or by losses of CREs due to ionization of neutral gas, or steepened at high radio frequencies by synchrotron or inverse Compton losses of CREs (Basu et al. 2015a).

The energy densities of cosmic rays (mostly relativistic protons + electrons), of magnetic fields and of turbulent gas motions, averaged over a large volume of the interstellar medium and averaged over time, are comparable *(energy equipartition)*:

$$W_{cr} \approx \frac{B^2}{8\pi} \approx \frac{\rho\, v^2}{2}$$

where $W_{cr}$ is the total energy density of cosmic rays,
$B^2 / 8\pi$ the total energy density of the magnetic field,
$\rho\, v^2/ 2$ the energy density of turbulent gas motions with density $\rho$ and velocity dispersion v.

On spatial scales smaller than the diffusion length of cosmic-ray electrons (typically a few 100 pc) or on time scales smaller than the acceleration time of cosmic rays (typically a few million years), energy equipartition is not valid. For a detailed discussion on the validity of the equipartition assumption, see Seta & Beck (2019).

Equipartition between the total energy densities of cosmic rays and magnetic fields allows us to estimate the total magnetic field strength:

$$B_{eq} \sim ((k+1)\, I_\nu / L)^{2/(5-\gamma)}$$

This revised formula by Beck & Krause (2005) (further improved by Arbutina et al. 2012) is based on integrating the energy spectrum of the cosmic-ray protons and assuming a ratio k between the number densities of protons and electrons in the relevant energy range. The revised formula may lead to significantly different field strengths than the classical textbook formula, which is based on integration over the radio frequency spectrum. Note that the exponent of 2/7 given in the minimum-energy formula in several textbooks is valid only for $\gamma$ = -2. The widely used *minimum-energy* estimate of the field strength is smaller than $B_{eq}$ by the factor $((1-\gamma) / 4)^{2/(5-\gamma)}$, hence equal to $B_{eq}$ for $\gamma \approx -3$. For cosmic-ray spectra flatter than $\gamma$ = -2, the above formula is not valid, and the integration over the energy spectrum of the cosmic rays to obtain $W_{cr}$ has to be restricted to a limited energy interval.

For electromagnetic mechanisms of particle acceleration, the cosmic-ray proton/electron number density ratio k for GeV particles is ≈ 40 – 100, as a result of their different rest masses (Bell 1978). If energy losses of the electrons are significant, e.g. in strong magnetic fields or far away from their places of origin, the spectral index of the cosmic-ray electron spectrum becomes steeper than that of the cosmic-ray protons, so that k increases and the equipartition estimate is a lower limit of the true field strength. Due to the small exponent in the above equipartition formula, the dependence on the input parameters is weak, so that even large uncertainties do not affect the result much. On the other hand, the nonlinear relation between $I_\nu$ and $B_\perp$ may lead to an overestimate of the true field strength when using the equipartition estimate if strong fluctuations in $B_\perp$ occur within the observed volume. Another systematic error occurs if only a small volume of the source is filled with magnetic fields.

In spite of these uncertainties, the equipartition assumption provides reasonable estimates. Independent measurements of the field strength by the Faraday effect in “magnetic arms” leads to similar values (section 4.2). Furthermore, estimates of the synchrotron loss time based on the equipartition assumption explains well the extent of radio halos around galaxies seen edge-on (see section 4.6). Finally, the magnetic energy density based on the equipartition estimate is similar to that of the turbulent gas motions (Fig. 20), as expected from turbulent field amplification. Note that the equipartition estimate also holds in starburst galaxies where secondary electrons contribute significantly (Lacki & Beck 2013).

Combination of data from radio waves and $\gamma$-rays that are emitted by the same cosmic-ray electrons via bremsstrahlung, also yields field strengths similar to equipartition in the Large Magellanic Cloud (Mao et al. 2012b) and in the starburst region of M82 (Yoast-Hull et al. 2013), but probably not in NGC253 (Yoast-Hull et al. 2014). In our Galaxy the accuracy of the equipartition assumption can be tested directly, because there is independent information about the energy density and spectrum of local cosmic rays from in-situ measurements and from γ-ray data. Combination with the radio synchrotron data yields a local strength of the total field of ≈ 6 μG and ≈ 10 μG in the inner Galaxy. These values are similar to those derived from energy equipartition.

*Linear polarization* is a distinct signature of synchrotron emission. The emission from a single cosmic-ray electron gyrating in magnetic fields is elliptically polarized. An ensemble of electrons shows only very low circular polarization, but strong linear polarization with the orientation of the E-vector normal to the magnetic field orientation. The intrinsic degree of linear polarization p is given by:

$$p_0 = \frac{1-\gamma}{7/3 \; -\gamma}$$

Considering galactic radio emission with $\gamma \approx -2.8$, a maximum linear polarization of $p_0$ = 74% is expected. In realistic observing situations the percentage polarization is reduced due to fluctuations of the magnetic field orientation within the volume traced by the telescope beam (section 2.3) or by Faraday depolarization (section 2.4). The observed degree of polarization is also reduced due to the contribution of unpolarized thermal emission, which may dominate in star-forming regions.

## 2.3 Magnetic field components

*Table 1: Magnetic field components and their observational signatures.*

| Field component | Notation | Geometry | Observational signatures |
|---|---|---|---|
| Total field | $B_{tot}^2 = B_{turb}^2 + B_{reg}^2$ | 3D | Total synchrotron intensity, corrected for inclination |
| Total field perpendicular to the line of sight | $B_{tot,\perp}^2 = B_{turb,\perp}^2 + B_{reg,\perp}^2$ | 2D | Total synchrotron intensity |
| Turbulent or tangled field (a) | $B_{turb}^2 = B_{iso}^2 + B_{aniso}^2$ | 3D | Total synchrotron emission, partly polarized, corrected for inclination |
| Isotropic turbulent or tangled field perpendicular to the line of sight | $B_{iso,\perp}$ | 2D | Unpolarized synchrotron intensity, beam depolarization; Faraday depolarization |
| Isotropic turbulent or tangled field along line of sight | $B_{iso,\parallel}$ | 1D | Faraday depolarization |
| Ordered field perpendicular to the line of sight | $B_{ord,\perp}^2 = B_{aniso,\perp}^2 + B_{reg,\perp}^2$ | 2D | Intensity and orientation of radio, optical, IR, or sub-mm polarization |
| Anisotropic turbulent or tangled regular field perpendicular to the line of sight (b) | $B_{aniso,\perp}$ | 2D | Intensity and orientation of radio, optical, IR, or sub-mm polarization; Faraday depolarization |
| Regular field perpendicular to the line of sight (b) | $B_{reg,\perp}$ | 2D | Intensity and orientation of radio, optical, IR, or sub-mm polarization; Goldreich-Kylafis effect |
| Regular field along line of sight | $B_{reg,\parallel}$ | 1D | Faraday rotation and depolari-zation; longitudinal Zeeman effect |

Notes: (a) Turbulent fields and fields tangled on small scales by small-scale gas motions have different power spectra that can possibly be distinguished by future observations with high spatial resolution. (b) Anisotropic turbulent fields, tangled regular fields, and large-scale regular fields cannot be distinguished by polarization observations with limited resolution; Faraday rotation helps here.

The intensity of synchrotron emission is a measure of the number density of cosmic-ray electrons in the relevant energy range and of the strength of the *total magnetic field* component in the sky plane. Polarized emission emerges from *ordered fields*. As polarization angles are ambiguous by multiples of 180°, they cannot be used to distinguish *regular fields* with a well-defined direction within the telescope beam from *anisotropic turbulent fields* (also called *ordered random fields* by Jaffe 2019 or *striated fields* by Jansson & Farrar 2012a). The latter are generated from isotropic turbulent magnetic fields by compression or shearing gas flows and have a preferred orientation, but frequently reverse their direction on scales smaller than the telescope beam. Anisotropic fields increase the diffusion coefficient of cosmic-ray electrons and hence decrease their residence time in the galactic disk (Giacinti et al. 2018). Unpolarized synchrotron emission is a signature of *isotropic turbulent fields* with random directions that are amplified and tangled by turbulent gas flows.

Magnetic fields preserve their direction over the *coherence scale*, i.e. the maximum scale of field tangling or of turbulence. The isotropic turbulent field $B_{turb}$ can be described by cells with the size of the maximum turbulence scale. If N is the number of cells within the volume observed by the telescope beam, wavelength-independent *beam depolarization* occurs, which in the case of constant coherence scale (Sokoloff et al. 1998) can be estimated as:

$$DP = p/p_0 = N^{-1/2}$$

If the medium is pervaded by an isotropic turbulent field $B_{turb}$ (unresolved field with randomly changing directions) plus an ordered field $B_{ord}$ (regular and/or anisotropic turbulent) with a constant orientation in the volume observed by the telescope beam, it follows for constant density of cosmic-ray electrons:

$$DP = 1 / (1 + q^2)$$

while for the equipartition case (Sokoloff et al. 1998):

$$DP = (1 + 3.5\, q^2) / (1 + 4.5\, q^2 + 2.5\, q^4)$$

where $q = B_{iso,\perp} / B_{ord,\perp}$ (the components in the sky plane, with $B_{iso,\perp} = B_{iso} \sqrt{2/3}$). The latter case gives larger DP values (i.e. less depolarization) than for the former case.

## 2.4 **Faraday rotation and Faraday depolarization**

The linearly polarized radio wave is rotated by the Faraday effect in the passage through a magneto-ionic medium (Fig. 1). Classically, a linear relation between the rotation $\Delta\psi$ of the polarization angle $\psi$ and the square of the wavelength $\lambda$ is assumed:

$$\Delta\psi = RM\, \lambda^2$$

The *Rotation Measure* RM (in rad $m^{-2}$, when $\lambda$ in m) can be estimated under simple conditions by:

$$RM = 0.812 \langle n_e B_{\|} \rangle L$$

with $n_e$ the thermal electron density (in $cm^{-3}$),
$B_{\|}$ the mean strength of the magnetic field component along the line of sight (in μG),
$L$ the total pathlength from the source to the observer (in parsec).

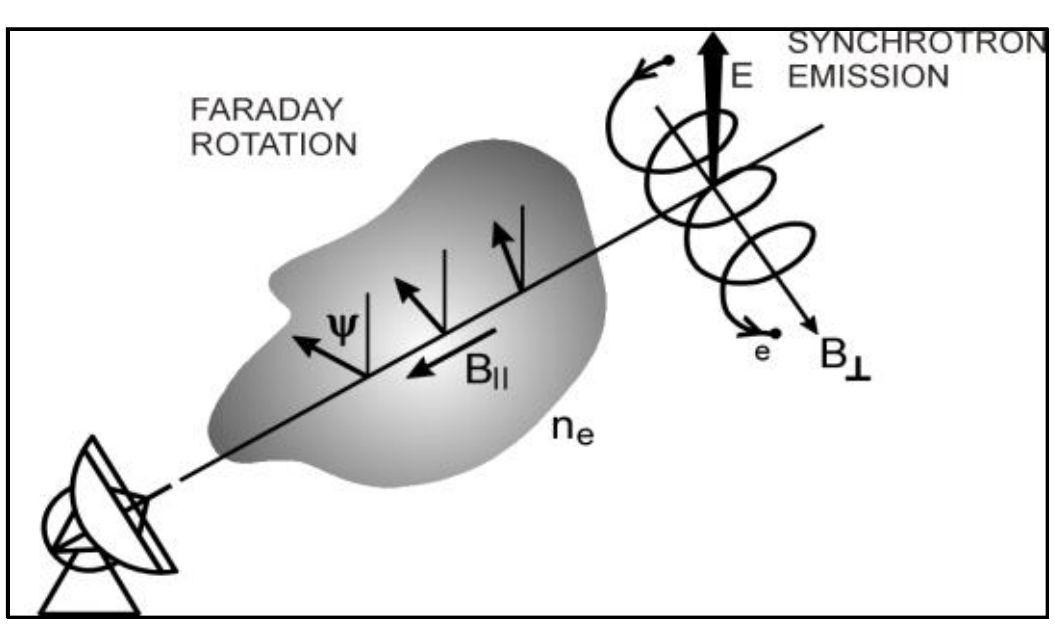

*Fig. 1: Synchrotron emission and Faraday rotation.*

For typical plasma densities and regular field strengths in the interstellar medium, Faraday rotation becomes significant at wavelengths larger than a few cm. In the central regions of galaxies, Faraday rotation is strong already at 1 − 3 cm wavelength.

By definition, RM is positive when the averaged line-of-sight component of the magnetic field (weighted by the density of thermal electrons) points towards the observer. As the Faraday rotation angle is sensitive to the sign of the field direction, only regular fields give rise to Faraday rotation, while the Faraday rotation of turbulent fields cancels out. Measurements of the Faraday rotation angle from multi-wavelength observations allow determination the average strength and direction of the regular field component along the line of sight. Its combination with the total intensity and the polarization orientations yields in principle a three-dimensional picture of galactic magnetic fields and the three basic field components: regular, anisotropic turbulent, and isotropic turbulent (Table 1).

The quantity $\langle n_e B_{\parallel} \rangle$ is the average of the product ($n_e B_{\parallel}$) along the line of sight, which generally is not equal to the product of the averages $\langle n_e \rangle \langle B_{\parallel} \rangle$ if fluctuations in $n_e$ and $B_{\parallel}$ are correlated or anticorrelated. As a consequence, the field strength $\langle B_{\parallel} \rangle$ determined from RM is uncertain even if additional information about $\langle n_e \rangle$ is available, e.g. from pulsar dispersion measures (section 3.3.2).

RM is a useful quantity only in the rare cases when polarization angle $\psi$ is a linear function of $\lambda^2$. if several emitting and Faraday-rotating sources are located within the volume traced by the telescope beam, $\psi$ is no longer a linear function of $\lambda^2$. For such cases, the physical quantity *Faraday Depth (FD)* (Burn 1966) needs to be computed by integration along the pathlength for each emitting source:

$$FD = \int n_e B_{\parallel} \, dl$$

If the rotating region is located in front of the emitting region (a *Faraday screen*), the numerical values of RM and FD are identical. In case of a single emitting and rotating region with a symmetric magnetic field profile and negligible Faraday depolarization (see below), RM is half of the FD value.

The combination of several emitting and rotating sources along the line of sight and/or within the telescope beam leads to a complex variation of $\psi$ with $\lambda^2$. To retrieve information on the behavior of the polarization angle $\psi$ over a broad range of $\lambda^2$, broad-band spectro-polarimetric observations are needed, so that *RM Synthesis* can be applied (Brentjens & de Bruyn 2005). RM Synthesis Fourier-transforms the complex polarization that provides a *Faraday spectrum* that gives the intensity of polarized emission (and its intrinsic polarization angle) as a function of FD. Imaging a sky region provides a data cube, the *Faraday cube*. The total span and the distribution of the frequency channels of the observations in $\lambda^2$ space define the resolution in the Faraday spectrum, given by the width of the *Rotation Measure Spread Function (RMSF)*. This allows cleaning of the Faraday spectrum, similar to cleaning of synthesis data from interferometric telescopes (Heald 2015). Each emitting source located along the line of sight has its specific FD in the Faraday spectrum. RM Synthesis is able to separate FD components, which are generated by several emitting and rotating regions within the telescope beam and/or the line of sight and are signatures of the 3D structure of the magnetized medium. If the RMSF is sufficiently narrow, magnetic field reversals and turbulent fields can be identified in the Faraday spectrum (Frick et al. 2011; Bell et al. 2011; Beck et al. 2012). However, even simplified models of galaxies reveal complicated Faraday spectra (Ideguchi et al. 2014). Still, the shape of the Faraday spectrum can be used to extract global parameters of the magnetic field (Ideguchi et al. 2017; Eguchi et al. 2020). Turbulent magnetic fields lead to many spikes in the Faraday spectrum that further complicate the interpretation (Basu et al. 2019).

Deviations from a linear $\lambda^2$ dependence of $\psi$ also occur in case of significant *Faraday depolarization*. In a region containing cosmic-ray electrons, thermal electrons, and purely regular magnetic fields, wavelength-dependent Faraday depolarization occurs because the polarization angles from the far side of the emitting layer are more rotated than those from the near side. This effect is called *differential Faraday rotation.* For the simplest case of one single layer with a symmetric distribution of thermal electron density and field strength along the line of sight (Burn 1966; Sokoloff et al. 1998):

$$DP = p/p_0 = | \sin (2 \, RM \, \lambda^2) / (2 \, RM \, \lambda^2) |$$

where RM is the observed rotation measure. RM is half of the Faraday depth FD through the whole layer. DP varies periodically with $\lambda^2$. With |RM| = 100 rad $m^{-2}$, typical for normal galaxies, DP has zero points at wavelengths of ($12.5 \sqrt{n}$) cm where n = 1, 2, … (Fig. 2).

At each zero point of the curves in Fig. 2 the observed polarization angle jumps by 90°. Observing at a fixed wavelength hits zero points at certain values of the intrinsic RM, giving rise to *depolarization canals* (Haverkorn et al. 2004). At wavelengths below that of the first zero point in DP, only the central layer of the emitting region is observed, because the emission from the far side and that from the near side cancel because their Faraday rotation angles differ by 90°. Beyond the first zero point, only a small layer on the near side of the disk remains visible.

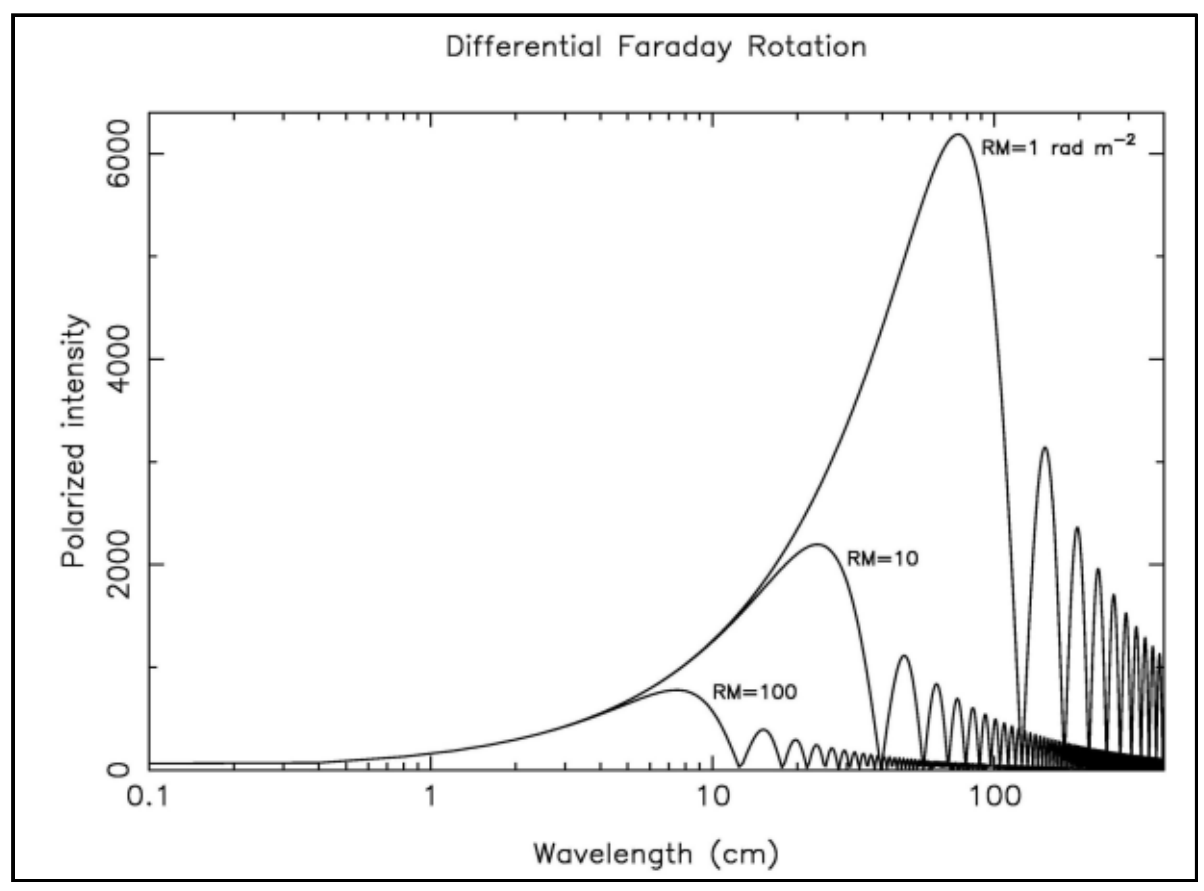

*Fig. 2: Spectrum of polarized emission (in arbitrary units) for a simple source with an intrinsic spectral index of $\alpha$ = -0.9 and depolarization by differential Faraday rotation (Arshakian & Beck 2011).*

Turbulent fields cause wavelength-dependent depolarization, called *Faraday dispersion* (Sokoloff et al. 1998). For an emitting and Faraday-rotating region (*internal* dispersion):

$$DP = p/p_0 = (1 - \exp(-S)) / S$$

where $S = 2\, \sigma_{RM}^2\, \lambda^4$ is valid at cm and dm wavelengths. $\sigma_{RM}^2$ is the dispersion in rotation measure, depending on the turbulent field strength along the line of sight, the maximum turbulence scale, the thermal electron density, and the pathlength through the medium. The main effect of Faraday dispersion is that the interstellar medium becomes “Faraday thick” for polarized radio emission beyond a wavelength, depending on $\sigma_{RM}$ (Fig. 3), and only a front layer remains visible in polarized intensity. Galaxy halos and intracluster media have typical values of $\sigma_{RM}$ = 1 – 10 rad $m^{-2}$, while for galaxy disks typical values are $\sigma_{RM}$ = 10 – 100 rad $m^{-2}$. Central regions of galaxies can have higher dispersions. Fig. 3 shows the optimum wavelength ranges to detect polarized emission for regions with different $\sigma_{RM}$.

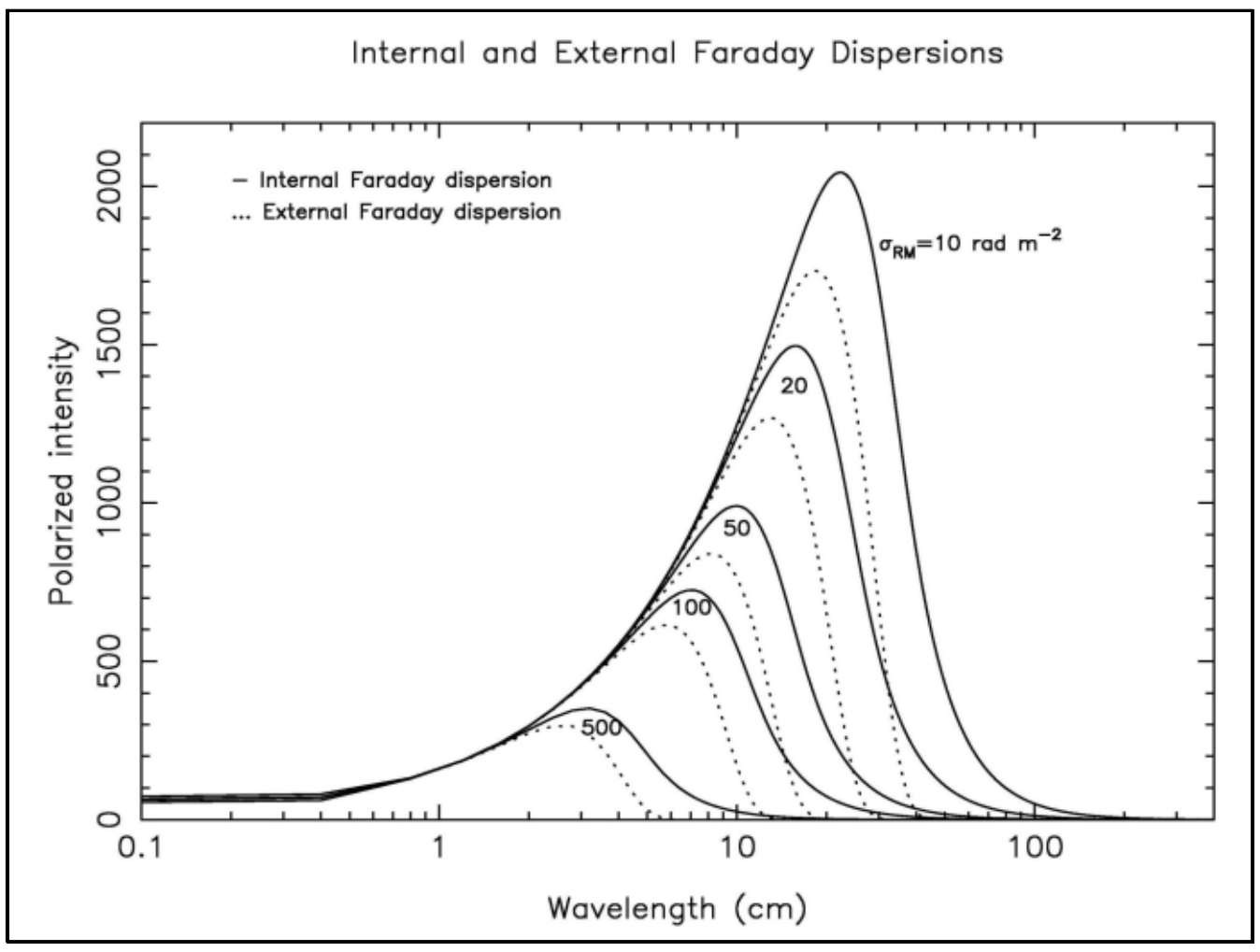

*Fig. 3: Spectrum of polarized emission (in arbitrary units) for a source with an intrinsic spectral index $\alpha$ = -0.9 and depolarization by Faraday dispersion parameterized by $\sigma_{RM}$. Solid curve: internal Faraday dispersion within an emitting source; dotted curve: external Faraday dispersion in a non-emitting object in the foreground (Arshakian & Beck 2011).*

Regular fields in a non-emitting foreground *Faraday screen* do not depolarize, while turbulent fields do (*external* Faraday dispersion), if the source has an angular extent larger than the telescope beam:

$$DP = \exp(-S)$$

At long wavelengths (small DP), the wavelength dependence of S becomes $S = \sigma_{RM}\, \lambda^2$ (Tribble 1991).

Finally, unresolved *RM gradients* within the beam also lead to depolarization (Sokoloff et al. 1998).

Wavelength-dependent Faraday depolarization can be classified as *depth depolarization* (differential Faraday rotation, Faraday dispersion along the line of sight) and *beam depolarization* (RM gradients, Faraday dispersion in the sky plane). Both types occur in emitting regions, while in non-emitting Faraday screens only beam depolarization occurs. A model including both types of depolarization in a multi-layer magneto-ionic medium was presented by Shneider et al. (2014) and Kierdorf et al. (2020).

## 2.5 **Zeeman effect**

The Zeeman effect is the most direct method of remote sensing of magnetic fields. It is used in optical astronomy since the first detection of magnetic fields in sunspots of the Sun. The radio detection was first made in the HI line. In the presence of a regular magnetic field $B_{reg,\|}$ along the line of sight, the spectral line at the frequency $\nu_0$ is split into two components (*longitudinal Zeeman effect,* Fig. 4). The two components are circularly polarized of the opposite sign. The frequency shift is minute, e.g. 2.8 MHz/Gauss for the HI line. More recent observation of the OH or $H_2O$ lines used the larger frequency shifts of these molecular line species (e.g. Heiles & Crutcher 2005; McBride & Heiles 2013).

In magnetic fields perpendicular to the line of sight, two shifted lines together with the main unshifted line are all linearly polarized. This *transversal Zeeman effect* is much more difficult to observe because for unresolved and symmetric lines no net polarization is observed. Detection of linearly polarized lines becomes possible for unequal populations of the different sublevels, a gradient in optical depth or velocity, or an anisotropic velocity field, called the *Goldreich-Kylafis effect* (Goldreich & Kylafis 1981). The orientation of linear polarization can be parallel or perpendicular to the magnetic field orientation. The effect was detected in molecular clouds, star-forming regions, outflows of young stellar systems and supernova remnants of the Milky Way, and in the ISM of the galaxy M33 (Li & Henning 2011).

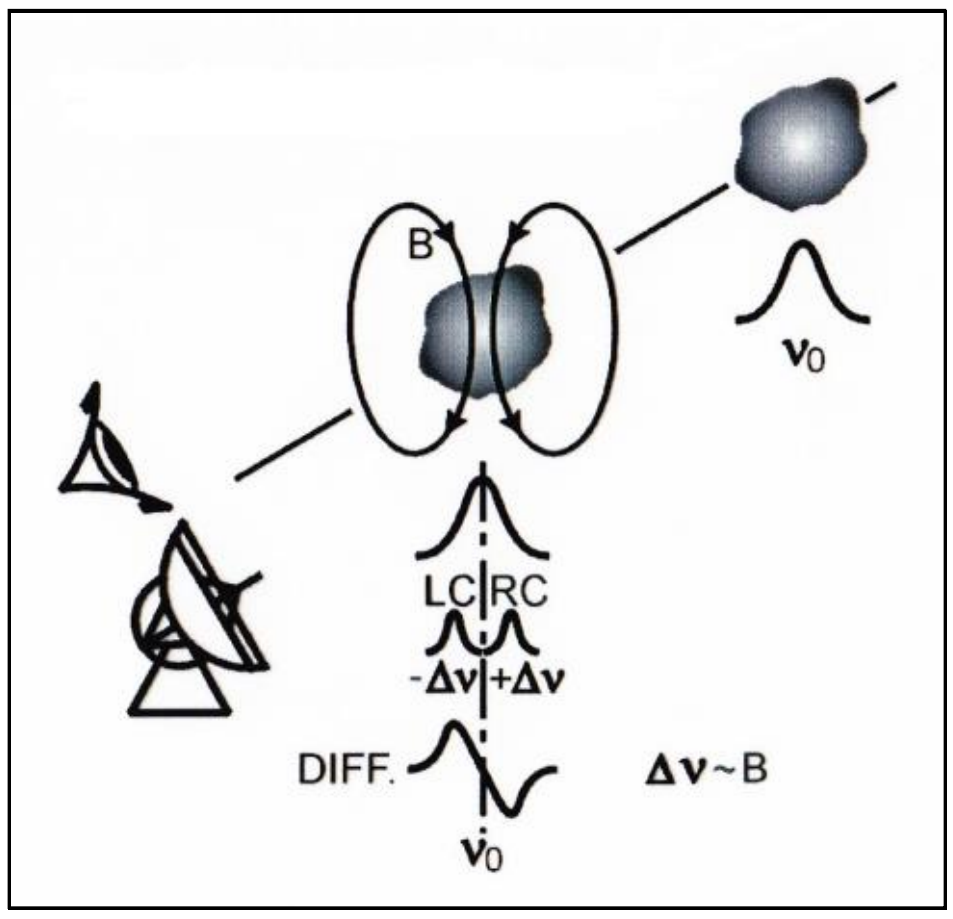

*Fig. 4: Longitudinal Zeeman effect: splitting into two components with opposite circular polarization.*

## 2.6 **Velocity Gradient Technique (VGT)**

Gradients of turbulent velocities in the spectral lines of atomic or molecular gas are dominantly perpendicular to the local magnetic field. Therefore, measuring the velocity gradients yields the orientations of the magnetic fields in the sky plane (Yuen & Lazarian 2017; Hu et al. 2019; 2022).

## 2.7 **Field origin and amplification**

The origin of the first magnetic fields in the Universe is still a mystery. The generation of the very first "seed" fields needs separation of electric charges. A large-scale intergalactic field of ≤ $10^{-12}$ G may be generated in the early Universe (Durrer & Neronov 2013). A lower limit of ≥ $10^{-16}$ G was derived from high-energy $\gamma$-ray observations with HESS and FERMI, assuming that the secondary particles are deflected by the intergalactic fields (Neronov & Vovk 2010). Strong primordial fields would affect the formation of the first galaxies and hence the reionization of the intergalactic medium (IGM). The observation that the IGM was mostly ionized at a redshift of about 7 leads to an upper limit of 2 – 3 nG (Schleicher & Miniati 2011). Analysis of the CMB power spectra from PLANCK data gave an upper limit of about 5 nG comoving strength on a scale of 1 Mpc (Ade et al. 2016a). Faraday rotation differences between close pairs of extragalactic radio sources places an upper limit of 4 nG co-moving strength on Mpc scales (O'Sullivan et al. 2020). RM data from LOFAR indicated typical field strengths in intergalactic filaments of 10 – 20 nG (Carretti et al. 2023).

Models of field amplification in galaxies from weak, large-scale primordial seed fields from early cosmological epochs (see Widrow 2002 for a review) are not supported by the data. A large-scale seed field wound up in a differentially rotating galaxy (Sofue et al. 1986) can generate only the even bisymmetric mode (S1) or the odd dipolar mode (A0), but for both modes there is no convincing evidence so far (see Tables A4-A6 in the Appendix). Furthermore, a large-scale primordial field is hard to maintain because galaxies rotate differentially, so that field lines get strongly wound up and are destroyed by turbulent diffusion, in contrast to the observations (Shukurov 2005). For example, the large-scale regular field observed in M31 (Fig. 29), cannot be explained by a primordial field model. Similar issues occur with kinematical models of field amplification by shearing and compressing gas flows that can generate fields with a coherence length of less than a kiloparsec and frequent reversals.

A seed field of ≤ $10^{-12}$ G could be generated also in protogalaxies, e.g. by the *Weibel instability* (Lazar et al. 2009) or fluctuations in the intergalactic thermal plasma after the onset of reionization (Schlickeiser 2012; Schlickeiser & Felten 2013). Magnetization of protogalaxies to ≥ $10^{-9}$ G could also be achieved by field ejection from the first stars or the first black holes (Fig. 5) or the first supernova remnants (Hanayama et al. 2005*)*, followed by dynamo action.

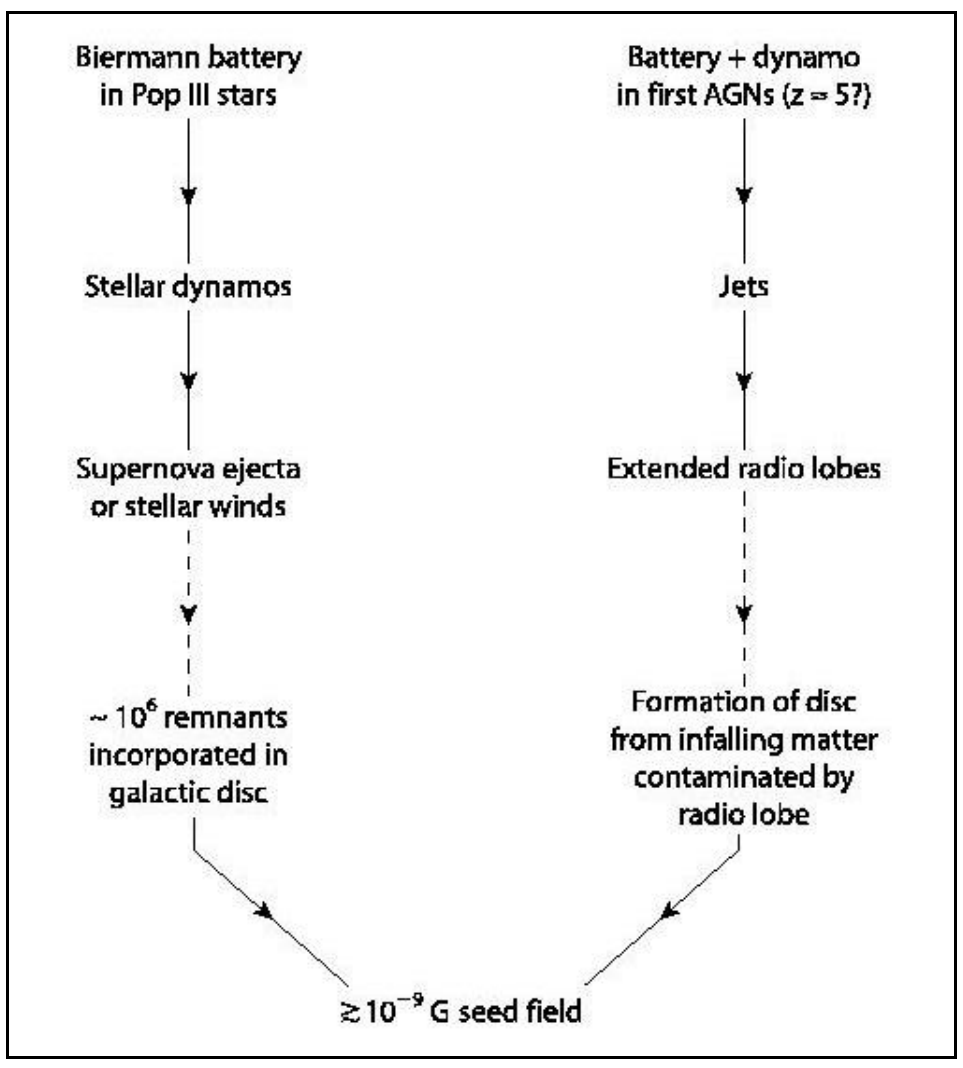

*Fig. 5: Origin of seed fields in protogalaxies (Rees 2005).*

The dynamo transfers mechanical into magnetic energy. It amplifies and orders a seed field. *The small-scale* (or *fluctuation dynamo*) does not need general rotation, only turbulent gas motions (Federrath 2016; Seta & Federrath 2022; Gent et al. 2023). The source of turbulence can be thermal virialization in protogalactic halos or supernovae in the disk or the *magneto-rotational instability (MRI)* (Rüdiger et al. 2013). Within a few $10^8$ yr, weak seed fields are exponentially amplified to about 10% of the energy density level of kinetic turbulence, reaching strengths of a few µG (Kulsrud et al. 1997; Schleicher et al. 2010; Schober et al. 2012; A. Beck et al. 2012; Machida et al. 2013; Rieder &

Teyssier 2016). However, due to the small viscosity and resistivity of the warm ionized gas, the magnetic Reynolds number and the magnetic Prandtl number are very large (Ferrière, 2020), much larger than what can be reached in numerical simulations, so that the amplification rate and possibly the saturation level of the small-scale field will be underestimated (Schober et al. 2012; Liu et al. 2022). In the cold gas, the dynamo is slower but field can attain a saturation level of around 100% (Gent et al. 2023). – In addition to the fluctuation dynamo, small-scale fields can also be generated and amplified by tangling of regular fields (Seta & Federrath 2020), but the role is still unclear.

*The large-scale dynamo* (or *$\alpha$-$\Omega$ dynamo* or *mean-field dynamo*) is driven by helical turbulence ($\alpha$), turbulent as motions twisted by the Coriolis force, differential rotation ($\Omega$), and magnetic diffusivity ($\eta$) (e.g. Parker 1979; Ruzmaikin et al. 1988; Beck et al. 1996; 2019). Turbulence can be excited by supernova explosions or cosmic-ray driven Parker loops. The “mean” quantities $\alpha$ and $\eta$ from the small-scale properties of the interstellar medium, which needs numerical modelling (see below). A large-scale regular field is generated from a turbulent field within a few $10^9$ yr in a typical spiral galaxy. If the small-scale dynamo already amplified turbulent fields to a few μG in protogalaxies, the large-scale dynamo is needed only for the organization of the field (“order out of chaos”). The field pattern is described by modes of order *m* in azimuthal symmetry in the galaxy plane and vertical symmetry (S) or antisymmetry (A) perpendicular to the plane. Several modes can be excited simultaneously.

In almost spherical, rotating bodies, like stars, planets, or galaxy halos, the strongest mode consists of a toroidal field component with a sign reversal across the equatorial plane (vertically antisymmetric or odd-symmetry mode A0 with azimuthal axisymmetry in the plane) and a poloidal field component of odd symmetry with field lines crossing the equatorial plane, a *dipolar* field (Fig. 6 top). The halo field can be oscillatory and reverse its symmetry on a long timescale. In flat, rotating objects like galaxy disks, the $\alpha$-$\Omega$ dynamo dominates. The strongest mode consists of a toroidal field component that is symmetric with respect to the galaxy plane and has the azimuthal symmetry of an axisymmetric spiral (m = 0) in the disk without a sign reversal across the plane (vertically symmetric or even-symmetry mode S0) and a weaker poloidal field component of even symmetry, also without a reversal of the vertical field component across the plane, a *quadrupolar* field (Fig. 6 bottom). The next higher azimuthal mode, the bisymmetric spiral (m = 1) has two reversals along azimuthal angle in the plane, followed by more complicated modes. For a forcing with m = 2 symmetry (e.g. a two-arm spiral or a bar), the m = 2 and m = 4 modes dominate over m = 1 (Chamandy et al. 2013).

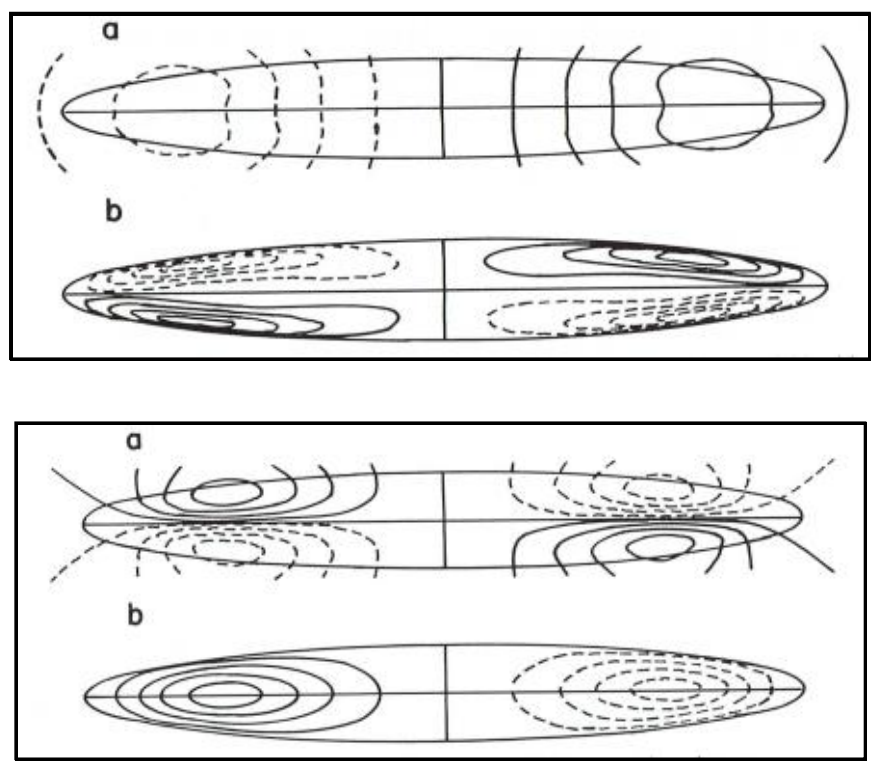

*Fig. 6: Poloidal field lines (a) and contours of toroidal field strength (b) for the simplest version of an odd-symmetry dipolar (top) and an even-symmetry quadrupolar (bottom) dynamo field (Stix 1975). More realistic dynamo fields have many “poles”.*

In principle, the halo and the disk of a galaxy may drive different dynamos and host different field modes. However, there is a tendency of “mode slaving”, especially in case of outflows from the disk into the halo. The more dynamo-active region determines the global symmetry, so that the halo and disk field should have the same symmetry (Moss et al. 2010), as observed (section 4.7).

The ordering time scale of the $\alpha$-$\Omega$ dynamo depends on the size of the galaxy (Arshakian et al. 2009). Very large galaxies may not have had sufficient time to build up a fully coherent regular field and may still host complicated field patterns, as observed in most galaxies. The process of field ordering may also be interrupted by tidal interactions or merging with another galaxy, which may destroy the regular field and significantly delay the development of coherent fields (section 5). Strong star formation as the result of a merger event or mass inflow amplifies the turbulent field and can suppress the α-Ω dynamo in a galaxy if the total star-formation rate is larger than about 20 solar masses per year. Continuous injection of small-scale fields by the small-scale dynamo in star-forming regions may also decelerate the α-Ω dynamo and allow initial field reversals to persist (Moss et al. 2012).

The α-Ω dynamo generates large-scale magnetic helicity with a non-zero mean in each hemisphere of a galaxy that can in principle be measured from observations (Oppermann et al. 2011). As total magnetic helicity is a conserved quantity, small-scale fields with opposite helicity are generated that suppress dynamo action, unless removed from the system (e.g. Vishniac et al. 2003). Hence, an outflow related to star formation in the disk is essential to drive an α-Ω dynamo (Sur et al. 2007; Chamandy et al. 2014). This effect may relate the efficiency of dynamo action to the star-formation rate in the galaxy disk (Rodrigues et al. 2015). For fast outflows the advection time becomes smaller than the dynamo amplification time, and the dynamo is no longer efficient (Chamandy et al. 2015).

Enhanced supply of turbulent magnetic fields by the small-scale dynamo in spiral arms may result in a concentration of large-scale regular magnetic fields between the material arms (Moss et al. 2013; 2015), as observed in many galaxies (section 4.4). A similar result was obtained by the inclusion of fast outflows from spiral arms that can suppress large-scale dynamo action (Chamandy et al. 2015).

There are theoretical issues with dynamo theory (e.g. Vishniac et al. 2003). The “mean-field” approximation is simplified because it assumes a dynamical separation between the small and the large scales. The large-scale field is assumed to be smoothed by turbulent diffusion that requires fast and efficient field reconnection. Still, the predictions of the α-Ω dynamo theory are basically consistent by present-day observations of spiral galaxies (sections 4.4 and 4.7). Dynamo models including outflows with moderate velocities can generate the observed X-shaped fields in the halo (Moss et al. 2010). The predicted pitch angles of the spiral field for the saturated dynamo state are reasonably consistent with the data (Chamandy et al. 2016). However, the quantitative comparison by Van Eck et al. (2015) did not reveal the expected correlations between observable quantities and predictions in most cases. Processes in addition to the dynamo are obviously amplifying and shaping the field.

Numerical high-resolution MHD simulations of dynamos in a box region driven by supernovae (SN) (Gent et al. 2013) supported the concept of the α-Ω dynamo theory. Global MHD models of large-scale dynamo action driven by the buoyancy of cosmic rays in spiral galaxies (Hanasz et al. 2009), in barred galaxies (Kulpa-Dybeł et al. 2011), and in dwarf galaxies (Siejkowski et al. 2014) achieved appreciable similarity to observations, but turbulence and small-scale dynamo action still need to be included.

Most global numerical MHD models of star-forming galaxies are based on ISM turbulence induced by SN shock fronts, accretion shocks, spiral shocks, or the magneto-rotational instability. These models have limited spatial resolution, so that large-scale dynamo action is not fully included (Wissing & Shen 2023). After the small-scale dynamo in the disk has reached saturation, the field strength is relatively low, but further increases by shear. The large-scale dynamo is not required to amplify the field from its primordial value, but it is needed at later evolutionary stages to maintain the field strength by compensating dissipative losses and to build up a large-scale field order.

The resulting scenarios are diverse and partly discrepant: Ponnada et al. (2022) found that the energy densities of magnetic fields, cosmic rays, and thermal gas are similar in the galaxy disk, but not in the halo, while in Machida et al. (2013) and in Wissing & Shen (2023) the magnetic energy density remains always smaller than the thermal one. In Pakmor et al. (2017), the magnetic energy density is always far below the turbulent one. In Wibking & Krumholz (2023) the thermal energy density is always subdominant, whereas the magnetic energy density finally reaches or even exceeds equipartition with the turbulent one, as observed (section 4.2). The strengths of the saturated fields at the end of the simulations are lower than observed in Machida et al. (2013), Steinwandel et al. (2020), Ponnada et al. (2022), and Wissing & Shen (2023), while those in Pakmor et al. (2017), Liu et al. (2022), and Wibking & Krumholz (2023) are consistent with observations. The field patterns in Machida et al. (2013), Rieder & Teyssier (2016), Pakmor et al. (2018), Steinwandel et al. (2020), and Wissing & Shen (2023) show (too) many reversals, also in the azimuthal field component, in contrast to observations (section 4.4.2), while Liu et al. (2022) and Wibking & Krumholz (2023) obtained large-scale azimuthal fields without reversals. Rapid progress in modelling galactic magnetic fields can be expected, triggered by future radio observations with higher resolution and sensitivity (section 5).

## 3. Magnetic fields in the Milky Way

### 3.1 Optical, infrared, and sub-mm polarization

The earliest optical polarization observations in 1949 were interpreted to be due to dust alignment in magnetic fields and hence a tracer of ordered magnetic fields in galaxies. The radio polarization observations (section 3.2) confirmed the magnetic explanation. The first large catalogue of the polarization of stars was made by Behr (1961). This work continued in the southern skies, culminating in an all-sky catalogue of Mathewson & Ford (1970a) with 1800 entries and Axon & Ellis (1976) with 5070 entries, showing a field aligned along the Galactic plane. A homogeneous region of alignment, with high polarization degrees, was seen towards Galactic longitude $l \approx 140°$. Well aligned magnetic field orientations are also seen along the North Polar Spur that extends in to the northern halo from $l \approx 30°$. These early observations were possible for nearby stars, a few at a maximal distance of 4 kpc. A more recent compilation of 9286 stars, collected by Heiles (2000) and discussed by Fosalba et al. (2002) (Fig. 7), included some stars out to ≈ 8 kpc. In view of these distance limitations, it is not possible to model the large-scale magnetic field of the Milky Way. Within a few kpc of the Sun, optical polarimetry combined with high-precision stellar distances from the GAIA catalogue allows magnetic tomography of local gas clouds (Panopoulou et al. 2019). The interaction between the interstellar field and the heliosphere can also be investigated with help of starlight polarization (Frisch et al. 2022).

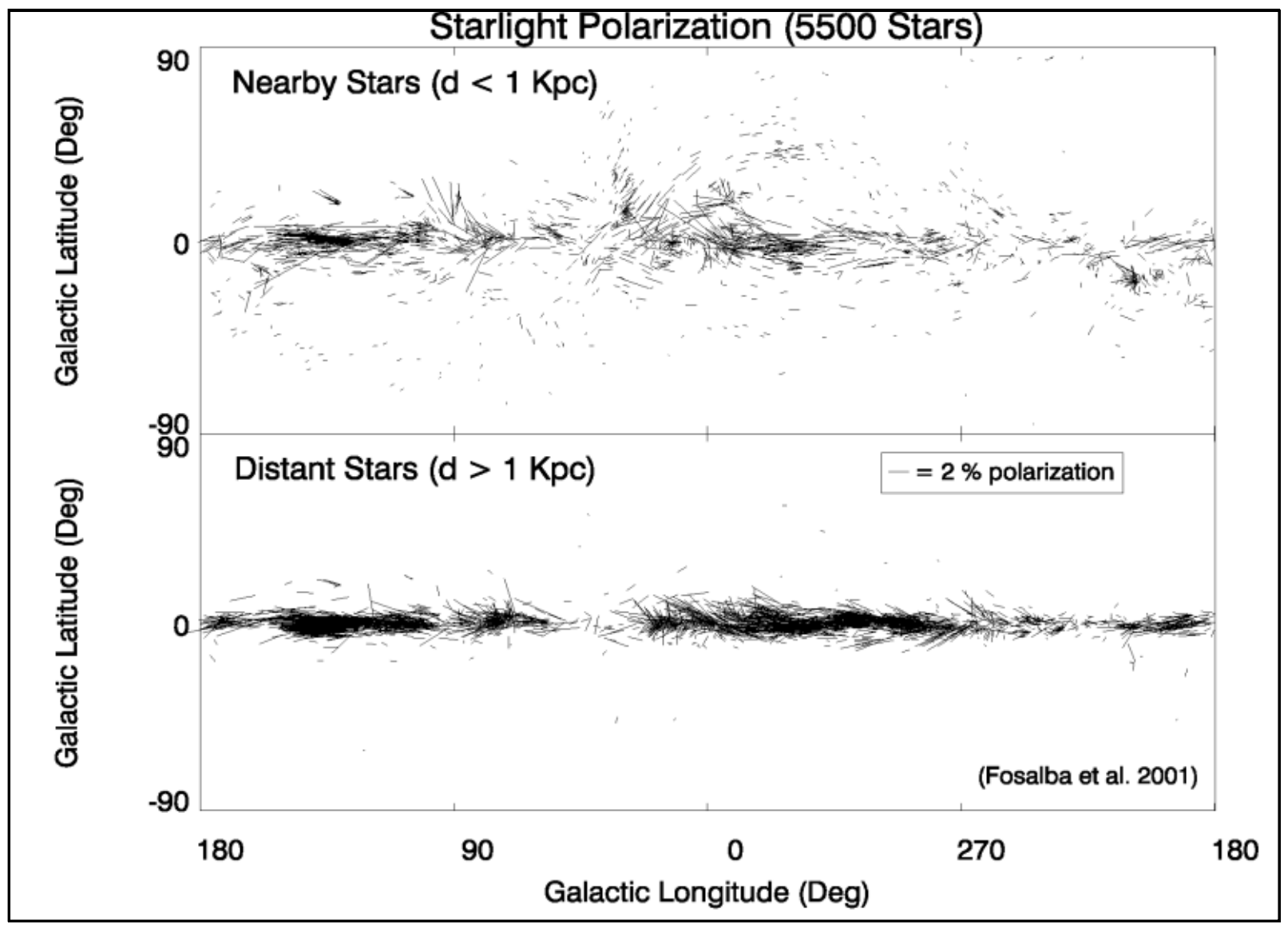

*Fig. 7: Starlight polarization in the Galactic plane for two distance intervals (Fosalba et al. 2002).*

Polarization observations of the diffuse far-infrared or sub-mm emission in the Milky Way trace dense molecular/dust clouds. The PLANCK satellite provided an all-sky map at 353 GHz with 1° resolution, showing much more details in the field structure (Fig. 8). For individual clouds observed with higher resolution, the Davis-Chandrasekhar-Fermi method (section 2.1) gives typical field strengths of a few mG (e.g. Houde et al. 2009), similar to Zeeman measurements of OH or CN lines in other dense clouds (section 3.4). At low column densities, the cloud structures measured with the BLASTPol sub-mm telescope with about 0.6 pc resolution align parallel to the magnetic field, but change to perpendicular orientation at higher column densities (Soler et al. 2017). Recent FIR polarization observations with SOFIA/HAWC+ with 0.04 pc resolution shows a transition back to parallel orientation in the densest gas (Pillai et al. 2020). Interferometric observations with the BIMA sub-mm array with about 0.1 pc resolution reveals hourglass morphologies in the envelopes of dust cores in the molecular clouds associated with ultra-compact HII regions (Tang et al. 2009). The supercritical cores seem to collapse in a subcritical envelope supported by strong magnetic fields, suggesting that ambipolar diffusion plays a key role in the evolution of the cloud. The correlation of the field orientation in the intercloud medium on a scale of several 100 pc, derived from optical polarization, with that in the cloud core on a scale of less than 1 pc, derived from sub-mm polarimetry, further indicates that the fields are strong and preserve their orientation during cloud formation (Li et al. 2009).

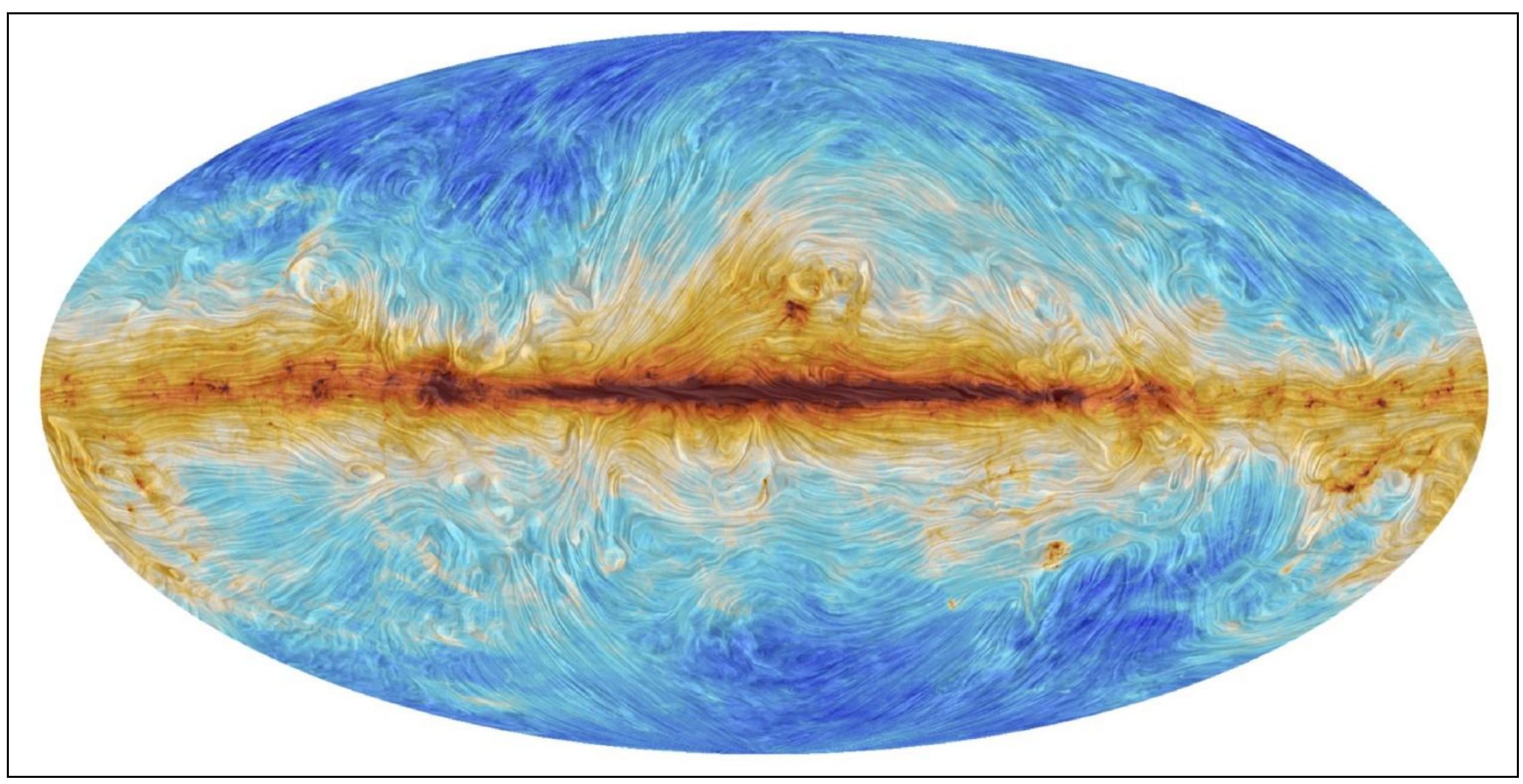

*Fig. 8: All-sky survey of dust polarization at 353 GHz (0.85 mm) observed with PLANCK with 10' resolution. The streamlines (applying the Line Integral Convolution, LIC) show the field orientations in the sky plane, the colours the total intensities at 353 GHz (Adam et al. 2016).*

### 3.2 Radio continuum

3.2.1 All-sky surveys in total intensity

The radio continuum emission of the Milky Way and star-forming galaxies at frequencies below 10 GHz mostly originates from the synchrotron process and hence traces the distribution of magnetic fields and cosmic rays. The contribution of thermal radio emission is generally small, except in bright star-forming regions. At frequencies higher than 10 GHz the thermal emission may dominate locally. At frequencies below about 300 MHz absorption of synchrotron emission by thermal gas can become strong. Hence, the observation of total radio continuum intensity in the frequency range of about 300 MHz − 10 GHz is a perfect method to investigate magnetic fields. Since the observed intensity is the integral from many emission areas along the line of sight, its interpretation is not always simple.

Numerous radio continuum surveys from the early days of radio astronomy (Appendix, Table A1) showed Galactic emission with a maximum towards the Galactic center, a band of emission along the Galactic plane, maxima in the tangential directions of the local spiral arm: Cygnus (Galactic longitude $l$ ≈ 80°) in the northern and Vela ($l$ ≈ 265°) in the southern skies and some “spurs” of emission. In addition, a few strong extragalactic sources were seen -superposed on the Galactic emission.

The analysis of total synchrotron emission gives an equipartition strength of the total field of 6 ± 2 μG in the local neighborhood and 10 ± 3 μG at 3 kpc distance from the Galactic Center (Berkhuijsen, in Beck 2001). The radial exponential scale length of the total field is about 12 kpc. These values are similar to those in external galaxies (section 4.2). Voyager 1 reached interstellar space in 2012 and measured a smooth increase in field strength to 4.8 ± 0.4 μG (Burlaga & Ness 2016). This value is very close to the equipartition estimate above and to those obtained with other methods (section 3.5).

The angular resolution has improved so that at present all-sky surveys with resolution of under 1° are available (see Table A3 of the Appendix). At 1.4 GHz the surveys delineated many extended Galactic sources (HII regions, SNRs) seen along the Galactic plane. Some extragalactic sources like Centaurus A, Virgo A, Cygnus A, and the Magellanic Clouds are also clearly seen in the surveys. Surveys at 45 MHz covered most of the sky with medium angular resolution. At these low frequencies, absorption of the synchrotron emission by ionized gas takes place near the Galactic plane.

The WMAP satellite surveys at frequencies 23 – 94 GHz (Bennett et al. 2003; Hinshaw et al. 2009) gave us a new view of the radio continuum sky at high radio frequencies. At the highest WMAP frequencies thermal emission originating in interstellar dust dominates. An additional component due to spinning dust has been postulated (Draine & Lazarian 1998), to be seen in the 10 – 100 GHz frequency range, and has been confirmed (Dobler et al. 2009) from the WMAP data. The PLANCK satellite provided further improved all-sky surveys between 20 GHz and 100 GHz (Ade et al. 2016b).

To fill the gap between the lower-frequency and the high-frequency data, surveys in the range 2 − 10 GHz with compatible angular resolution are needed. The southern sky has been observed at 2.3 GHz with the Parkes telescope (S-PASS; Carretti et al. 2019). At 4.7 GHz, part of the northern sky has been observed at 5 GHz with a 6.1-m telescope at OVRO (C-BASS, Dickinson et al. 2019); observations of the southern sky with a 7.6-m MeerKAT telescope are underway. – Table A2 in the Appendix lists the all-sky surveys.

### 3.2.2 All-sky surveys in linear polarization

Linear polarization of the continuum emission is a more direct indicator of magnetic fields, because there is no confusing thermal component. However, linear polarization is subject to Faraday effects (section 3.3). After the first detections of polarized Galactic radio waves (Wielebinski et al. 1962) several all-sky or all-hemisphere polarization surveys were made (Appendix, Table A2). The early polarization surveys did not have sufficient angular resolution to elucidate many details and were made at 408 MHz where Faraday effects are considerable. A multi-frequency collection of polarization data for the northern sky was published by Brouw & Spoelstra (1976), albeit not fully sampled.

Major progress was achieved by Wolleben et al. (2006) and Testori et al. (2008), who mapped the whole sky in linear polarization at 1.4 GHz with the 26-m DRAO and 30-m Villa Elisa telescopes with an angular resolution of 36 arcminutes (Fig. 10). Several polarization maxima are seen, e.g. towards the “Fan region” at $l \approx +140°$, $b \approx +10°$, where the line of sight is oriented perpendicular to the local spiral arm. The “North Polar Spur” (NPS) emerges from the Galactic plane at $l \approx 30°$, as well as additional spur-like features that are the results of magnetic fields compressed by expanding supernova remnants. In polarization, the NPS can be followed to the southern sky. Towards the inner Galaxy (Galactic longitude $+90° > l > -90°$, Galactic latitude $|b| < 30°$) strong turbulence in polarized intensity is seen, due to Faraday effects on small scales (section 2.4). Diffuse polarized in the NRAO VLA sky survey (NVSS) has been analyzed by Rudnick & Brown (2009). All-sky polarization data at 23 GHz was published by the WMAP team (Kogut et al. 2007; Hinshaw et al. 2009). There is good agreement between the 23 GHz and the 1.4 GHz polarization maps in the polarization features away from the Galactic plane, but the high-frequency map shows less Faraday depolarization towards the inner Galaxy and near the plane.

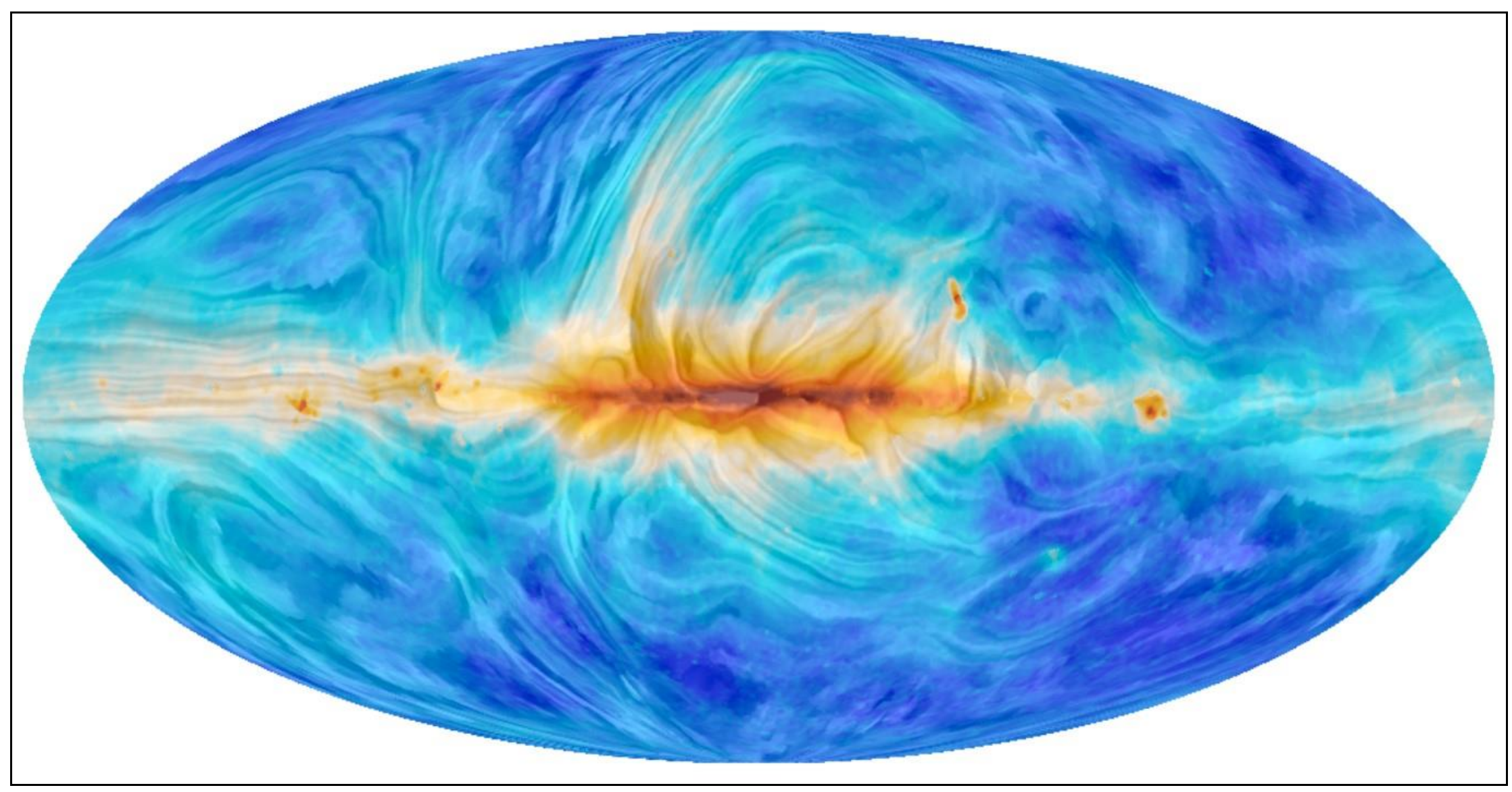

*Fig. 9: All-sky survey of synchrotron polarization at 30 GHz (1 cm) observed with PLANCK with 1° resolution. The streamlines show the field orientations in the sky plane, the colours the total intensities at 30 GHz (Adam et al. 2016).*

The PLANCK satellite has provided six polarization surveys between 30 and 345 GHz (Ade et al. 2016b). The 30 GHz refers to synchrotron polarization (Fig. 9). The 345 GHz data refers to the dust polarization (Fig. 8), which is crucial to separate foreground synchrotron emission, spinning dust radiation, and the Cosmic Microwave Background contribution. A combination of PLANCK and WMAP data allowed the construction of all-sky polarized synchrotron emission maps above a few GHz. Most of the polarized emission, and indicator of magnetic fields, are associated with distant large-scale loops and filaments of Galactic emission that imply organized magnetic fields. The brightest polarized features in the sky can be explained by magnetic filaments that surround the local spiral arm and/or the Local Bubble (West et al. 2021).

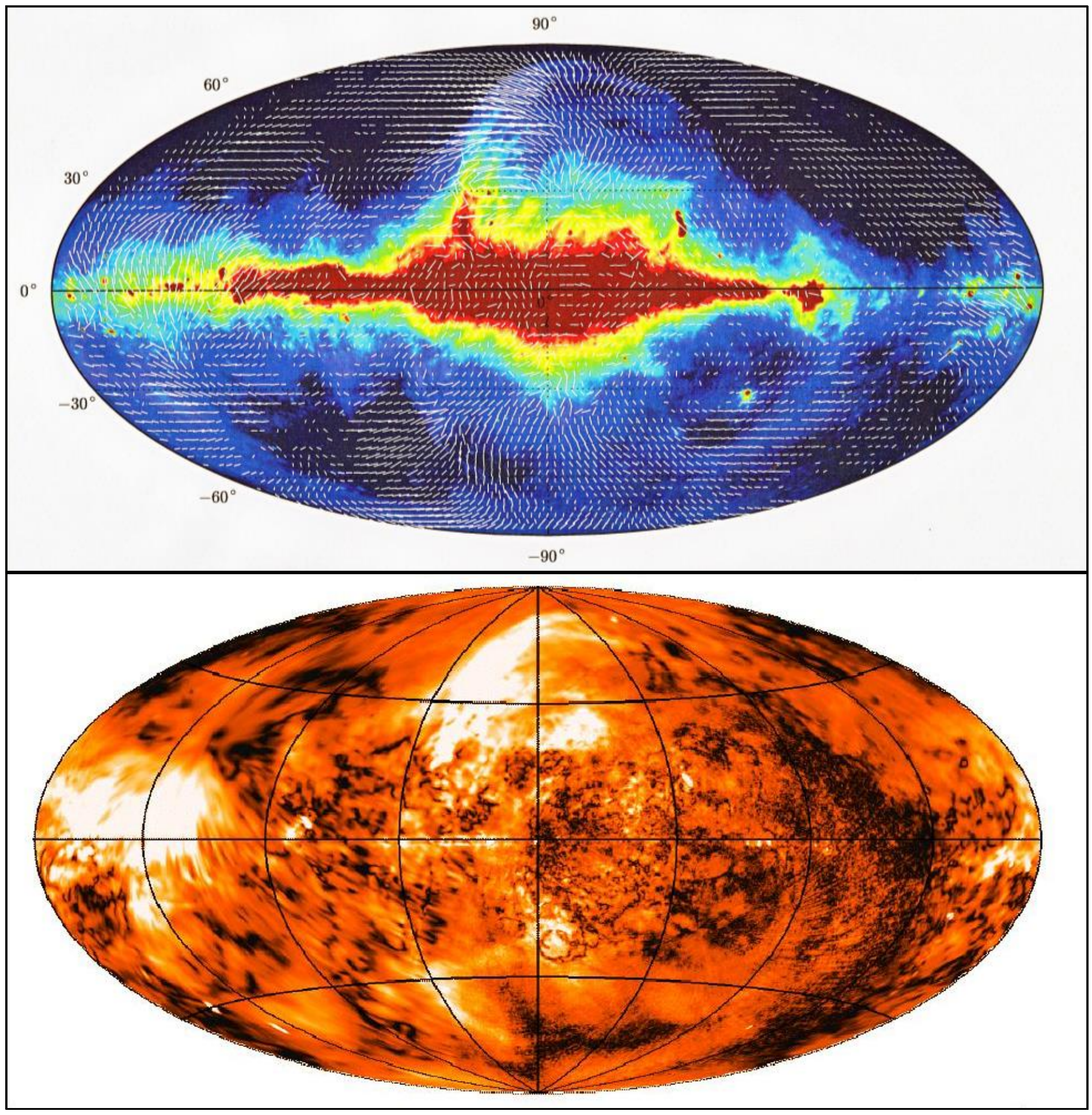

*Fig. 10: All-sky surveys in total intensity with superimposed polarization orientations (top) and polarized intensity (bottom) at 1.4 GHz (20 cm) with 36' resolution (Reich 1982; Wolleben et al. 2006; Testori et al. 2008). Image courtesy: Maik Wolleben and Wolfgang Reich.*

Another major survey, called the Global Magneto-Ionic Medium Survey (GMIMS), covers the whole sky at frequencies between 300 MHz and 1.8 GHz and allows measurement of the Faraday depth of the diffuse emission over the whole sky (Wolleben et al. 2009; 2021). First science results from the northern sky at 1280 – 1750 MHz (GMIMS-HBN) were presented by Wolleben et al. (2010b), Sun et al. (2015), Hill et al. (2017), Wolleben et al. (2021), and Dickey et al. (2022), and from the southern sky at 300 – 480 MHz by Wolleben et al. (2019). The southern hemisphere was surveyed with the Parkes telescope at 2.3 GHz (S-PASS, Carretti et al. 2019). Compared to 1.4 GHz (Fig. 10, lower panel), the S-PASS polarization map suffers less from Faraday depolarization and shows signals extending closer to the Galactic plane. Combination with WMAP and PLANCK polarization data at 23 and 30 GHz yielded a detailed RM map of the diffuse emission that is similar to the RM map derived from extragalactic sources (Fig. 13). Iacobelli et al. (2014) used S-PASS polarization data and found indications for interstellar turbulence in the transonic regime. A concentration of large polarization gradients was found above the Galactic Center, associated with the giant magnetic outflows known as the Fermi bubbles (Robitaille et al. 2017). Another all-sky polarization survey at 4.8 GHz (C-BASS) is complete in the northern and underway in the southern hemisphere (Dickinson et al. 2019).

The first polarization data from the Murchison Widefield Array (MWA) at 189 MHz with 15.6’ resolution was presented by Bernardi et al. (2013). Observations with the Low Frequency Array (LOFAR) around 150 MHz added new information of polarized synchrotron emission at low frequencies with higher angular resolution. The power spectrum of interstellar turbulence was measured in the Fan region ($l$ = +137°; $b$ = +7°) by Iacobelli et al. (2013). Jelić et al. (2014; 2015) showed the ability of LOFAR to unravel regions that show a wide range of morphological features, e.g. straight filaments parallel to the Galactic plane that are correlated with cold filamentary structures in neutral hydrogen (Kalberla & Kerp 2016; Kalberla et al. 2017). Two remarkable polarized features extending over several degrees were assigned to almost neutral gas clouds in the local interstellar medium that are visible only in low-frequency polarization due to their small Faraday rotation and depolarization (Van Eck et al. 2017).

The whole northern sky will be covered at low frequencies (120 – 168 MHz) by the LOFAR Two-meter Sky Survey (LoTSS). A test region of 568 square degrees at Galactic latitudes between about +55° and +70° demonstrates the power of low-frequency polarization (Van Eck et al. 2019). Many new features with up to 15° length were discovered, which are not correlated with tracers of neutral or ionized gas. One polarized filament reveals a large-scale gradient in Faraday rotation, possibly a magnetic loop. A larger off-plane region around RA = 12h covering about 3100 square degrees highlights the high level of complexity in the nearby magnetoionic medium (Erceg et al. 2022; Fig. 11). – A summary of all-sky or all-hemisphere polarization surveys is given in Table A2 of the Appendix.

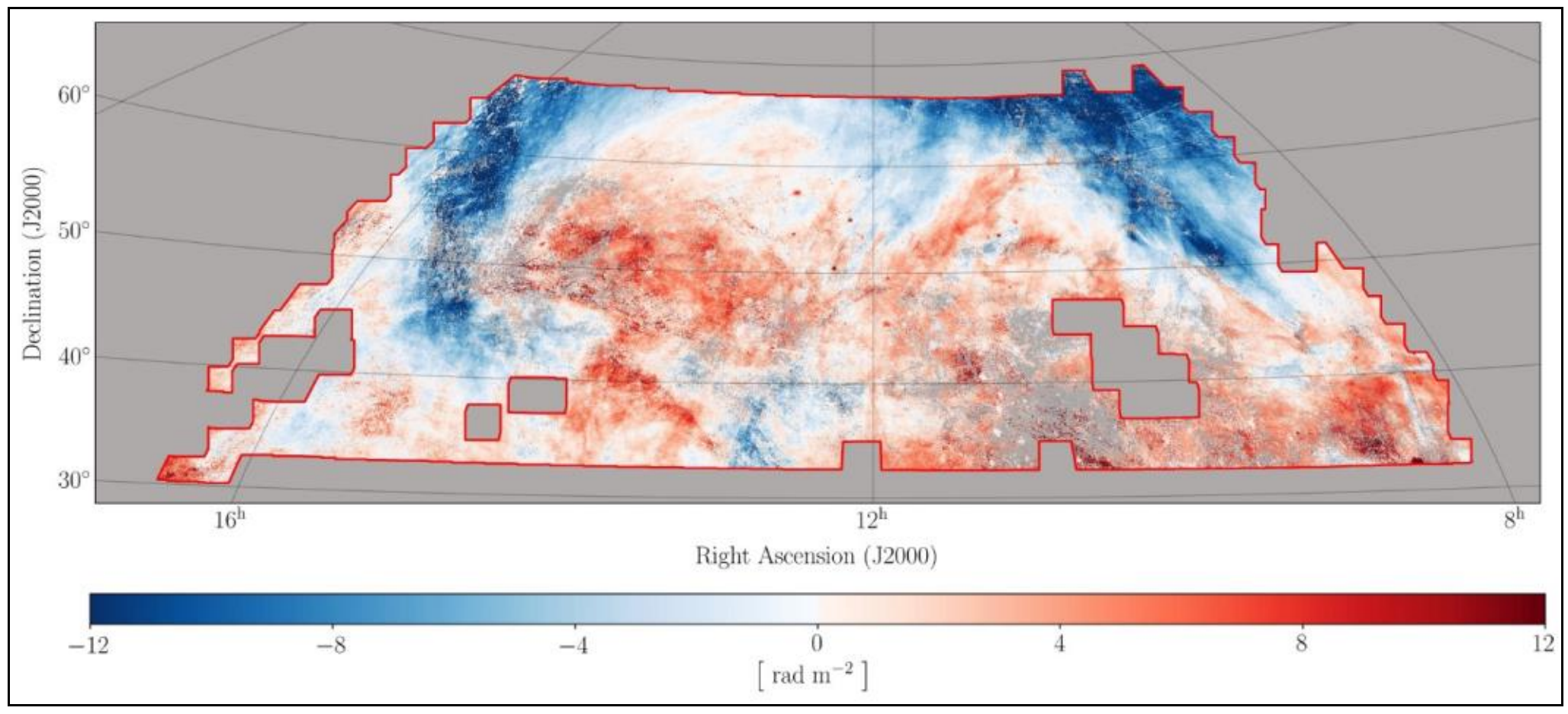

*Fig. 11: Mean Faraday depth of the Faraday cube, weighted by polarized intensity, observed with LOFAR at 120 – 167 MHz with 40’ resolution (Erceg et al. 2022).*

*Galactic plane* surveys were performed from the earliest days of radio astronomy to delineate the extended Galactic sources like supernova remnants and HII regions, usually with no linear polarization data (Appendix, Table A3). Many of the published Galactic plane surveys between 22 MHz and 10 GHz cover only a narrow strip along the Galactic plane in the inner Galaxy. Total intensity surveys at several frequencies were used to separate the thermal HII regions (with a flat radio spectrum) from the steep-spectrum non-thermal sources (supernova remnants). From the total intensity surveys numerous previously unknown supernova remnants could be identified.

Since most non-thermal sources are polarized, the Galactic plane was observed also in linear polarization. The first step was the 2.7 GHz survey by Junkes et al. (1987), followed by the surveys of the southern Galactic plane at 2.3 GHz (Duncan et al. 1995) and the northern counterpart at 2.7 GHz (Duncan et al. 1999), which covered a relatively wide strip ($|b| < 5°$) around the plane. Early high-resolution observations by Wieringa et al. (1993) showed that a lot of small-scale polarization is present in the diffuse Galactic emission that is unrelated to any structures in total intensity. The next major development is the Effelsberg Medium Latitude Survey (EMLS) at 1.4 GHz that covers ± 20° from the plane (Uyanıker et al. 1999; Reich et al. 2004). A section of the southern Galactic plane was mapped at 1.4 GHz (Gaensler et al. 2001; Haverkorn et al. 2006), complemented on the northern sky by the CGPS survey (Taylor et al. 2003; Landecker et al. 2010). A new southern survey with MeerKAT in L-band (900 – 1670 MHz) is close to completion, to be followed by S-band (1.75 – 3.5 GHz).

In all these surveys Faraday effects (section 2.4) play an important role. At frequencies of 1.4 GHz and below, Faraday rotation generates small-scale structures in polarization that are not related to physical structures. Even at 5 GHz Faraday rotation occurs near the Galactic plane ($|b| < 5°$). Even with high angular resolution, Faraday rotation leads to complete depolarization at certain values of Faraday rotation measure (RM) (Fig. 2), showing up as "canals" in the maps of polarized intensity (e.g. Haverkorn et al. 2003; 2004; Schnitzeler et al. 2009). Hence, a careful determination of the extended background emission is necessary for a reliable determination of polarized intensity, polarization angles, and Faraday depth (FD). After adding the large-scale emission to the maps of Stokes Q and U ("absolute calibration") at 1.4 GHz, most of the "canals" disappear. Some of the canals, especially those observed with LOFAR around 150 MHz (e.g. Jelić et al. 2015), are probably still depolarization structures. They mark the edge of filaments where FD changes rapidly. A method devised to distinguish canals made by irregularities in the medium from canals due to instrumental effects is the "polarization gradient" method introduced by Gaensler et al. (2011).

As a second new phenomenon, "Faraday screens" were discovered (e.g. Gray et al. 1998; Uyanıker et al. 1999; Wolleben & Reich 2004; Schnitzeler et al. 2009). These are foreground clouds of diffuse thermal gas and magnetic fields that Faraday-rotate or depolarize the extended polarized emission from the background. In addition to the well-known polarized supernova remnants (SNRs) and unpolarized HII regions, ionized envelopes of molecular clouds, pulsar-wind nebulae, and planetary nebulae were identified as Faraday screens. Depending on their Faraday rotation angle and the polarization angle of the background emission, such screens may appear bright or dark. Such observations can trace magnetic structures on sub-parsec scales.

Surveys of the Galactic plane are listed in Table A3 of the Appendix. Higher radio frequencies allow a systematic study of the Faraday Screen phenomenon, e.g. the Sino-German survey at 5 GHz with the Urumqi telescope (Sun et al. 2007; 2011; 2014; Gao et al. 2010; Xiao et al. 2011). Sun et al. (2014) investigated absolutely calibrated radio polarimetry data of the inner Galactic plane at 2.3 and 4.8 GHz and concluded that there is still considerable depolarization in the 2.3 GHz emission.

### 3.2.3 The Galactic Center

The Galactic Center region is unique with unusual radio features. Mapping the Galactic Center region by Yusef-Zadeh et al. (1984) showed several features vertical to the plane. The radio continuum emission has a flat spectral index (Reich et al. 1988), often accounted to thermal emission. However, this intense emission is highly polarized (e.g. Seiradakis et al. 1985) and was interpreted to be due to mono-energetic electrons (Lesch et al. 1988). The polarization "strings" imply vertical magnetic structures, different from the azimuthal directions of the magnetic fields seen along the Galactic plane.

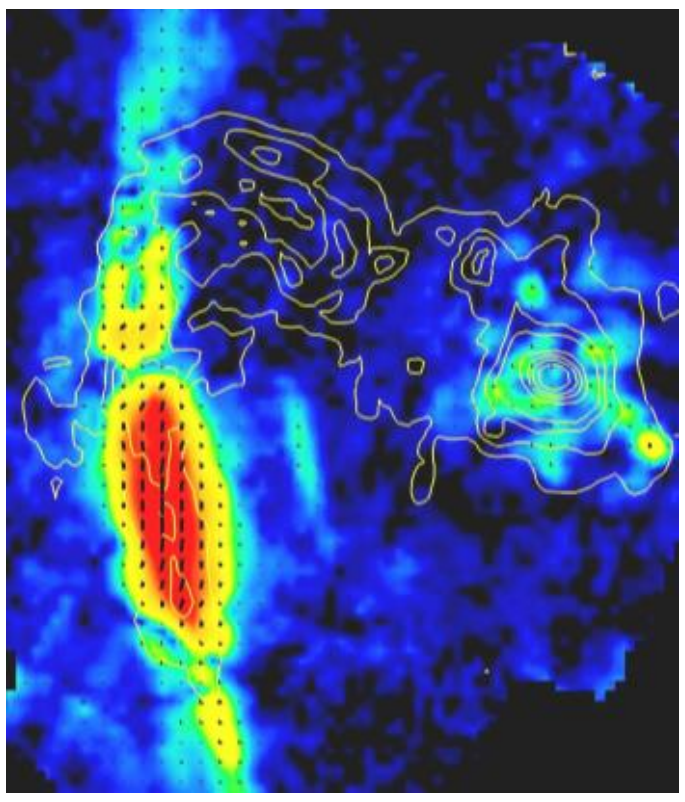

*Fig. 12: Galactic Center region. Total intensity (contours), polarized intensity (colours), and magnetic field orientations at 32 GHz (9 mm), observed with the Effelsberg telescope with 27" resolution. The map size is about 23' x 31' along Galactic longitude and latitude. Courtesy: Wolfgang Reich (MPIfR).*

Mapping of the Galactic Center at 32 GHz (Reich 2003; Fig. 12) showed RMs in the vertical filaments in excess of ±1600 rad $m^{-2}$, suggesting strong fields. Based on Zeeman splitting observations of OH masers (e.g. Yusef-Zadeh et al. 1996), field strengths are high (2 − 4 mG), while other authors (e.g. Crocker et al. 2010) found lower values of 50 − 100 μG, based on the radio synchrotron spectrum. Fields of such strength could affect the rotation curve in the inner Galaxy and explain the observed excess rotation velocity at 5 − 50 pc distance from the Galactic Center (Chan & Del Popolo 2022).

Detailed high-resolution studies also brought controversial results. High-resolution radio maps of the Galactic Center (e.g. Nord et al. 2004) showed a spiral structure at the position of Sgr A* and thin vertical radio continuum "strings". Polarimetric observations at sub-mm wavelengths suggest a stretched magnetic field (Novak et al. 2000), as expected in sheared clouds, while the large-scale ordered field is mostly toroidal (Novak et al. 2003). Near-IR polarimetry suggests a smooth transition from toroidal to poloidal field configuration in the Galactic Center at $|b| \approx 0.4°$ (Nishiyama et al. 2010). If our Galaxy does not differ much from nearby galaxies, the vertical field detected close to the center is a phenomenon local to the Galactic Center (Ferrière 2009).

Pulsars located very close to the Galactic Center show very large RMs (Eatough et al. 2013; Schnitzeler et al. 2016). The resulting line-of-sight field components of ≈ 16 − 33 μG are relatively low because of the extremely high electron densities. This implies that the degree of field regularity is low, e.g. due to field reversals and/or a small inclination angle with respect to the plane of the sky.

### 3.3 Faraday Rotation of extragalactic radio sources and pulsars

Faraday rotation is another powerful tool for studying magnetic fields. First, ionospheric rotation was detected. Soon after the discovery of linear polarization of Galactic radio waves, Faraday rotation of diffuse emission due to the Galactic ISM were detected. Increasing samples of polarized extragalactic sources and Galactic pulsars, most of which are concentrated to the Galactic plane, allowed us to model the Galactic magnetic field.

#### 3.3.1 Extragalactic radio sources (EGRS)

Faraday rotation measures (RMs) towards polarized EGRS behind the Galaxy originate in the source itself and in the various magneto-ionic media in the foreground. The contributions from intergalactic space, intervening galaxies, and interplanetary space are generally small. The contribution from the ionosphere of the Earth can be subtracted with help of calibration sources with known polarization angles, leaving the RM from the Milky Way and the intrinsic RM. Interestingly, the intrinsic RM dispersion of extragalactic sources is smaller than the RM dispersion in the Galactic foreground for sightlines that are separated by at least 1°, so that the contribution by the Galactic foreground becomes important at low and intermediate Galactic latitudes (Schnitzeler 2010). To separate the intrinsic RM from that in the foreground, observations at many adjacent frequency channels across a wide band helps (*RM Synthesis*, section 2.4).

Here a word of caution should be given. Polarization and RM of EGRS may originate in the nucleus of a radio galaxy or in the jets or lobes. When combining observations at various frequencies to obtain RM, care must be taken that the same source structure is measured. Problems may arise when combining data from single dish observations with those of an interferometer.

The earliest all-sky catalogue of RMs towards EGRS by Simard-Normandin & Kronberg (1980) included only a few sources with measured RMs along the Galactic plane, where the Galactic magnetic fields are concentrated. The interpretation of this data gave indication of a local magnetic field only. In recent years additional data on the RM of sources in the Galactic plane were obtained (Brown et al. 2003; 2007; Van Eck et al. 2011; Ma et al. 2020; Van Eck et al. 2021). As these surveys cover the Galactic plane only partly, interpretations were difficult. The highest values were |RM| ≈ 1000 rad $m^{-2}$ towards the Galactic Center (Roy et al. 2008). The RMs towards EGRS near the very center are mainly positive, suggesting a magnetic field aligned with a central bar.

A statistical method to visualize the RM distribution over the sky, developed by Johnston-Hollitt et al. (2003) and based on 800 sources, showed several areas of consistent RM values (of the same sign) as well as structures above and below the plane. A major extension of the data set was achieved by Taylor et al. (2009), who re-analyzed the NRAO VLA Sky Survey (NVSS). This study involved 37543 sources and added a huge number of new RMs, but the rather close frequency separation of the two frequency bands leads to large RM errors and RM ambiguities, especially near the Galactic plane. A new VLA Sky Survey (VLASS) at 2 – 4 GHz is underway (Lacy et al. 2020).

Another project to increase the number of RMs over the whole sky was undertaken at the Effelsberg 100-m radio telescope. The combination of polarization data in 8 channels in two bands around 1.4 and 1.6 GHz enables an accurate determination of the RMs of the polarized sources. Some 1600 new RMs were added and a preliminary result was given in Wielebinski et al. (2008). The data for 2469 sources were used to model the Galactic magnetic field (Sun et al. 2008; section 3.5). An update by Xu & Han (2014) reassessed the accuracy of published RMs. The coverage of the southern sky was improved by a survey of compact sources in the southern hemisphere carried out with the Australia Telescope Compact Array (Schnitzeler et al. 2019). These sightlines fill the gap below declination -40° that is not covered by the RM catalogue from Taylor et al. (2009).

A well-sampled view of the entire Milky Way and its halo has been achieved (Fig. 13). The averaged RMs towards EGRS reveal the same mostly positive signs above and below the plane in the Galactic longitude range between -160° and -120° (Fig. 13, right part) and mostly negative in the range +110° to +150° (Fig. 13, left part): The local disk field seems to be part of a large-scale even-symmetric field structure. A survey of RMs towards 641 EGRS near the Galactic plane in the Galactic longitude range +100° to +117° confirmed this symmetry (Mao et al. 2012a). RMs towards 2234 EGRS towards the outer Galaxy, derived from the Canadian Galactic Plane Survey (CGPS; Van Eck et al. 2021), were also found to be even-symmetric (mostly negative) in an even larger longitude range, between about +80° and about +160°, with a smooth trend from negative RMs to zero RMs from +130° to +180° longitude. In the inner Galaxy, from +39° to +52° longitude, RMs towards 127 EGRS near the plane are mostly positive and even-symmetric, with very large RM values concentrated around the tangent of the Sagittarius arm around +48° longitude (Shanahan et al. 2019).

Towards the inner Galaxy (between about -60° and +60° longitude) and away from the Galactic plane, the RM signs are opposite above and below the plane and opposite west and east of the center, with the pattern $\begin{smallmatrix} + & - \\ - & + \end{smallmatrix}$ . The distance to the magnetic fields generating this pattern is not known; they may be embedded in local gas clouds (Wolleben et al. 2010b) or they may be part of a large-scale antisymmetric field in the Milky Way's halo (section 3.5). Remarkably, the region of this RM pattern approximately coincides with the two huge bubbles of hot gas seen in X-rays that extend to about 14 kpc above and below the Galactic Center (Predehl et al. 2020), which calls for a detailed investigation.

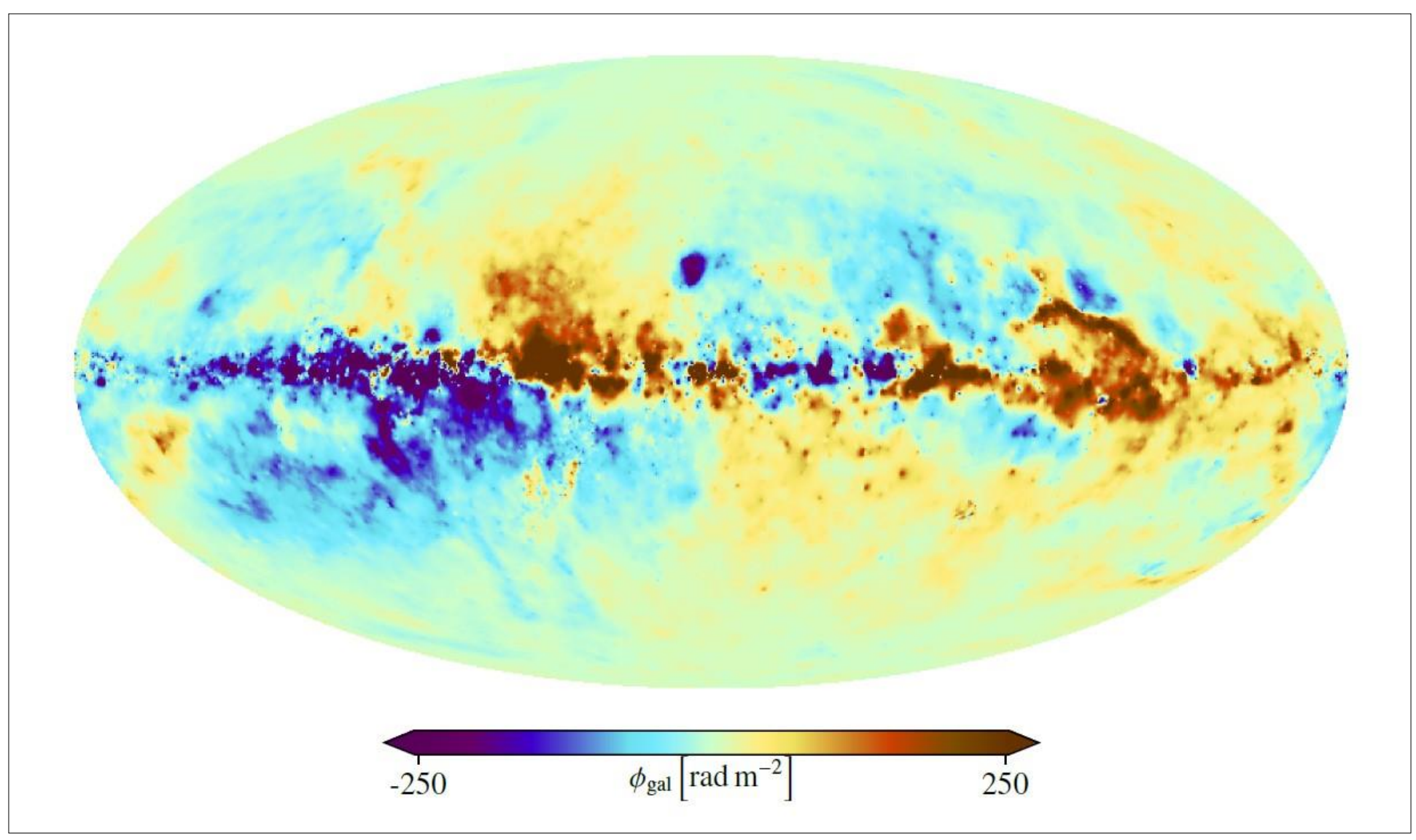

*Fig. 13: All-sky map of rotation measures in the Milky Way, constructed from the RM data of 55190 polarized extragalactic sources of the VLA NVSS survey and many other catalogs. Red: positive RMs, blue: negative RMs (Hutschenreuter et al. 2022). Galactic longitude 0° (Galactic center) is in the map center. Galactic longitude increases from right to left, Galactic latitude increases from bottom to top.*

Another note of caution: Numerical MHD simulations were found to be broadly consistent with Fig. 13 (Pakmor et al. 2018). However, when varying the position of the observer within along the Solar radius, the synthetic RM sky changes completely. The Local Bubble may dominate the RM signal over large regions of the sky (Reissl et al. 2023). The field morphology is much more complicated than a simple symmetric or antisymmetric pattern (section 3.5). Support comes from comparing the |RMs| from EGRS with the corresponding ones of the diffuse polarized emission. While a factor of two is expected for perfectly regular fields, the ratios are between one and two, indicating field complexity along the line of sight, e.g. field reversals (Dickey et al. 2022).

The advent of RM observations of EGRS with help of RM Synthesis opened a new way to study magnetic fields in three dimensions. From 1105 sources behind the Smith high-velocity cloud observed with the VLA, Betti et al. (2019) constrained the detailed geometry and the strength of the magnetic field in the cloud.

3.3.2 Pulsars

In principle, pulsars are the ideal sources to probe the magnetic fields through the Faraday effect. Pulsars have no measurable angular structure and are highly polarized. However, pulsars with accurately measured distances are still rare. Pulsars are Galactic objects and hence their distribution is close to the Galactic plane towards the inner Galaxy. In fact, very few pulsars are known towards the anti-center of the Galaxy. Hence a combination of pulsars and EGRS is optimal for studies of the Galactic magnetic field. Pulsars also allow measurement of the Dispersion Measure (DM) that follows from the signal delay occurring in the foreground medium. Together with the Rotation Measure (RM) the value of the average regular magnetic field in the line of sight can be deduced:

$$< B_{\parallel} > = 1.232 \ \mu\text{G} \ (\text{RM}/\text{DM})$$

Application gives an average strength of the regular field in the local spiral arm of 1.4 ± 0.2 μG. In the inner Norma arm, the average strength of the regular field is 4.4 ± 0.9 μG. However, this formula is only valid if variations in the field strength and in thermal electron density are *not correlated*. If they are correlated, the above formula gives an overestimate of $<B_{\parallel}>$ and an underestimate for anti-correlated variations (Beck et al. 2003). MHD simulations of driven turbulence revealed that local field strength and thermal electron density are uncorrelated only for subsonic and trans-sonic turbulent flows, while for supersonic flows the field strength can be severely overestimated by using the above equation. Observational data indicate that field strength and electron density are largely uncorrelated over kpc scales, so that the equation is safe when applied to such large scales (Seta & Federrath 2021).

Compilations of pulsar rotation measures were given in Han et al. (2006; 2009; 2018), additional results in Mitra (2003), Noutsos et al. (2008), and Van Eck et al. (2011). RMs with lower uncertainties are now available from pulsars observed at low frequencies with LOFAR (Sobey et al. 2019) and with CHIME (Ng et al. 2020).

The distribution of rotation measures, as given by Han (2018), shows a huge variation of signs and magnitudes (Fig. 14). This indicates a large-scale regular magnetic field with multiple reversals (section 3.5) or many local field distortions, e.g. by HII regions (Mitra et al. 2003; Nota & Katgert 2010).

|RM| increases for distant objects, but very few pulsars were found beyond the Galactic Center (Fig. 14). Except for a few pulsars very close to the Galactic center (section 3.2.3), the limit of |RM| ≈ 1000 rad $m^{-2}$ for EGRS holds also for pulsars. This indicates that the RMs towards EGRS are partly averaged out in passage through the Galaxy. The large-scale regular field of the outer Milky Way is either weak or frequently reversing its direction. As distances to most pulsars are uncertain, this result should be taken with some caution. A larger sample of pulsar RM data and improved distance measurements to pulsars are needed.

Pulsars in the Galactic halo are rare. The catalogues by Sobey et al. (2019), Ng et al. (2020), and Xu et al. (2022) include a significant number of faint pulsars in the halo. The strength of the regular halo field was found to be around 2 μG and its scale height 2.7 ± 0.3 kpc (Xu et al. 2022).

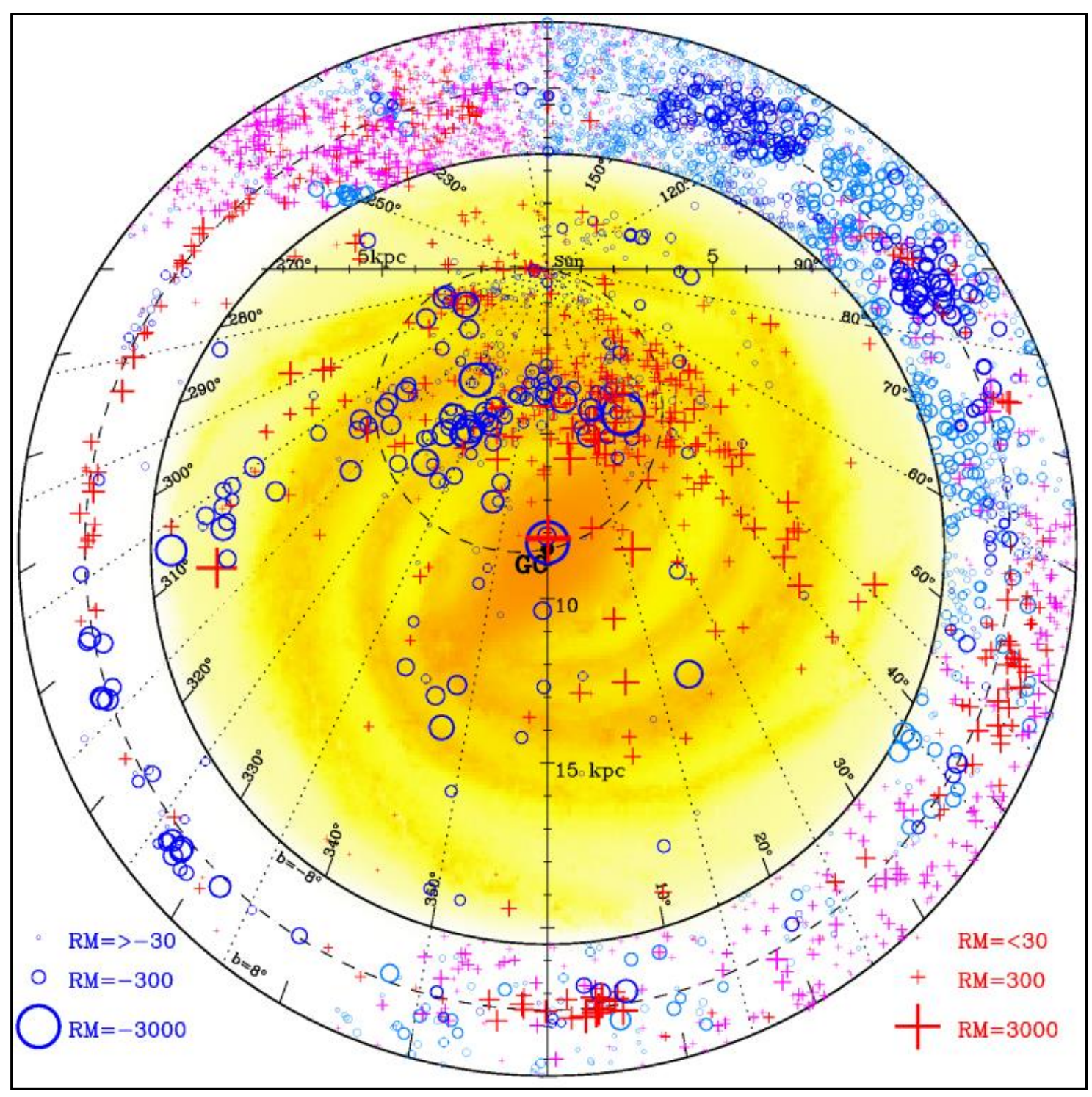

*Fig. 14: Faraday rotation measures (RM) towards pulsars in the Milky Way at latitudes |b| < 8° (shown in the inner circle) and towards extragalactic radio sources at latitudes |b| < 8° (shown between the inner and outer circles). Red "+" signs indicate positive RMs towards pulsars, blue circles negative RMs. Pink and light blue symbols indicate RMs towards extragalactic sources compiled by Taylor et al. (2009); dark red and dark blue symbols are from the compilation of Xu & Han (2014). Our Sun is located at the crossing of coordinate lines. The dashed circle gives the locus of tangent points, assuming a spiral pitch angle of 11°. The background image is an artist's impression of the Galactic spiral structure (Han et al. 2018).*

### 3.4 Zeeman effect

The Zeeman effect is the most direct method of measuring magnetic fields. It has been used in the optical range for detecting magnetic fields in the Sun and in stars. At radio wavelength the use of the Zeeman effect proved to be more difficult. For one, the frequency shifts caused by the weak magnetic fields are minute and require sophisticated instrumentation. The HI line gave the first definitive detections, usually in absorption towards strong Galactic sources (Verschuur 1968). The technique was refined so that at present magnetic fields as weak as ≈ 5 μG can be detected with the Arecibo telescopes (Heiles & Crutcher 2005). The observation of the Zeeman effect in the OH molecule (e.g. Crutcher et al. 1987) advanced the field further. Even stronger fields (≈ 80 mG) were detected in interstellar $H_2O$ masers (Fiebig & Güsten 1989). Millimeter-wavelength astronomy gave us additional results for high recombination lines (Thum & Morris 1999) and in such molecules as CN (Crutcher et al. 1999) or CCS (Levin et al. 2000).

A compilation of present-day Zeeman measurements of the magnetic field in HI gas clouds gives a mean total field in the cold neutral interstellar gas of 6 ± 2 μG, so that the magnetic field dominates thermal motion, but is in equipartition with turbulence, as also found on much larger scales in external galaxies (section 4.2). Beyond cloud densities of ≈ 300 $cm^{-3}$ the field strength scales with $n^{0.65 \pm 0.05}$ (Crutcher et al. 2010; Crutcher 2012).

The importance of magnetic fields in the star-formation process is obvious. Diffuse clouds are subcritical with respect to collapse and probably balanced by magnetic fields, while dense molecular are supercritical and collapse. The transition from subcritical to supercritical state may be the result of ambipolar diffusion or turbulence. Zeeman observations in the HI and OH lines can measure the ratio of mass to magnetic flux in the cloud envelope and the core. A smaller ratio in the core may indicate that supersonic turbulence plays a similarly important role as ambipolar diffusion (Crutcher et al.

2009), but effects of the field geometry also have to be accounted for (Mouschovias & Tassis 2009). More and higher-quality data are needed.

The use of the Zeeman data for the investigation of a large-scale regular magnetic field of the Galaxy was attempted by several authors (e.g. Fish et al. 2003). The number of detected sources was rather small and the interpretation in terms of Galactic magnetic fields rather inconclusive. Han & Zhang (2007) collected a large data set of Zeeman results and studied the question if the magnetic fields in molecular clouds preserve information of the direction of the large-scale magnetic fields in the spiral arms. In spite of a larger data set all that the conclusion offered was that clouds “may still remember the directions of regular magnetic fields in the Galactic ISM to some extent”.

### 3.5 Modelling the magnetic field of the Milky Way

Based on all the data described in previous sections models of the magnetic fields of the Milky Way have been repeatedly made. At first the low frequency all-sky data was used to describe the Galactic non-thermal emission (e.g. Yates 1968) produced in magnetic fields. The all-sky survey of Haslam et al. (1982) has been interpreted by Phillips et al. (1981) and Beuermann et al. (1985). Using the data on the HII regions (e.g. Georgelin & Georgelin 1976) of the Galaxy it could be shown that the spiral structure is also seen in the diffuse total intensity radio emission.

The RM data first for pulsars and later for extragalactic radio sources (EGRS) led to more detailed modelling of the magnetic fields of the Milky Way. Pulsars are concentrated in the direction of the inner Galaxy; only few pulsars are known outside the inner quadrants. The data on EGRS gave information about the Faraday effects over much of the sky and in particular in the inner Galactic plane.

The catalogue of 543 rotation measures of EGRS distributed across the sky by Simard-Normandin & Kronberg (1980) showed areas with similar RM directions, suggesting organized magnetic fields in the Galactic disk over large scales. Pulsar observers (e.g. Han & Qiao 1994) were the first to point out that in addition to large areas of similar magnetic field directions there were some regions where the field *reverses* along Galactic radius. These results were analyzed with wavelets (e.g. Stepanov et al. 2002) and confirmed the existence of at least one large-scale reversal. Since most of the EGRS investigated were away from the Galactic plane they did not trace the large-scale Galactic magnetic field in the disk, but more likely some local magnetic features. The number of RMs has been steadily increasing (e.g. Taylor et al. 2009), which increased the sampling of the Galaxy considerably. The same general conclusions were reached as in the earlier work – organized magnetic structures in sections of the Galaxy and highly disorganized magnetic fields towards the central region.

The analysis of data from radio continuum all-sky surveys at 1.4 and 23 GHz, from RMs towards EGRS, the best available thermal electron model and an assumed cosmic ray distribution (Sun et al. 2008; Sun & Reich 2010; Jaffe et al. 2010; Jansson & Farrar 2012a; 2012b; Ordog et al. 2017) constrained the average field strength of the Galaxy to ≈ 1 – 2 μG for the regular field and ≈ 3 – 5 μG for the random field in the solar neighborhood, similar to the results from pulsar RMs (section 3.3.2). An axisymmetric spiral (ASS) magnetic field configuration (section 2.7) with a pitch angle of ≈ 12° fits the observed data best, which is much smaller than the pitch angle of ≈ 30° derived from the data of polarized emission only (Fauvet et al. 2011). This demonstrates the problems of the models to obtain a reliable estimate for the pitch angle.

Including the polarized thermal dust emission from WMAP at 94 GHz (Fig. 8) complicates the analysis (Jaffe et al. 2013). The magnetic fields in the dust-emitting regions seem to be more ordered than those in the synchrotron-emitting regions. The isotropic random field components peak in the gaseous arms, while the ordered (probably anisotropic random) fields are shifted, similar to the observations in external galaxies (section 4.4.1). The *local magnetic field* is distorted by the Local Bubble (Alves et al. 2018).

Outside of the Local Bubble, the local field is mostly oriented parallel to the Galactic plane (Fig. 8) and its direction is symmetric (even symmetry) with respect to the Galactic plane (Fig.13) while the toroidal component of the halo field may have different directions above and below the Galactic plane (odd symmetry, see section 2.7), to account for the different signs of the observed RMs. However, the asymmetry can be partly explained by distorted field lines around a local HI bubble (Wolleben et al. 2010b). Observations with better sampling of the sky are needed (section 5).

Near-infrared starlight polarization data around +150° Galactic longitude are inconsistent with an odd-symmetry disk field and are better compatible with an even-symmetry axisymmetric spiral field with a pitch angle of about 6° (Pavel et al. 2012), smaller than that derived from radio data.

Pulsars are promising objects to deduce the Galactic magnetic field because their RMs can provide a 3-D map of field directions in the Galaxy, if distances to pulsars are sufficiently accurate. Since most pulsars are concentrated along the Galactic plane, they sample the field in the Galaxy's disk. Analysis of the pulsar and EGRS data led to several attempts to model the Galactic magnetic field (Brown et al. 2007; Van Eck et al. 2011; Han et al. 2018).

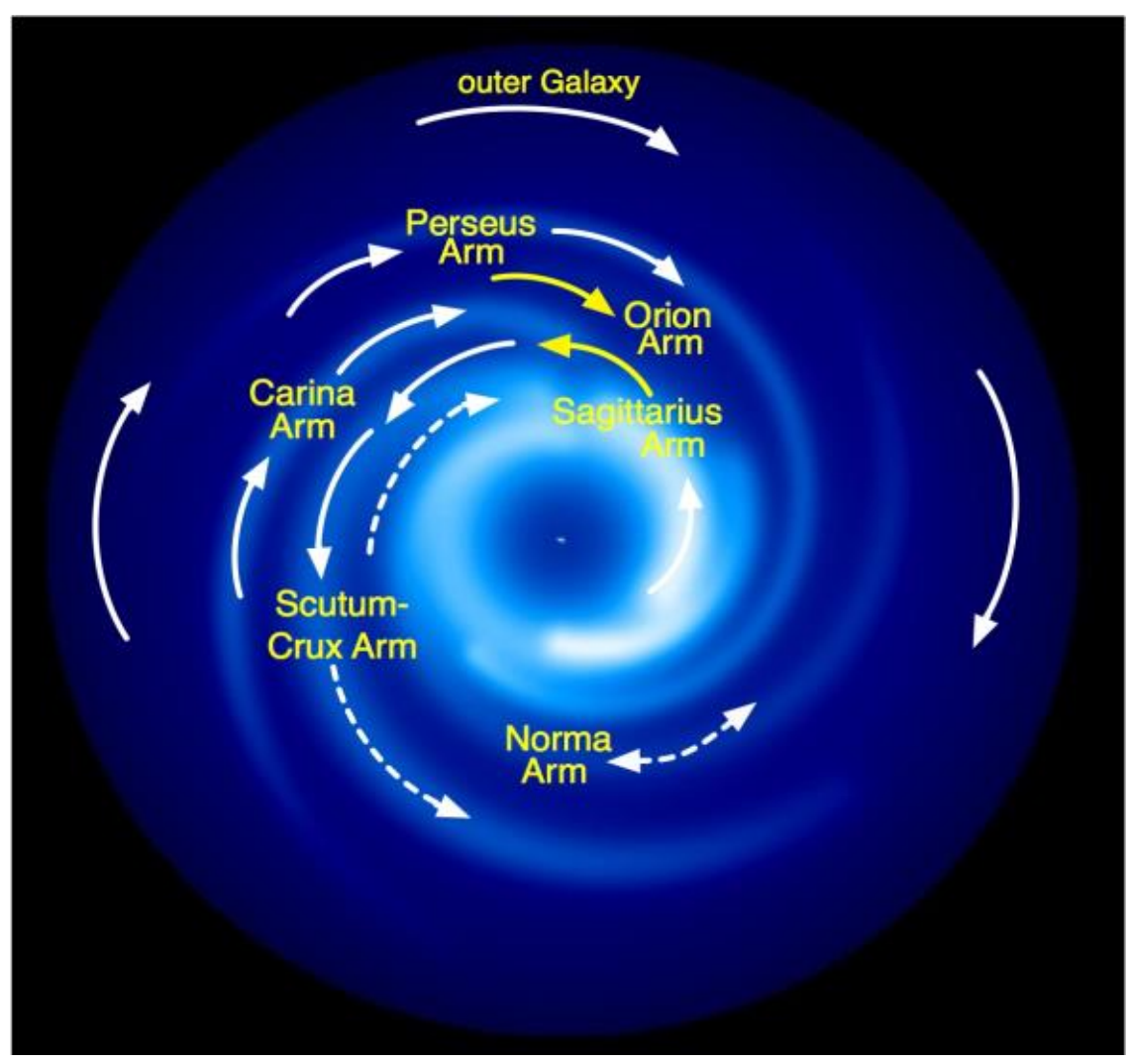

*Fig. 15: Model of the large-scale magnetic field in the Galaxy's disk, derived from Faraday rotation measures of pulsars and extragalactic sources. Yellow arrows indicate confirmed results, while white and dashed arrows still need confirmation. Courtesy: Jo-Anne Brown (University of Calgary).*

A large-scale magnetic field reversal is present between the clockwise directions of the local Carina-Orion arm (and the outer Perseus arm) and the counter-clockwise direction of the inner Scutum-Crux-Sagittarius arm (Fig. 15), located at about 1 – 2 kpc inside the solar radius. Comparison of rotation measures of pulsars in the disk at different distances as well as with rotation measures of background radio sources beyond the disk reveals between six and eight large-scale reversals of the field directions between spiral arms and interarm regions (Han et al. 2018; Fig. 16).

The reversals were often used as an argument in favor of a bisymmetric spiral (BSS) field structure, but a detailed analysis (e.g. Vallée 1996; Noutsos et al. 2008) showed that the concept of a single large-scale field mode is not compatible with the data. The analysis of the previous interpretations by Men et al. (2008) gave no proof of either a BSS or an ASS configuration, so that the large-scale field structure is more complicated.

Studies of the effects of large HII regions on the RMs of pulsars beyond these Faraday screens showed that some of the claimed field reversals are only local phenomena (Mitra et al. 2003). Furthermore, the comparison of the RMs of pulsar and EGRS towards the Galactic Center (Brown et al. 2007) revealed similar values of RM, as if there were no other half of the Galaxy. This result suggests that the RMs are dominated by local ISM features and that the large-scale field is weak and cannot be delineated from the available data. Only RM data free from the effects of HII regions should be used, as was demonstrated by Nota & Katgert (2010).

The existence of at least one large-scale field reversal in the Milky Way is puzzling. Very few large-scale reversals have been detected so far in external spiral galaxies, and only one along the radial direction (section 4.4.2). The different observational methods may be responsible for this discrepancy between Galactic and extragalactic results. RMs in external galaxies are averages over the line of sight through the whole disk and halo and over a large volume traced by the telescope beam, and they may miss field reversals, e.g. if these are restricted to a thin region near to the galaxy's plane, while the results in the Milky Way are mostly based on RMs of pulsars that trace the magneto-ionic medium near the plane. Alternatively, the Milky Way may be “magnetically young” and may still not have generated a coherent large-scale field over the whole disk. The timescale for fully coherent fields can be longer than the galaxy age, especially if interactions with other galaxies occur (section 2.7). Another model to explain reversals is the continuous generation of small-scale fields; these may disturb the action of the $\alpha$-$\Omega$ dynamo and allow reversals of the initial field to persist (Moss et al. 2012).

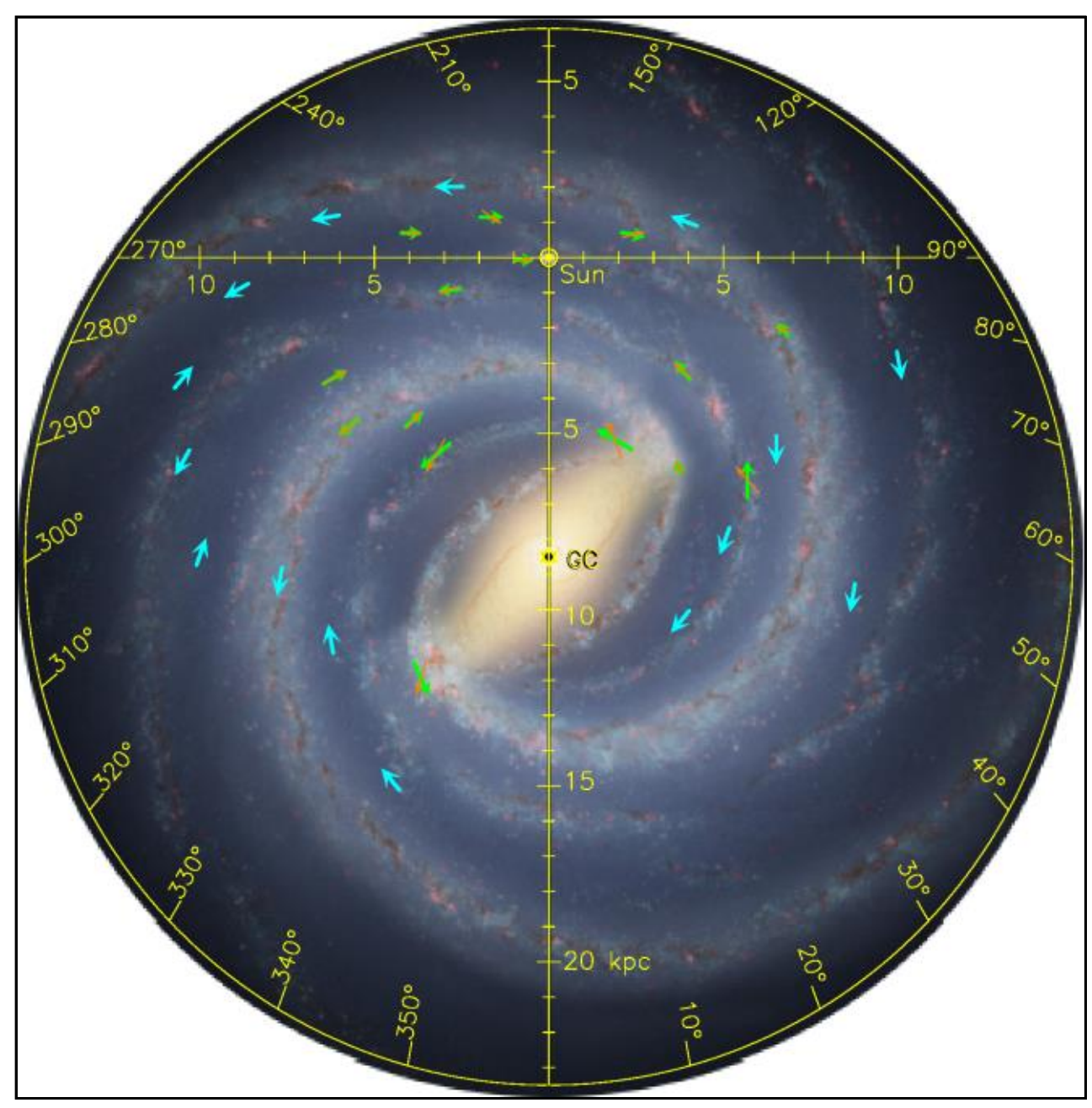

*Fig. 16: Alternative model of the large-scale magnetic field in the Milky Way’s disk, derived from Faraday rotation measures of pulsars and extragalactic sources. The orange arrows give the derived line-of-sight magnetic field components and the green arrows show the inferred spiral field. The blue arrows give the inferred spiral field direction derived from the comparison of RMs for EGRS with RMs for distant pulsars and are placed at the distance of the central axis of the next outer spiral feature (arm or interarm). The background image shows an artist’s impression of the structure of our Galaxy (Han et al. 2018).*

Localized field reversals are common in numerical MHD simulations of galaxies (section 2.7). In galaxies with an imposed spiral potential, reversals could be associated with velocity changes across the spiral shock of at least 20 km/s (Dobbs et al. 2016). Such signatures should be searched for in the velocity fields of interstellar gas. Many MHD models of evolving galaxies show reversals in all field components on small and medium scales (Rieder & Teyssier 2016; Pakmor et al. 2018; Steinwandel et al. 2020; Wissing & Shen 2023), other models show reversals mostly in the radial field (Liu et al. 2022; Wibking & Krumholz 2023).

Little is known about the large-scale field in the Milky Way’s *halo*. In the Galactic Center vertical magnetic fields apparently extend into the halo (section 3.2.3). RMs towards 137 pulsars in the northern sky observed with LOFAR over a large range of Galactic latitudes yield a scale height of the halo regular field along the sightline of 2.7 ± 0.3 kpc (Xu et al. 2022). This value may be a lower limit if the field geometry changes with height. The field geometry around the Smith High Velocity Cloud, located about 3 kpc below the Galactic plane, requires a surprisingly high strength of the regular halo field of +5 μG (Betti et al. 2019), larger than the average regular field in the Galactic disk (section 3.3.2), but similar to that in the halo of the galaxy NGC4631 (section 4.7).

The RM antisymmetry in the inner Galaxy (Fig. 13) may be a local phenomenon or indicate a halo field with a dipolar (odd-symmetric) pattern, in contrast to that found in several external galaxies (section 4.7). The field near the Sun seems to be neither even- nor odd-symmetric (Dickey et al. 2022). From a survey of RMs towards more than 1000 EGRS near the Galactic poles, Mao et al. (2010) derived a local large-scale field perpendicular to the plane of +0.31 ± 0.03 μG towards the south Galactic pole, but no significant field towards the north Galactic pole. This is neither consistent with an odd-symmetry nor with an even-symmetry halo field. Observations of 194 RMs towards EGRS near the Galactic plane in the first Galactic quadrant cannot be reproduced by any of the major field models, again calling for more advanced models (Ma et al. 2020). Numerical MHD simulations of evolving galaxies give an idea about the complexity of the field structure (section 2.7).

Jansson & Farrar (2012a; 2012b; further improved by Kleimann et al. 2019) modelled the diffuse polarized emission and RMs of EGRS and found evidence for X-shaped vertical field components, similar to those in external galaxies. A vertical halo field is also needed to explain the shape of a sample of shell-type axisymmetric supernova remnants expanding into the ambient field (West et al. 2016). A visualization of the Jansson & Farrar model is shown in Fig. 17.

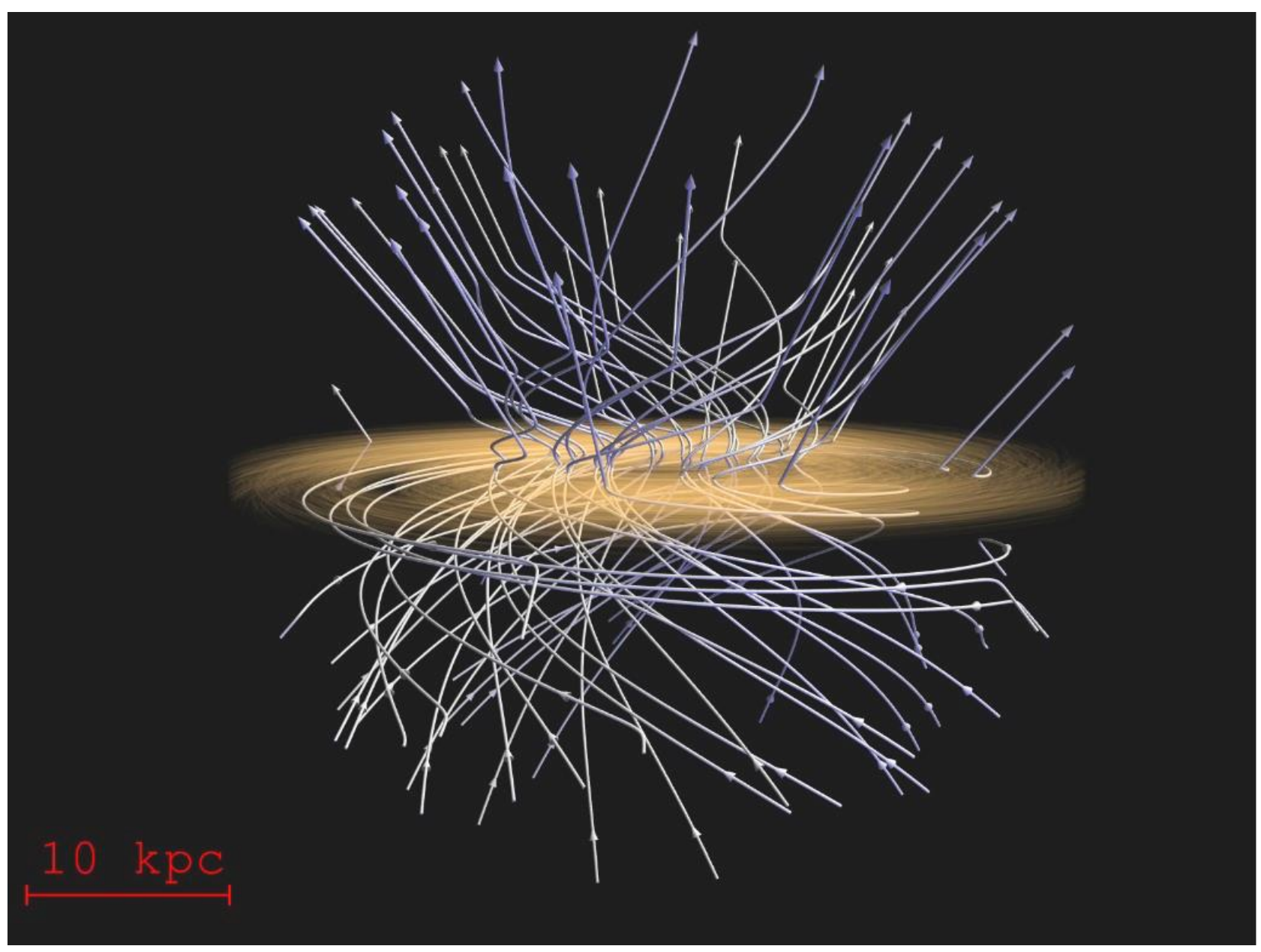

*Fig. 17: Model of the outward-spiraling halo field of the Milky Way (Farrar 2016). Visualization: T. Sandstrom (NASA).*

Terral & Ferrière (2017) compared synthetic all-sky RM maps derived from their analytical models with the observed RMs of EGRS and found a preference for a bisymmetric halo field that can generate an X-shaped pattern. However, the halo field is probably more complicated than predicted by $\alpha$-$\Omega$ dynamo models.

From a combined analysis of all-sky radio polarization and RM surveys, West et al. (2020) searched for signatures of magnetic helicity in the Milky Way. They found evidence for a quadrupolar field with left-handed helicity in the northern Galactic hemisphere and right-handed helicity in the southern hemisphere. The vertical field is pointing away from the observer in both Galactic hemispheres.

In summary, the magnetic field structure in the disk of the Milky Way is quite complex. Although observations in the Milky Way can trace magnetic structures to much smaller scales than in external galaxies, details can only partly be resolved. The large-scale field in the Milky Way is even more difficult to measure due to the Sun's location inside the disk. These information gaps will be closed with help of future radio telescopes (section 5). Many new pulsars, together with more accurate distance measurements, and a large extension of the EGRS data base, especially in the southern hemisphere, will yield an improved image of the Milky Way's magnetic field (Haverkorn et al. 2015). Next-generation modelling of the Galactic magnetic field can take advantage of IMAGINE, a software package based on Bayesian sampling (Steininger et al. 2018).

A review of the various models of magnetic field configurations in the Milky Way is given in Haverkorn (2015). Prospects for future work are discussed by Jaffe (2019).

# 4. Galaxies

Magnetic fields in external galaxies can be observed with the same methods as in the Milky Way, except for extragalactic pulsars that have been found so far only in the Magellanic clouds. Naturally the spatial resolution of the telescopes is much worse in galaxies, and the detailed structure of extragalactic fields on scales below about 100 pc is still invisible. On the other hand, the large-scale field properties, like the overall pattern and the total extent, can best be measured in external galaxies. Hence, observations in the Milky Way and in external galaxies are complementary.

### 4.1 Optical polarization, infrared polarization, and Zeeman effect

Weak linear polarization (generally below 1%) is the result of extinction by elongated dust grains in the line of sight that are aligned in the interstellar magnetic field (section 2.1). Optical polarization surveys yielded the large-scale structure of the ordered field in the local spiral arm of our Milky Way (section 3.1). The first extragalactic results by Hiltner (1958), based on starlight polarization of globular clusters in M31, revealed a magnetic field aligned along the galaxy's major axis. Polarization of starlight in the LMC gave evidence for ordered fields near 30 Dor (Mathewson & Ford 1970b), along the Magellanic Bridge (Schmidt 1976). From a polarimetric catalogue of 7207 stars, Gomes et al. (2015) found an ordered field aligned with SMC’s bar with a strength of about 0.9 μG. By comparison with spatially filtered HI data, Ma et al. (2023) found evidence for magnetic alignment of HI filaments.

Polarization from diffuse optical light was used to search for large-scale magnetic fields, though some unknown fraction of the polarized light could be due to scattering on dust particles. Correction of the diffuse optical polarization for scattering effects is difficult and has not been attempted so far. A survey of 70 nearby galaxies revealed degrees of polarization of ≤1% (Jones et al. 2012). Indications for ordered fields along the spiral arms were found in M82 (Elvius 1962), M51 and M81 (Appenzeller 1967; Scarrott et al. 1987), in NGC1068 (Scarrott et al. 1991), and in NGC6946 (Fendt et al. 1998). The pattern in M51 (Scarrott) agrees well with the radio polarization results (see Fig. 22) in the inner spiral arms, but scattered light leads to major differences in the outer arms and in the companion galaxy that is unpolarized in the radio image. In the Sa-type edge-on Sombrero galaxy M104 and the Sb-type edge-on NGC4545, optical polarization indicated a field along the prominent dust lane and vertical fields above the plane (Scarrott et al. 1990), in agreement with the results from radio polarization (section 4.7). The polarization of the Sc-type edge-on galaxies NGC891, NGC5907, and NGC7331 indicated magnetic fields that are predominantly oriented perpendicular to the plane (Fendt et al. 1996), possibly aligned along the vertical dust filaments observed in these galaxies. On the other hand, near-infrared polarimetry of NGC891 revealed magnetic fields almost aligned with the plane (Montgomery & Clemens 2014), similar to the radio data (Fig. 41). Radio continuum polarization traces the magnetic fields in the diffuse medium that are mostly oriented parallel to the plane in the inner disk and have significant vertical components beyond some height above the plane (section 4.7).

Polarization techniques were developed in the infrared range where scattering is negligible. Near-IR (NIR) polarization in the edge-on galaxies NGC891 and NGC4565 indicate plane-parallel fields (Montgomery & Clemens 2014; Jones 1989), similar to that seen at radio wavelengths. In the far-IR (FIR) and sub-mm ranges, the degrees of polarization reach several %. The galaxy M82 was observed at 850 μm (Greaves et al. 2000), but the derived bubble-type field pattern is in contrast to the radio data indicating a field that is oriented radially outwards (Reuter et al. 1992; Adebahr et al. 2017), while NIR and FIR polarimetry show a vertical field (Jones 2000; Jones et al. 2019).

Polarimetry in the FIR with high resolution and sensitivity became possible with the HAWC+ polarimeter on board of SOFIA (Lopez-Rodriguez et al. 2022a; 2022b). Many nearby galaxies were observed so far (Fig. 22; Table A7 in the Appendix). FIR and radio polarimetry reveal generally similar structures of the ordered field, but also significant differences, giving insight into the magnetic field structures in the cold and warm gas components (section 4.4.1).

Zeeman measurements in external galaxies are still very rare. Field strengths of 6 – 28 mG were detected in 14 distant starburst galaxies by the Zeeman effect in the OH megamaser emission line at 1667 MHz (McBride & Heiles 2013). These values refer to highly compressed gas clouds and are not typical for the interstellar medium. HI Zeeman measurements in nearby spiral galaxies will become possible with the Square Kilometre Array (section 5).

### 4.2 Magnetic field strengths

The dynamical importance of the total magnetic field $B_{tot}$ may be estimated by its energy density that is proportional to $B_{tot}^2$. The average strength of the component $B_{tot,\perp}$ of the total field and $B_{ord,\perp}$ of the resolved ordered field in the plane of the sky can be derived from the total and polarized radio synchrotron intensity, respectively, if energy-density equipartition between total cosmic rays and total magnetic field $B_{tot}$ is valid (section 2.2). The field strengths $B_{tot,\perp}$ are given by the mean surface brightness (intensities) of synchrotron emission, hence do not depend on the distance to the galaxy.

The observed radio continuum emission from galaxies has a contribution of thermal emission from ionized gas (and at frequencies above about 50 GHz also from dust) that needs to be subtracted to obtain the pure synchrotron part. The mean thermal fraction is about 10% at 20 cm and about 30% at 3 cm, but may increase to ≥ 50% in star-forming regions. Subtraction of the radio thermal intensity needs an independent thermal template, e.g. the H$\alpha$ intensity corrected for extinction with help of a dust model based on far-infrared data (Tabatabaei et al. 2007) or of the ratio of hydrogen recombination lines. Modeling the radio spectrum observed over a wide frequeny range also allows a separation of thermal and synchrotron intensity components (Tabatabaei et al. 2017).

The average equipartition strength of the total fields (corrected for inclination) for a sample of 31 spiral galaxies is $B_{tot}$ = 13.5 ± 5.5 μG (Tabatabaei et al. 2017). The average strength of 9 bright galaxies observed with high spatial resolution is $B_{tot}$ = 13.3 ± 3.8 μG (Table 2). Dwarf galaxies host fields of similar strength as spirals if their star-formation rate per volume is similarly high. Blue compact dwarf galaxies are radio bright with equipartition field strengths of 10 – 20 μG (Klein et al. 1991). Spirals with moderate star-forming activity and moderate radio surface brightness, like M31 (Fig. 26) and M33 (Fig. 37), have $B_{tot}$ ≈ 6 μG. In "grand-design" galaxies with massive star formation like M51 (Fig. 18), M83 (Fig. 24), and NGC6946 (Fig. 25), $B_{tot}$ ≈ 15 μG is a typical average strength of the total field.

In the spiral arms of M51 the total field strength $B_{tot}$ is 25 – 30 μG (Fig. 18). Field compression by external forces like interaction with other galaxies may amplify the fields (section 4.8). The strongest total fields in spiral galaxies (50 – 300 μG) are found in starburst galaxies like M82 (Adebahr et al. 2013; 2017; Lacki & Beck 2013), the "Antennae" NGC4038/9 (Fig. 44), in nuclear starburst regions, like the central ring of NGC1097 (Tabatabaei et al. 2018; Fig. 35), and in nuclear jets (Fig. 49).

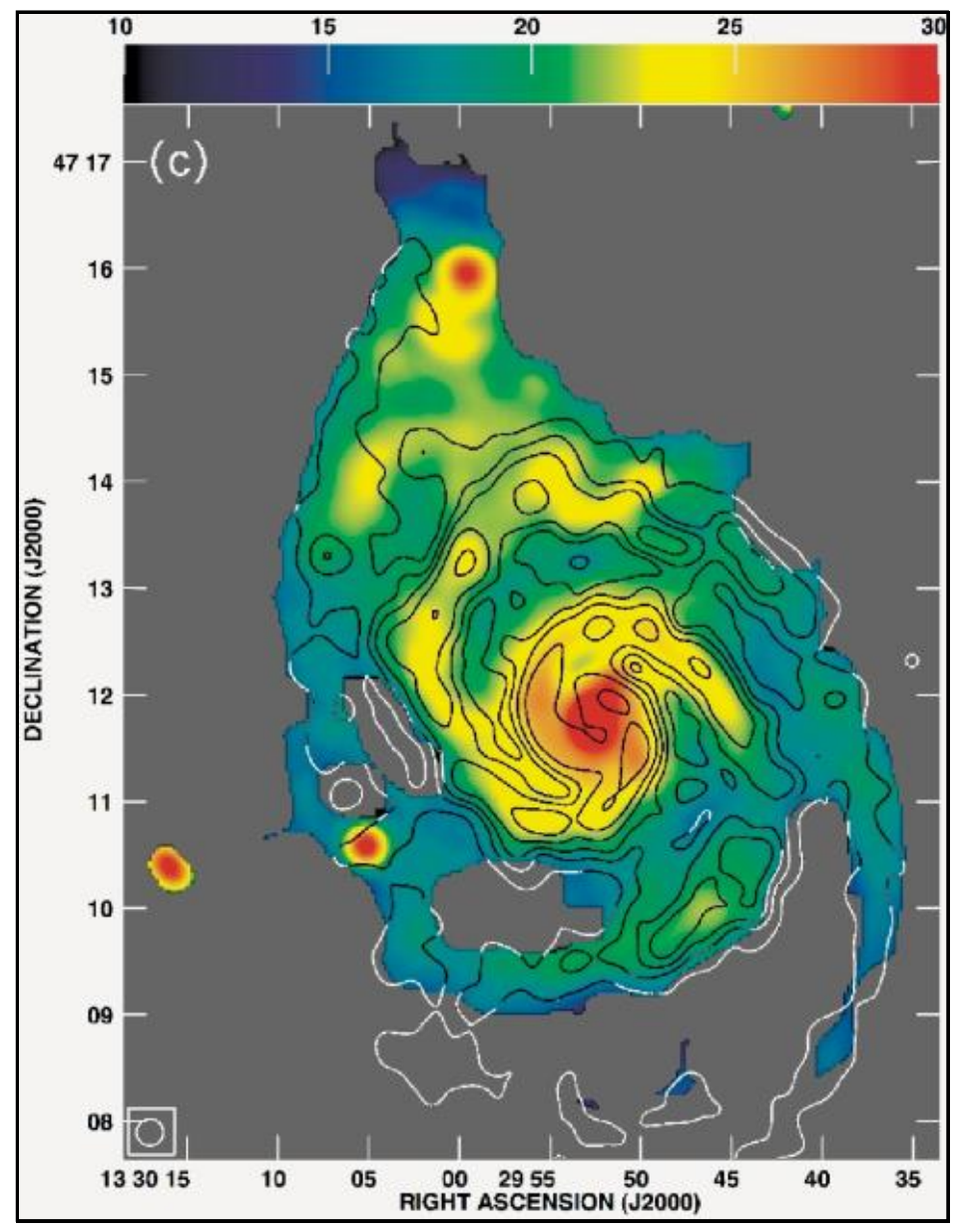

*Fig. 18: Spiral galaxy M51. Total equipartition magnetic field strengths (in μG), corrected for the inclination of the galaxy (Fletcher et al. 2011).*

If energy losses of cosmic-ray electrons are significant in starburst regions or massive spiral arms, the equipartition values are lower limits (section 2.2). The average equipartition field strength in normal spirals is proportional to the average gas surface density, but this relation is no longer valid for starburst galaxies (Thompson et al. 2006). Due to strong energy losses of the cosmic-ray electrons or even protons, the equipartition field strength is probably underestimated by a factor of a few.

The relative importance of various competing forces in the interstellar medium can be estimated by comparing the corresponding energy densities. In the local Milky Way, the energy densities of the stellar radiation field, turbulent gas motions, cosmic rays and total magnetic fields are similar (Boulares & Cox 1990). The mean energy densities of the total magnetic field and of the cosmic rays are $\approx 10^{-11}$ erg cm$^{-3}$ in NGC6946 (Beck 2007), M63, M83, NGC4736 (Basu & Roy 2013), and IC342 (Beck 2015), and $\approx 10^{-12}$ erg cm$^{-3}$ in M33 (Tabatabaei et al. 2008), in all cases similar to that of the kinetic energy of the cold, neutral gas with density $\rho$ across the star-forming disk (Fig. 19), as predicted by dynamo models (section 2.7). The kinetic energy may be underestimated if $v_{turb}$ is larger than 7 km/s. On the other hand, the turbulent velocity tends to decrease with radius (Tamburro et al. 2009), which would enhance the ratio of magnetic to kinetic energy.

The thermal energy density of the warm ionized gas $E_{th}$ is one order of magnitude smaller than that of the total magnetic field $E_B$ , which means that the ISM in spiral galaxies is a *low-$\beta$ plasma* ($\beta = E_{th}/E_B$), similar to that in the Milky Way (Boulares & Cox 1990). The energy density of hot gas in the ISM, neglected in Fig. 19, is similar or somewhat larger than that of warm gas, depending on its volume filling factor (Ferrière 2001), so that its inclusion would not change the above result significantly. The dominance of turbulent energy over thermal energy (supersonic turbulence) was also derived from numerical ISM simulations (de Avillez & Breitschwerdt 2005; Wibking & Krumholz 2023).

The radial distribution of synchrotron intensity in many spiral galaxies can be well described by an exponential decrease with a scale length $l_{syn}$ of about 4 kpc. In case of equipartition between the energy densities of magnetic fields and cosmic rays, the scale length of the total field is $(3 - \alpha)\ l_{syn} \approx$ 15 kpc (where $\alpha \approx -0.8$ is the synchrotron spectral index). The scale length of the ordered field is even larger (Fig. 19). These are still lower limits because energy losses of cosmic-ray electrons increase with increasing distance from their origin in the galaxy's star-forming regions, and a lower density of cosmic-ray electrons needs a stronger field to explain the observed synchrotron intensity. Fields in the outer disk of galaxies can be amplified by the $\alpha$-$\Omega$ dynamo (Mikhailov et al. 2014), even without star-formation activity because turbulence can be generated by the magneto-rotational instability (MRI, section 2.7). The typical scale lengths of the density of neutral and ionized gas are only about 3 kpc, so that the magnetic field energy dominates over the turbulent energy in the outer region of galaxies if a constant turbulent velocity is assumed (Fig. 19). The speculation that magnetic fields may affect the global gas rotation (Battaner & Florido 2000; Ruiz-Granados et al. 2010; Elstner et al. 2014) needs testing by future radio observations with higher sensitivity.

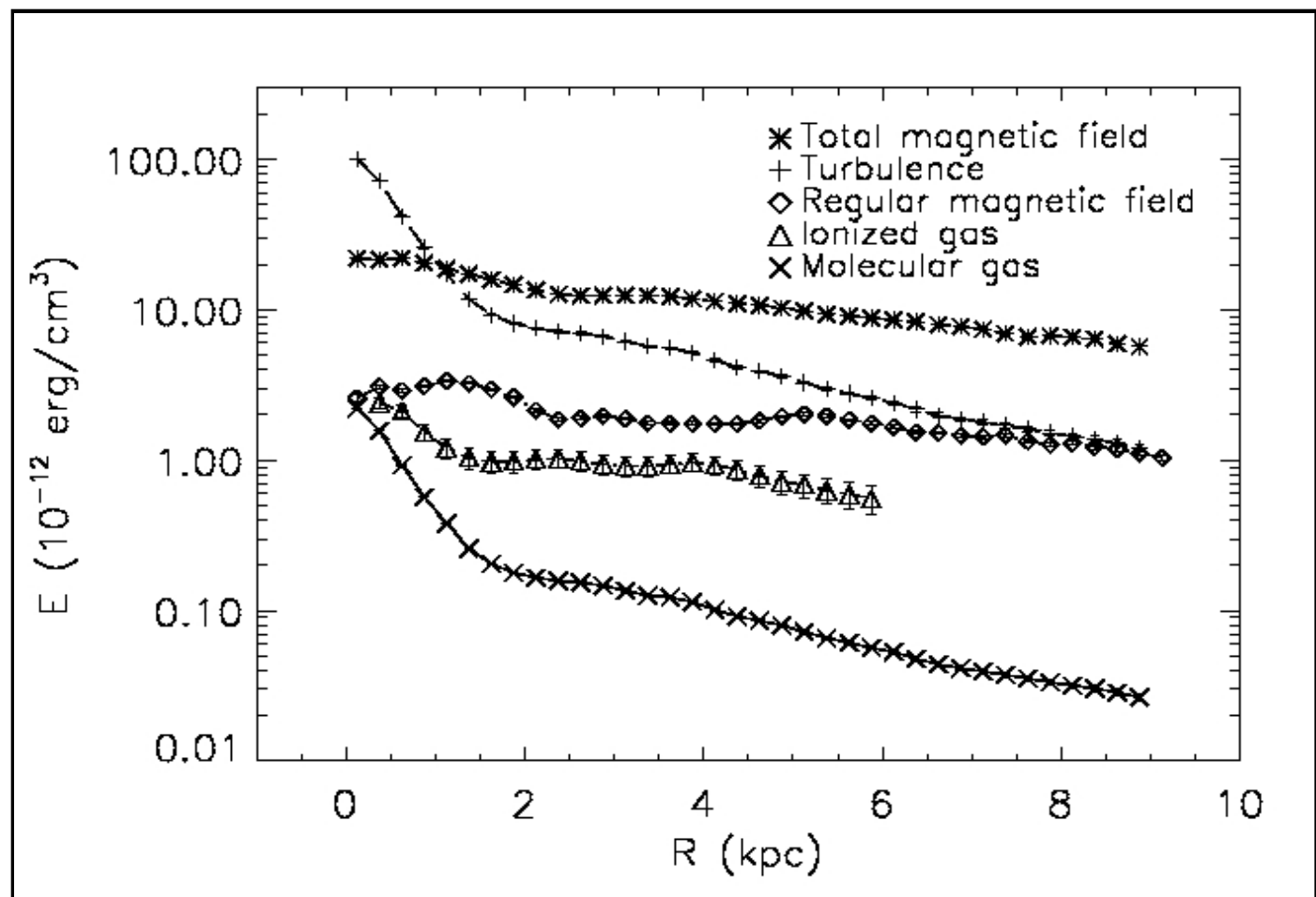

*Fig. 19: Spiral galaxy NGC6946. Radial variation of the energy densities of the total magnetic field $E_B$ ($B^2/8\pi$), the ordered (mostly regular) magnetic field ($B_{reg}^2/8\pi$), the turbulent motion of the neutral gas $E_{turb}$ ($0.5\ \rho_n\ v_{turb}^2$, where $v_{turb} \approx 7$ km/s), the thermal energy of the ionized gas $E_{th}$ ($1.5\ n_e\ k\ T_e$), and the thermal energy of the neutral molecular gas $E_n$ ($1.5\ \rho_n\ k\ T_n$), determined from observations of synchrotron and thermal radio continuum, and the CO and HI line emissions (Beck 2007).*

In spiral arms the typical degree of radio polarization is only a few %. The total field $B_\perp$ in the spiral arms is mostly *isotropic turbulent* with random orientations within the telescope beam (typically a few 100 pc at the distances to nearby galaxies). The ratios of isotropic turbulent fields to resolved ordered fields are ≥ 5 in spiral arms and circum-nuclear starburst regions, 0.5 – 2 in interarm regions, and 1 – 3 in radio halos. Turbulent fields in spiral arms can be generated by gas turbulence due to supernovae (de Avillez & Breitschwerdt 2005) or due to accretion shocks by gravitational collapse (Pfrommer et al. 2022), driving a small-scale dynamo (section 2.7), or due to spiral shocks (Dobbs & Price 2008).

Magnetic turbulence occurs over a large spectrum of scales. The maximum scale of the turbulence spectrum in the Milky Way derived from the dispersion of rotation measures of pulsars is $d \approx 50$ pc (Rand & Kulkarni 1989). This scale can also be estimated from beam depolarization by the superposition of emission from turbulent fields at centimeter wavelengths (section 2.3): For a typical degree of polarization of 3% in spiral arms, 500 pc resolution in nearby galaxies, and 1 kpc pathlength through the turbulent medium, $d \approx 70 \text{ pc } f^{1/3}$ where f is the filling factor of the magneto-ionic medium. At decimeter radio wavelengths the same turbulent field causes internal Faraday dispersion (section 2.4). Typically, $DP \approx 0.1$ at 20 cm ($\sigma_{RM} \approx 50 \text{ rad m}^{-2}$). An average electron density of the thermal gas of $0.03 \text{ cm}^{-3}$ and an average strength of the turbulent field of 10 μG yields $d \approx 130 \text{ pc} \cdot f$. The two estimates agree for $d \approx 50$ pc and $f \approx 0.4$, consistent with the results derived with other methods.

Faraday dispersion can also be used to measure the strength of isotropic turbulent magnetic fields. However, the achievable accuracy is limited because the ionized gas density has to be determined from independent measurements. The increase of the mean degree of polarization at 1.4 GHz with increasing distance from the plane of edge-on galaxies can constrain the parameters and, for NGC891 and NGC4631, yields strengths of the isotropic turbulent magnetic fields in the plane of 11 μG and 7 μG, and scale heights of 0.9 kpc and 1.3 kpc, respectively (Hummel et al. 1991).

The strength of the *resolved ordered* (large-scale regular and/or anisotropic turbulent and/or tangled regular) fields $B_{ord}$ in spiral galaxies is determined from the total equipartition field strength and the degree of polarization of the synchrotron emission. Present-day observations with typical spatial resolutions of a few 100 pc give an average value of $B_{ord} = 5.8 \pm 1.9$ μG (Table 2). $B_{ord}$ is almost linearly correlated with the rotational velocity (Tabatabaei et al. 2016). More rapidly rotating/more massive galaxies have stronger magnetic fields than more slowly rotating / less massive galaxies, possibly due to higher streaming velocities of the gas, to which the magnetic field is coupled.

*Table 2: Average strengths and ratios of the different magnetic field components for the sample of the 9 best-observed spiral galaxies, calculated from Table 3 in Beck et al. (2019). For the definition of the field components see Table 1.*

| $\langle B_{tot} \rangle$ (μG) | $\langle B_{ord} \rangle$ (μG) | $\langle B_{reg} \rangle$ (μG) | $\langle B_{an} \rangle$ (μG) | $\langle B_{ord} / B_{tot} \rangle$ | $\langle B_{reg} / B_{tot} \rangle$ | $\langle B_{reg} / B_{ord} \rangle$ |
|---|---|---|---|---|---|---|
| 13.3 ± 3.8 | 5.8 ± 1.9 | 1.7 ± 0.6 | 5.5 ± 2.0 | 0.45 ± 0.13 | 0.14 ± 0.09 | 0.32 ± 0.14 |

The ordered field is generally strongest in the regions between the optical spiral arms with peaks of about 12 μG, e.g. in NGC6946 (Fig. 25), is oriented parallel to the adjacent optical spiral arms and can be stronger than the tangled field. In several galaxies like in NGC6946 the field forms coherent *magnetic arms* between the optical arms (section 4.4.1). In galaxies with strong density waves some of the ordered field is concentrated at the inner edge of the spiral arms, e.g. in M51 (Fig. 21). Density waves enhance anisotropic turbulent fields that contribute mostly to the ordered field (section 4.4.1).

The strength of the *regular* (coherent) component of the ordered field can in principle be determined from Faraday rotation measures (section 2.4), if the mean electron density is known. In the Milky Way, the pulsar dispersion measure DM is a good measure of the thermal electron content along the pathlength to the pulsar. Only 19 extragalactic radio pulsars have been found so far, all in the LMC and SMC. In all other galaxies, the only source of information on electron densities of the warm ionized medium comes from thermal emission, e.g. in the Hα line. However, thermal emission is dominated by the HII regions that have a small volume filling factor, while Faraday rotation is dominated by the diffuse ionized emission with a much larger filling factor. Assuming the average thermal electron density of the diffuse ionized medium to be proportional to the average star-formation rate and hence to the average strength of the total field in a galaxy (section 4.3), the Faraday rotation measures of the 9 best-observed galaxies yield an average strength of the large-scale regular field of $B_{reg}$ = 1.7 ± 0.6 μG (Table 2). With the average strengths of the ordered and total fields (see above), the average ratio of the strengths of large-scale regular to ordered fields is 0.32 ± 0.14, which means that more than 90% of the ordered field is anisotropic turbulent (or tangled regular), and the average ratio of the strengths of large-scale regular to total fields is 0.14 ± 0.09. This ratio is consistent with the prediction of $\alpha$-$\Omega$ dynamo models.

The locally strongest regular field of about 8 μG was found in interarm regions of NGC6946 (Beck 2007). The similarity between the average regular (RM-based) and the ordered (equipartition-based) field strengths in NGC6946 and several other galaxies demonstrates that both methods are reliable and hence no major deviations from equipartition occur in this galaxy on scales of a few kpc.

### 4.3 The radio – infrared correlation

The highest total radio intensity (mostly synchrotron emission, tracing the total, mostly turbulent field) generally coincides with the strongest emission from dust and gas in the spiral arms. The total radio and far-infrared or mid-IR intensities are highly correlated within star-forming galaxies. The radio – IR correlation also holds between the integrated luminosities of galaxies (e.g. Bell 2003; Basu et al. 2015b; Gürkan et al. 2018; Smith et al. 2021). It is one of the tightest correlations known in astronomy. The tightness even increases when the total IR luminosity is used (Sinha et al. 2022). The tightness needs multiple feedback mechanisms that are not yet fully understood (Lacki et al. 2010).

Synchrotron intensity depends on the density of cosmic-ray electrons (CREs) that are accelerated in supernova remnants and diffuse into the interstellar medium, and on about the square of the strength of the total magnetic field (section 2.2). Infrared intensity between wavelengths of about 20 μm and 70 μm (emitted from warm dust particles in thermal equilibrium, heated mainly by UV photons) is a measure of the star-formation rate. Below about 20 μm wavelength, large PAH particles and stars contribute; emission beyond about 70 μm comes from cold dust heated by the general radiation field.

The exponent of the correlation within galaxies is different in the spiral arms and the interarm regions (Dumas et al. 2011; Basu et al. 2012). Deviations from the correlation may occur in interacting galaxies, where the magnetic field is compressed by shear or shock fronts, and in galaxies with strong magnetic arms where B is enhanced in interarm regions with low star-formation rate (Tabatabaei et al. 2013a; section 4.4.1). The scale-dependent radio – IR correlation (using wavelet functions) is strong at large spatial scales, but breaks down below a scale of a few 100 pc, which can be regarded as a measure of the electron diffusion length (Tabatabaei et al. 2013b).

The radio synchrotron – IR correlation requires that magnetic field amplification and star formation are connected. In the “electron calorimeter” model, valid for galaxies with very high star-formation rates (SFR) and strong fields where energy losses of the cosmic-ray electrons are strong, $B^2$ increases with the infrared luminosity and a linear radio – IR correlation is obtained (Lisenfeld et al. 1996). On the other hand, galaxies with low or medium SFR are no calorimeters because CREs can escape from the galaxy. A combination of several processes with self-regulation is needed to explain the correlation. If the dust is warm and optically thick to UV radiation, the IR intensity is proportional to the local SFR. The *non-calorimetric* scenario proposed by Niklas & Beck (1997) is based on two relations: (1) the coupling of magnetic fields to the neutral gas clouds ($B \sim \rho^a$, where $\rho$ is the neutral gas density, $a \approx$ 0.3…0.5; Niklas & Beck 1997; Heesen et al. 2023a; Manna & Roy 2023) and (2) the Schmidt-Kennicutt law ($SFR \sim \rho^b$, $b \approx 1.4$). Depending on the values of the exponents *a* and *b*, and whether or not equipartition between the energy densities of magnetic fields and cosmic rays is valid, a sub-linear, linear, or super-linear synchrotron – IR correlation is obtained (Dumas et al. 2011; Basu et al. 2012). Present data favour a super-linear relation with slope ≈ 1.1 (Li et al. 2016; Gürkan et al. 2018; Sinha et al. 2022), consistent with the model of a small-scale magnetic dynamo by Pfrommer et al. (2022). Real galaxies are probably semi-calorimeters, with correlations being between linear and non-linear. The two competing factors are magnetic field strength, regulating synchrotron lifetime, and advection speed, regulating CRE escape (Heesen et al. 2022). For FIR and radio intensities within galaxies, the correlation becomes sub-linear because propagation of CREs smears out the radio emission (Basu et al. 2012; Berkhuijsen et al. 2013; Heesen et al. 2019; Roy & Manna 2021).

In the non-calorimeter model, the correlation suggests the strength of the total (mostly isotropic turbulent) field is related and to the gas density (Basu et al. 2012) or the star-formation rate (Chyży 2008; Heesen et al. 2014) or the energy rate injected by supernovae (Li et al. 2016). The tightest relation in a sample of 55 dwarf and spiral galaxies is that between the average total field and the average star formation rate per galaxy surface area, with an exponent of 0.33 ± 0.03 (Chyży et al. 2017). Similar exponents were found by Heesen et al. (2014; 2019) and Tabatabaei et al. (2017). An exponent of 1/3 and an exponent of 4/3 for the synchrotron – IR correlation is predicted by field amplification by the small-scale dynamo (Schleicher & Beck 2016). In dwarf galaxies, star-formation is no longer steady, so that the timescale for energy losses of CREs becomes shorter than the injection timescale and the exponent of the synchrotron – IR correlation should steepen to 5/3, to be tested with future observations. The correlation may break down in regions with very strong magnetic fields due to negative feedback on star formation, like in the central region of NGC1097 (Tabatabaei et al. 2018).

The detection of strong radio emission in distant galaxies (that is at least partly of synchrotron origin) demonstrates that magnetic fields existed already in the early Universe. The correlation holds for starburst galaxies up to redshifts of at least 4, hosting fields of several 100 μG strengths (Murphy 2009). A steepening and finally a breakdown of the radio – IR correlation is expected when the inverse Compton losses of CREs interacting with CMB photons become stronger than the synchrotron losses; the redshift of this breakdown gives information about the field evolution in young galaxies (Schleicher & Beck 2013). Detailed predictions about the investigation of magnetic fields in young galaxies with present and future radio telescopes were given by Schober et al. (2016).

### 4.4 Magnetic field structures in spiral galaxies

4.4.1 Ordered fields

At wavelengths ≤ 6 cm, Faraday rotation of the polarized synchrotron emission is generally small (except in central regions), so that polarization angles (rotated by 90°) reveal the orientations of the ordered field (that can be regular or anisotropic turbulent, section 2.3). Spiral patterns were found in almost every galaxy, with M51 as the most prominent example (Fig. 21), even in those lacking optical spiral structure, like the ringed galaxy NGC4736 (Fig. 20) and flocculent galaxies. Nearby irregular galaxies show at most some patches of spiral structure (sections 4.6 and A.2). Spiral fields are also observed in the nuclear starburst regions of barred galaxies (section 4.5).

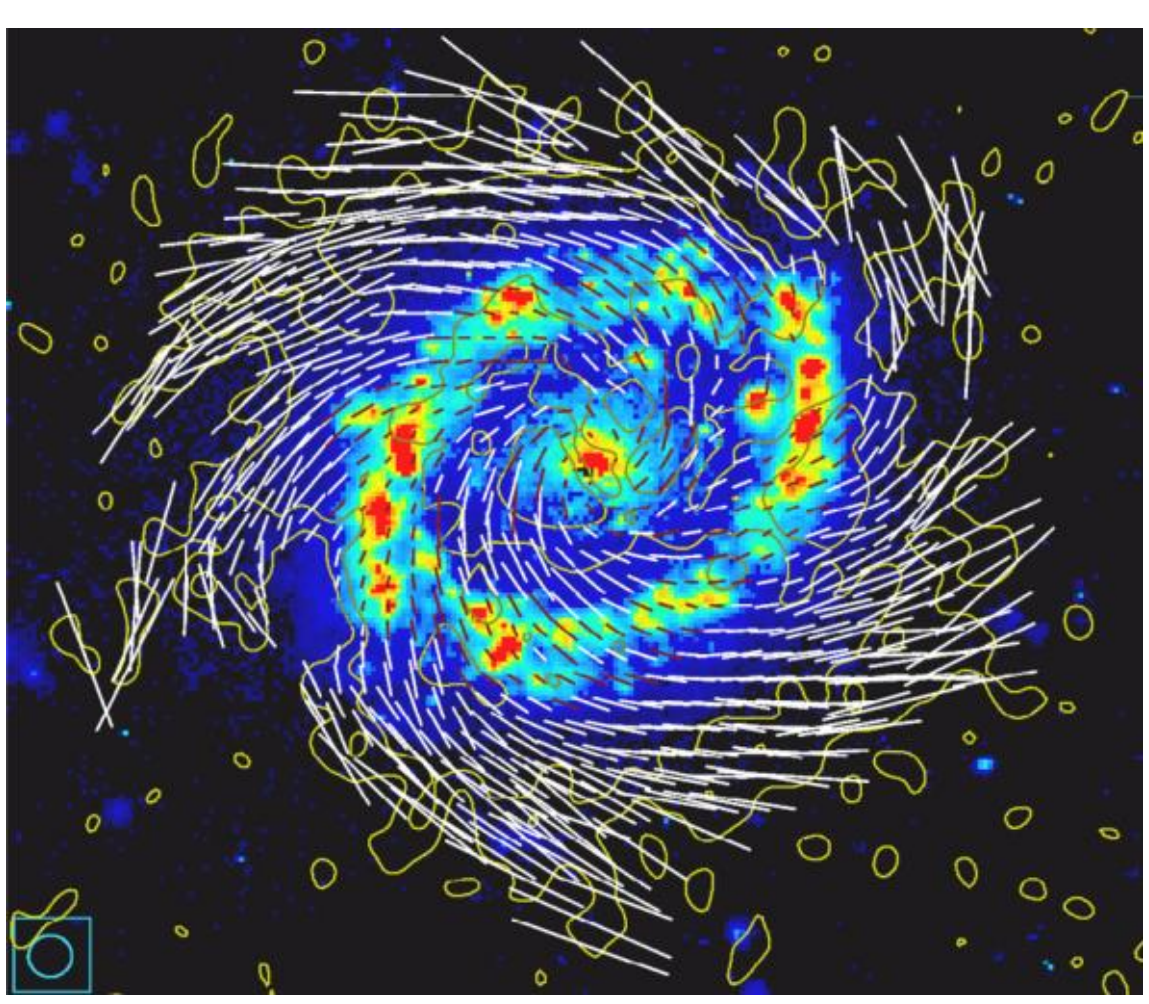

*Fig. 20: Ring galaxy NGC4736. Magnetic field orientations at 8.46 GHz (3.5 cm), observed with the VLA (Chyży & Buta 2008). The lengths of the lines are proportional to the degree of polarization. The background Hα image is from Johan Hendrik Knapen (Inst. Astr. de Canarias).*

FIR polarization (Fig. 22) and CO velocity gradients (VGT, section 2.6) trace the magnetic field in the cold molecular gas and are not affected by Faraday rotation. The VGT field orientations measured in M51 (Hu et al. 2022) as well as the FIR field orientations (Borlaff et al. 2021) are generally in good agreement with those in the diffuse ionized gas as measured in radio continuum (Fig. 21). FIR and radio polarization reveal similar spiral patterns also M83, NGC1068, and NGC4736 (Lopez-Rodriguez et al. 2022a; 2022b). A detailed comparison of spiral pitch angles revealed significant differences, especially in the inner and outer regions of the galaxies (Fig. 23). These authors also defined a magnetic alignment parameter, measuring how well the magnetic field orientations FIR and radio follow a spiral with constant pitch angle. In the galaxies M51 and M83, the field is mostly random in the inner region but well aligned along a spiral pattern at medium and large radii (Fig. 24). The analysis of field morphologies in FIR images showed spiral fields with significant deviations and mostly smaller FIR pitch angles than the radio ones (Surgent et al. 2023). – Redshifted FIR polarization showed ordered fields even in a young lensed star-forming galaxy at redshift z=2.6 (Geach et al. 2023).

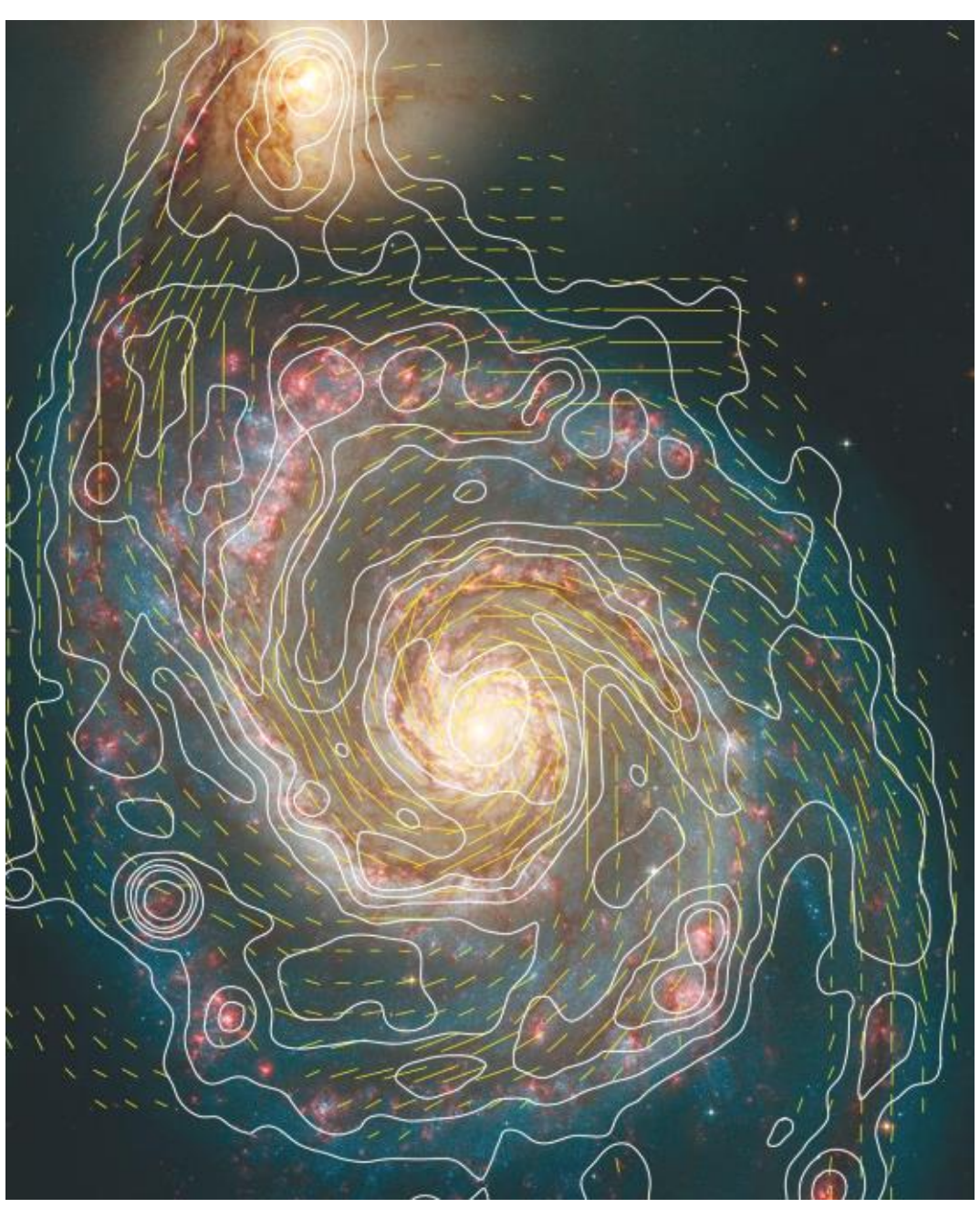

*Fig. 21: Spiral galaxy M51. Total radio intensity (contours) and magnetic field orientations at 4.86 GHz (6.2 cm), combined from observations with the VLA and Effelsberg 100-m telescopes (Fletcher et al. 2011). The lengths of the lines are proportional to the polarized intensity. The background optical image is from the HST (Hubble Heritage Team). Graphics: Sterne und Weltraum.*

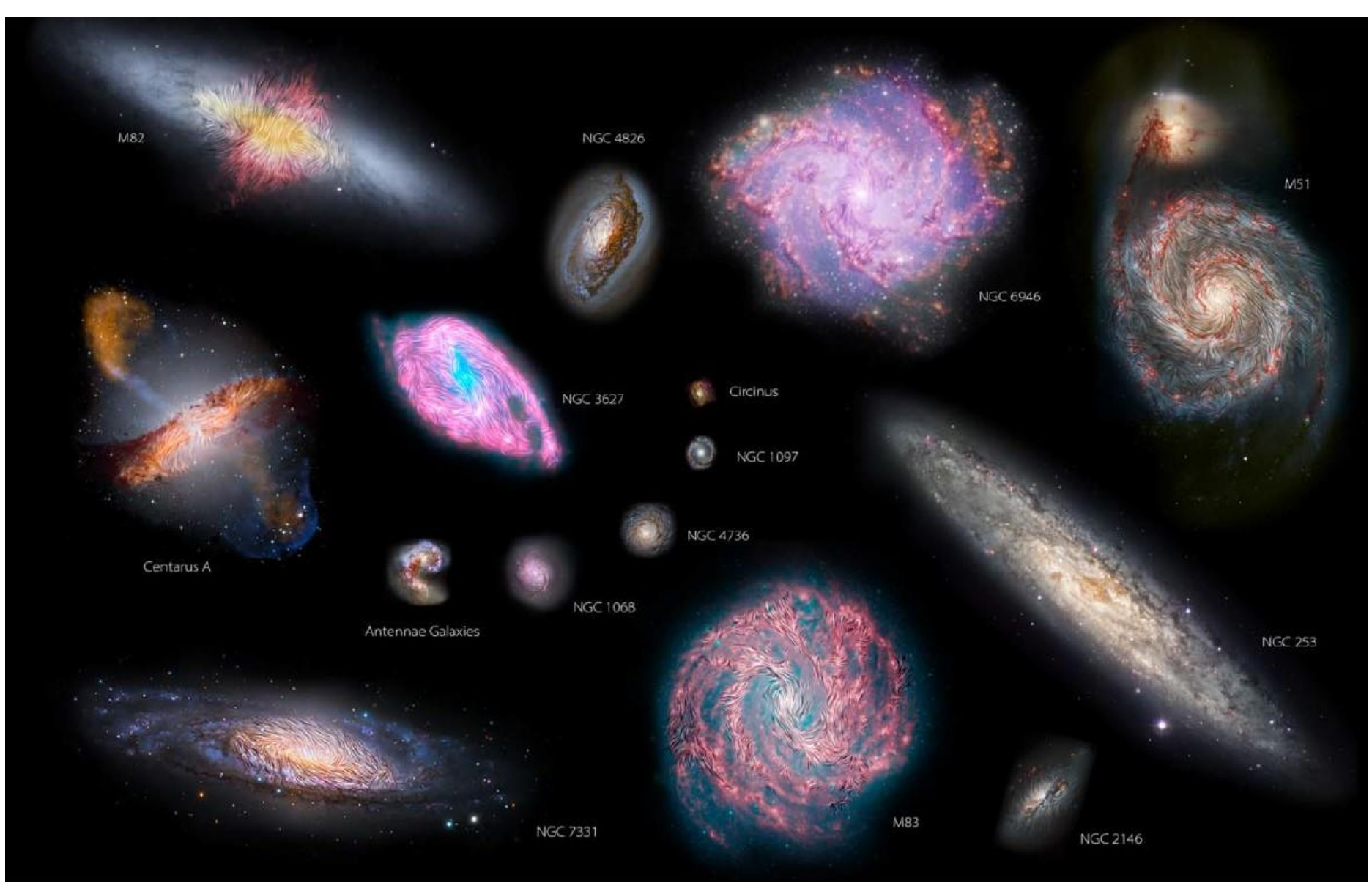

*Fig. 22: FIR magnetic field orientations in the plane of the sky, observed with SOFIA/HAWC+, represented over the galaxies in the SALSA sample, using the Line Integral Convolution technique, with RGB background images shown in their relative angular sizes (Borlaff et al. 2023).*

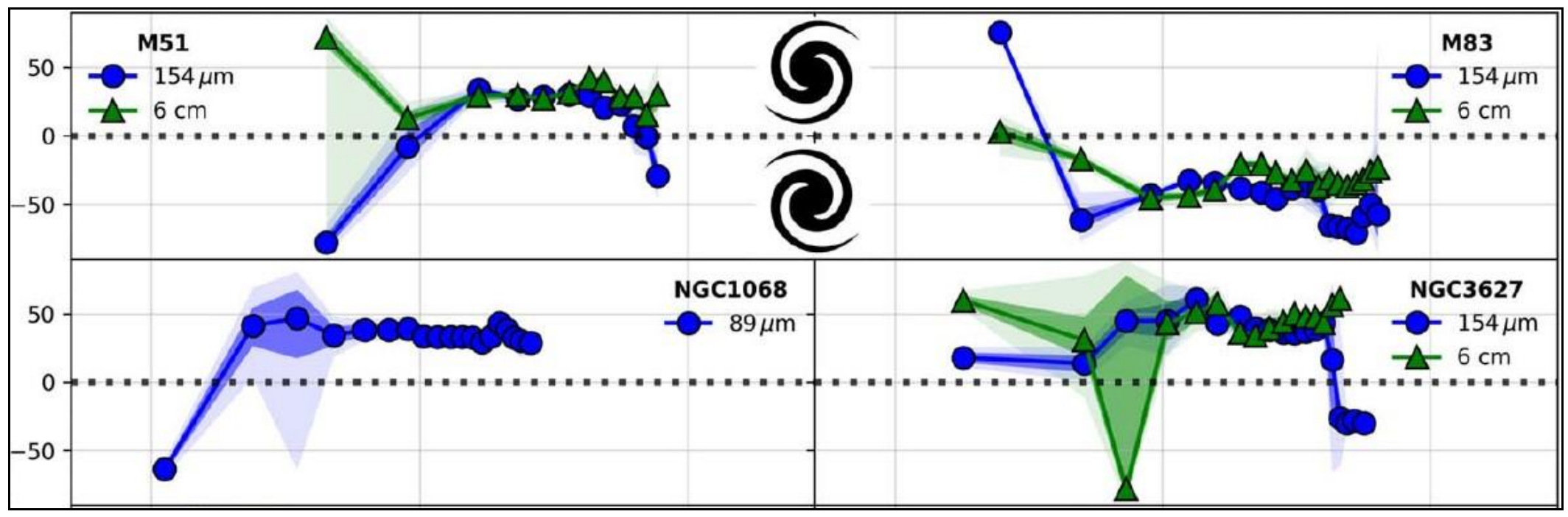

*Fig. 23: Radial profiles of the magnetic pitch angle in the spiral galaxies M51 and M83 at FIR (blue) at 154 μm and at radio (green) 6 cm wavelength (Borlaff et al. 2023).*

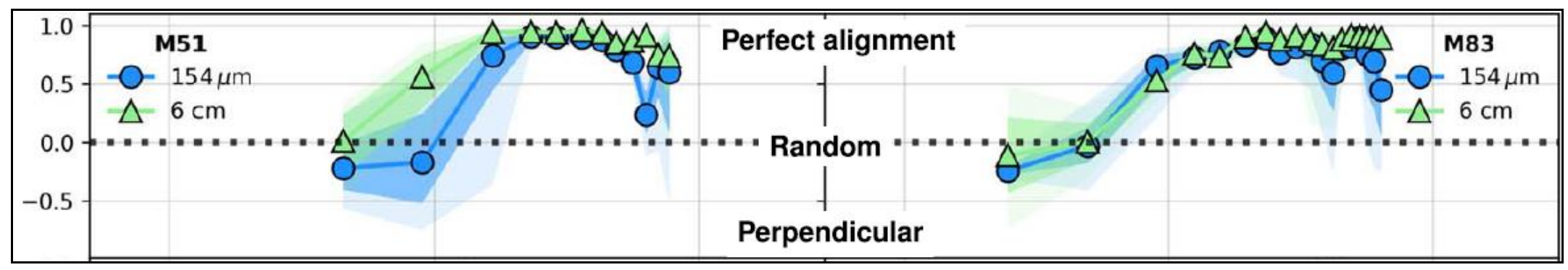

*Fig. 24: Radial profiles of the magnetic alignment parameter in M51 and M83 at FIR (blue) and radio (green). +1 corresponds to a perfect alignment along an axisymmetric spiral, 0 to a completely random B-field, and -1 to a perpendicular alignment of the average B-field (Borlaff et al. 2023).*

In gas-rich galaxies with strong density waves, the magnetic spiral pattern generally follows the spiral pattern of the gas arms. The gas flow in “smooth” galaxies (no bar, no tidal interaction, no strong density wave) is almost circular, while the field lines are spiral. If large-scale magnetic fields were frozen into the gas, differential rotation would wind them up to very small pitch angles. The observed smooth spiral patterns with significant pitch angles (10° – 40°, see Fletcher 2010) indicate a general *decoupling* between ordered magnetic fields and the gas flow due to magnetic diffusivity, which is indication for the action of α-Ω dynamos (section 2.7).

A detailed analysis revealed that the magnetic pitch angles in several spiral galaxies are similar to the pitch angles of the gaseous spiral arms, though systematically larger, by about 20° in M83 (Frick et al. 2016), by about 8° in M101 (Berkhuijsen et al. 2016), and by 12° − 23° in M74 (Mulcahy et al. 2017). This is another evidence for the action of a large-scale dynamo, where the magnetic field is not coupled to the gas flow and obtains a significant radial component.

However, the spiral field patterns cannot be solely explained by α-Ω dynamo action. In the prototypical density-wave galaxy M51, for example, the pitch angles of the magnetic lines are similar to those of the gaseous arms in the inner galaxy, but major deviations occur in the outer parts of the galaxy, where the tidal effects of the companion galaxy are strong (Patrikeev et al. 2006). In dynamo theory, the pitch angle of the magnetic lines depends on global parameters (Van Eck et al. 2015; Chamandy et al. 2016), so that it is difficult to obtain a pitch angle similar to that of the gas.

If the beautiful spiral pattern of M51 seen in radio polarization (Fig. 21) were due to a dominating regular field, it should be accompanied by a large-scale pattern in Faraday rotation, which is not observed (Kierdorf et al. 2020). This means that most of the ordered field is *anisotropic turbulent*. From an analysis of dispersions of the radio polarization angles at 6.2 cm in M51, Houde et al. (2013) measured a ratio of the correlation lengths parallel and perpendicular to the local ordered magnetic field of 1.83 ± 0.13. The anisotropic field is strongest at the positions of the prominent dust lanes on the inner edge of the inner gas spiral arms, due to compression and/or shear of isotropic turbulent fields in a density-wave shock (Patrikeev et al. 2006). Regular fields also exist in M51, but are much weaker (section 4.4.2). In the outer parts of M51, ordered fields coincide with the outer spiral arms in the south and south-west; these are possibly tidal arms with strong shear. The field in the north-east deviates from the gas arm and points towards the companion, a signature of interaction.

M74 (Mulcahy et al. 2017), M83 (Frick et al. 2016), IC342 (Fig. 27), and NGC2997 (Han et al. 1999) are cases similar to M51, with enhanced ordered (probably anisotropic turbulent) fields at the inner edges of the inner optical arms, ordered fields in interarm regions and ordered fields coinciding with the outer optical arms. Density-wave galaxies with less star-formation activity, like M81 (Krause et al. 1989b) and NGC1566 (Ehle et al. 1996), show little signs of field compression and their ordered fields occur mainly in the interarm regions.

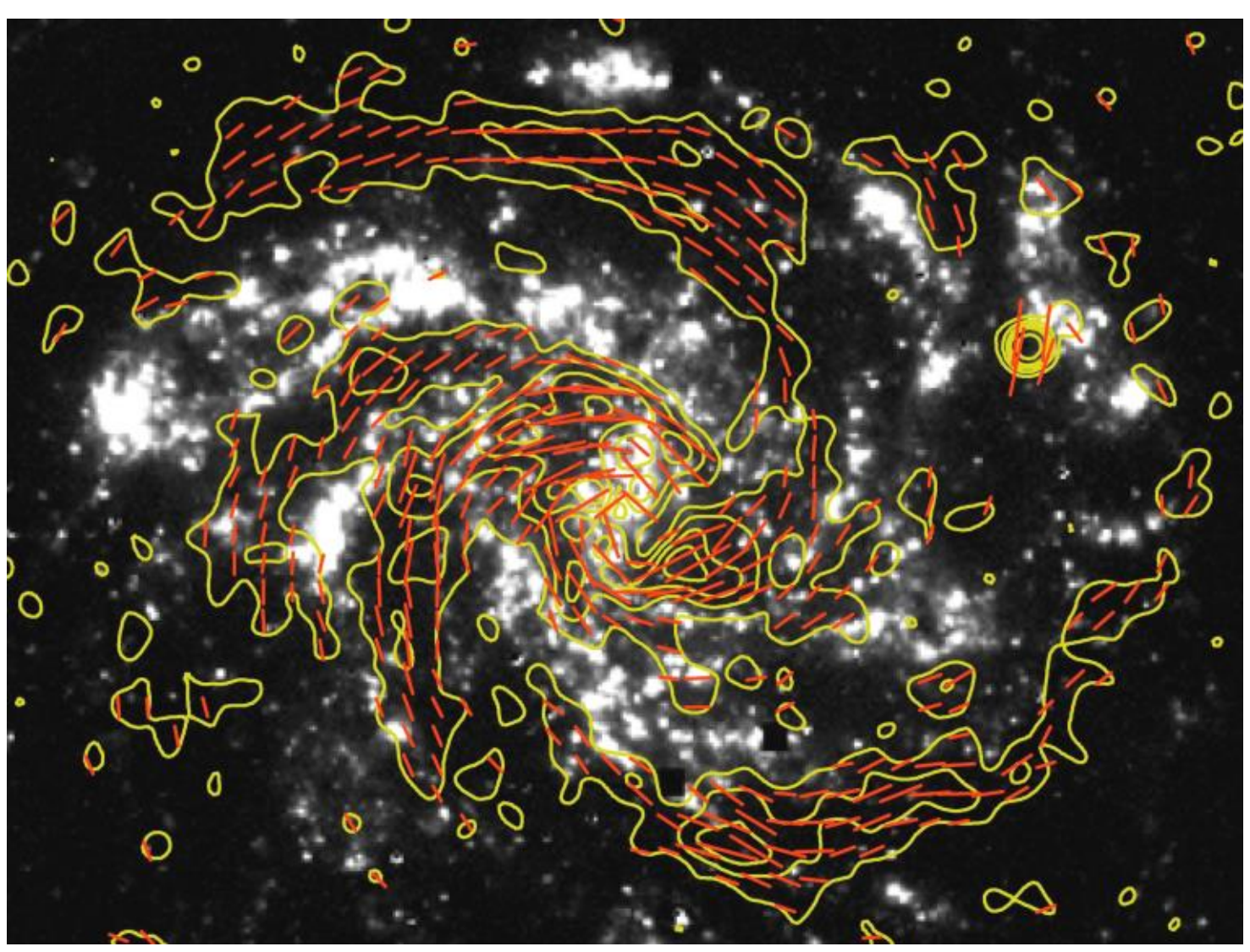

*Fig. 25: Spiral galaxy NGC6946. Polarized radio intensity (contours) and magnetic field orientations at 4.86 GHz (6.2 cm), combined from observations with the VLA and Effelsberg 100-m telescopes (Beck 2007). The lengths of the lines are proportional to the polarized intensity. The background H$\alpha$ image is from Anne Ferguson. Graphics: Sterne und Weltraum.*

Observations of another gas-rich spiral galaxy, NGC6946, revealed an amazing pattern of polarized emission with two symmetric *magnetic arms* located in interarm regions, with orientations parallel to the adjacent optical spiral arms and no signs of compression at the inner edges of the gas arms (Fig. 25). These are seen at several radio wavelengths and hence cannot be explained by reduced Faraday depolarization in the interarm regions. The degree of polarization is exceptionally high (up to 50%): the field is highly ordered and mostly regular, as indicated by Faraday rotation measures (Fig. 32). With the higher sensitivity achieved at 20 cm wavelength, more magnetic arms appear in the northern half of NGC6946, extending far beyond the optical arms, located between outer HI arms. FIR polarization shows that the field in the cold gas is restricted to the optical spiral arms (Borlaff et al. 2023).

Magnetic arms have also been found in the western part of M74, the north-west region of IC342 (Fig. 27), in NGC2997, and in several other gas-rich spiral galaxies (see compilation in Beck et al. 2019). Magnetic arms can be explained in the framework of $\alpha$-$\Omega$ dynamo models as a superposition of dynamo modes (see discussion in Beck 2015b). Alternatively, magnetic reconnection in the interarm regions may contribute to the field ordering (Weżgowiec et al. 2016; 2020).

In the Andromeda galaxy, M31 (Fig. 26), the spiral arms are hard to distinguish due to the high inclination. Star formation activity is concentrated to a limited radial range, the “ring”, at around 10 kpc distance from the center. The fields are strongest in the massive dust lanes, where the degree of polarization reaches about 40%. The field follows the “ring” with a coherent direction (Fig. 29) and hence forms a large-scale regular field.

Ordered magnetic fields may form spiral features that are disconnected from the optical spiral pattern. For example, long, highly polarized filaments were discovered in the outer regions of IC342, where only faint arms of HI line emission exist (Krause et al. 1989a). Sensitive observations at 20 cm revealed a system of such features extending to large distances from the center (Beck 2015a). In contrast to NGC6946, there is only a rudimentary magnetic arm in an interarm region in the north-west of IC342 (Fig. 27), probably because of weaker dynamo action compared to NGC6946. A polarization narrow arm of about 300 pc width, displaced inwards with respect to the inner arm east of the central region of IC342, indicates that magnetic fields are compressed by a density wave, like in M51. A broad polarization arm of 300 – 500 pc width around the northern optical arm shows systematic variations in polarized emission, polarization angles and Faraday rotation measures on a scale of about 2 kpc, indicative of a helically twisted flux tube generated by the Parker instability (Beck 2015a).

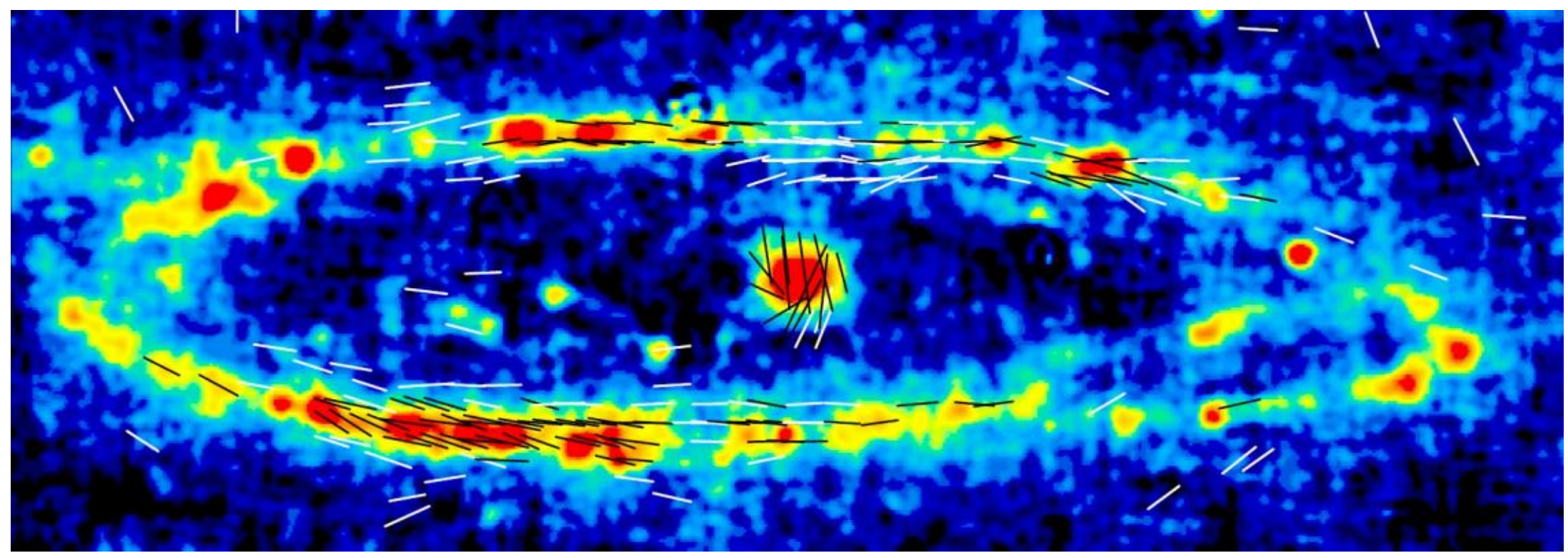

*Fig. 26: Spiral galaxy M31. Total radio intensity (colour) and magnetic field orientations at 8.35 GHz (3.6 cm), observed with the Effelsberg telescope (Beck et al. 2020). The lengths of the lines are proportional to the polarized intensity.*

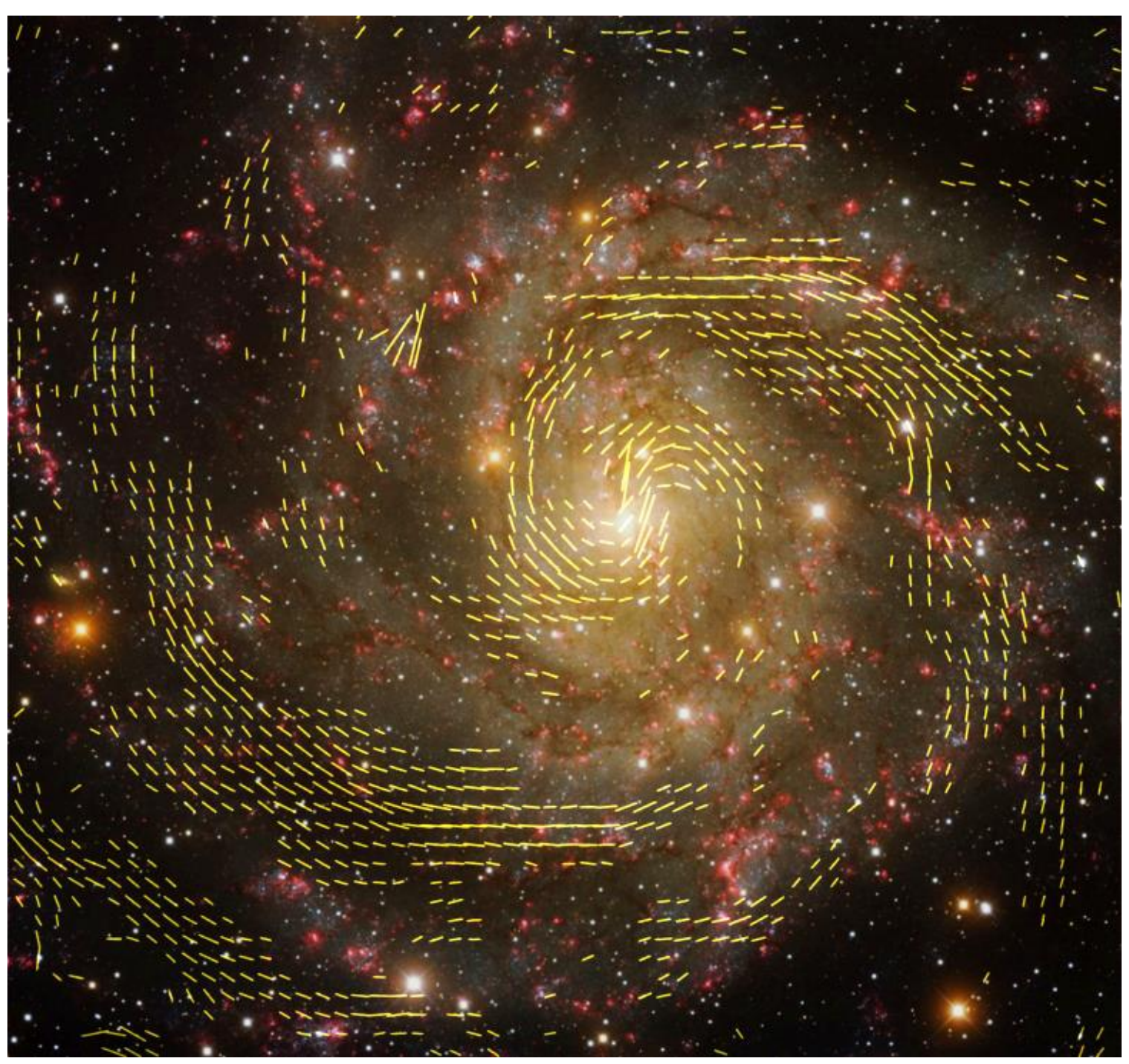

*Fig. 27: Spiral galaxy IC342. Magnetic field orientations at 4.86 GHz (6.2 cm), combined from observations with the VLA and Effelsberg telescopes (Beck 2015a). The lengths of the lines are proportional to the polarized intensity. The background colour image is from the Kitt Peak Observatory (credit: T.A. Rector, University of Alaska Anchorage, and H. Schweiker, WIYN and NOAO/AURA/NSF). A region of 16′ × 16′ (about 16 × 16 kpc) is shown. Graphics: Ulrich Klein.*

At wavelengths of around 20 cm, a striking asymmetry of the polarized intensity occurs along the major axis of all 12 spiral galaxies observed so far with sufficiently high sensitivity that have inclinations of more than about 20° (Urbanik et al. 1997; Braun et al. 2010). The emission is almost completely depolarized on one side of the major axis that is always the kinematically *receding* one (positive radial velocities). The asymmetry disappears at smaller wavelengths. This tells us that, in addition to spiral fields in the disk, large-scale regular fields in the halo exist, as predicted by $\alpha$-$\Omega$ dynamo models (section 4.7). The effect of such halo fields becomes prominent at 20 cm because most of the polarized emission from the disk is Faraday-depolarized.

Weaker asymmetries of the polarized intensity are observed at 6 cm in 18 edge-on galaxies, of which 13 show less emission on the receding side and only 4 on the following side (Krause et al. 2020). This may be the effect large-scale halo fields, as in less inclined galaxies, or to depolarization on the far side of the disk in connection with the field geometry of a trailing spiral (Stein et al. 2020).

At even longer wavelengths, Faraday effects depolarize the synchrotron emission almost completely. With help of RM Synthesis applied to 90 cm data from the Westerbork telescope, an extremely low average degree of polarization of 0.21 ± 0.05 % could be measured in the star-forming “ring” of M31 (Gießübel et al. 2013).

### 4.4.2 Regular fields

Ordered magnetic fields, as observed by polarized emission, can be anisotropic turbulent or large-scale regular or tangled regular fields (section 2.3). *Faraday rotation measures* (RMs) are signatures of regular fields with known directions. RM is determined from multi-wavelength radio polarization observations (section 2.4). Spiral dynamo modes (section 2.7) can be identified from the periodicity of the large-scale azimuthal variation of RMs in inclined galaxy disks (Fig. 28), where the RM can be determined from diffuse polarized emission (Krause 1990) or from RMs of polarized background sources (Stepanov et al. 2008). As several dynamo modes can be superimposed, mode fitting of polarization angles at several frequencies or Fourier analysis of the azimuthal RM variation is needed. The resolution of present-day radio observations is sufficient to identify 2 – 3 modes (Fletcher 2010, update in Table 3). Major progress can be expected from broad-band spectro–polarimetric data of spiral galaxies and the application of RM Synthesis (section 2.4).

The disks of a few spiral galaxies reveal large-scale RM patterns giving strong evidence for modes generated by the $\alpha$-$\Omega$ dynamo. There is presently no other model to explain large-scale regular fields with spiral patterns. M31 is the prototype of a large-scale, dynamo-generated regular magnetic field (Beck 1982; Fig. 29). The discovery became possible thanks to the large angular extent and the high inclination of M31. The polarized intensity is largest near the minor axis, where the field component $B_\perp$ is largest (Fig. 30a), while the maxima in |RM| are observed near the major axis, where the line-of-sight field component $B_\parallel$ is strongest (Fig. 30b). This single-periodic RM variation is a signature of a dominating axisymmetric spiral (ASS) disk field (dynamo mode m = 0) (Fletcher et al. 2004) that extends to at least 25 kpc distance from the center of M31 when observed with an *RM grid* (see below) (Han et al. 1998). A weaker bisymmetric spiral (BSS) field is superimposed (Beck et al. 2020).

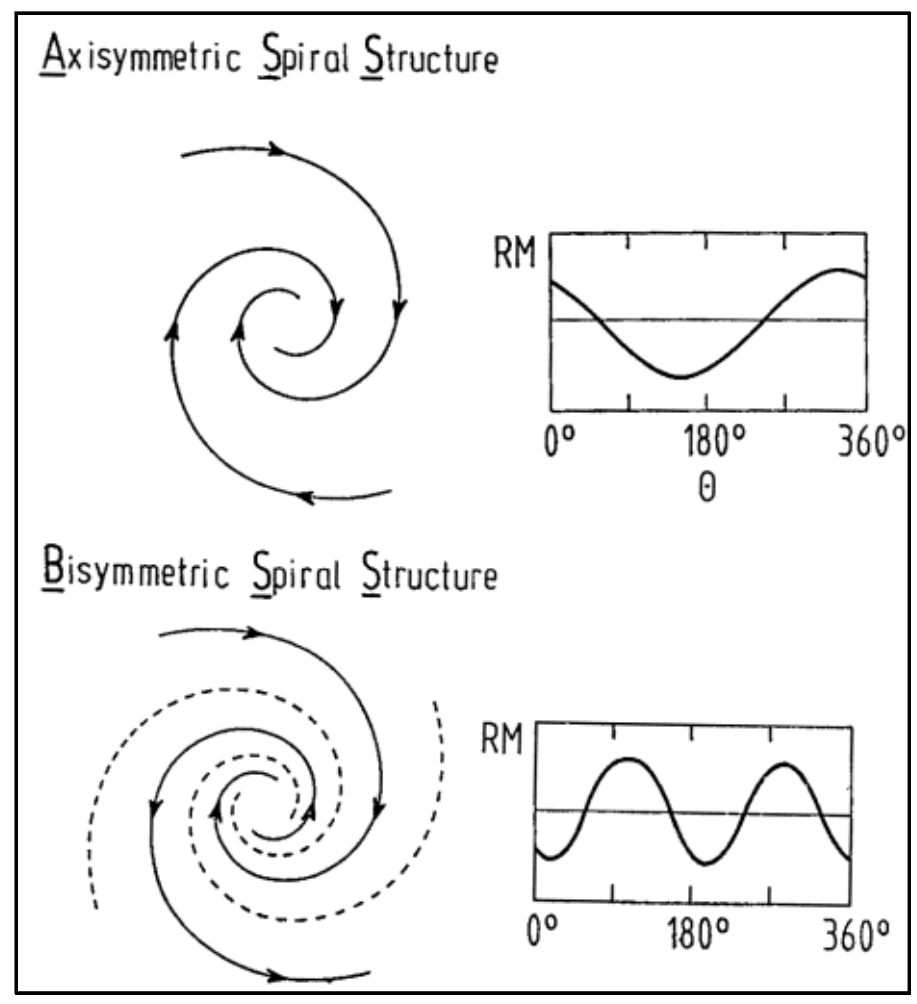

*Fig. 28: Azimuthal RM variations (measured from the major axis) for axisymmetric spiral (ASS) and bisymmetric spiral (BSS) fields in inclined galaxies (Krause 1990).*

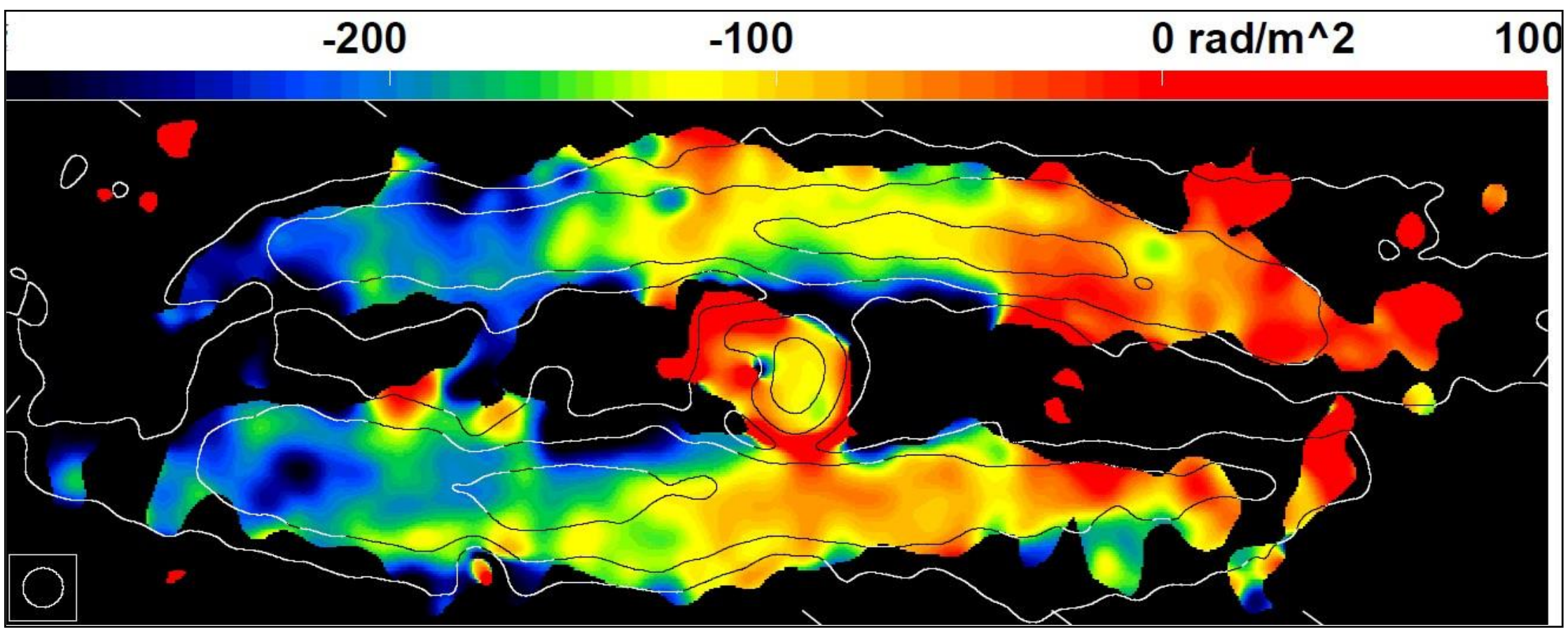

*Fig. 29: Spiral galaxy M31. Polarized radio intensity at 4.85 GHz (6.2 cm) (contours) and Faraday rotation measures between 8.35 GHz (3.6 cm) and 4.85 GHz (colours), derived from observations with the Effelsberg telescope (Beck et al. 2020). The average rotation measure of about -100 rad $m^{-2}$ is caused by the foreground medium in the Milky Way. The regular field points away from us on the left (northeast) side and towards us on the right (southwest) side of the "ring".*

Other galaxies with a dominating ASS disk field are IC342, the Virgo cluster galaxy NGC4254, the almost edge-on galaxies NGC253, NGC891, and NGC5775, the irregular Large Magellanic Cloud (LMC), and a few further candidates (Table 3 and Tables A4-A6 in the Appendix).

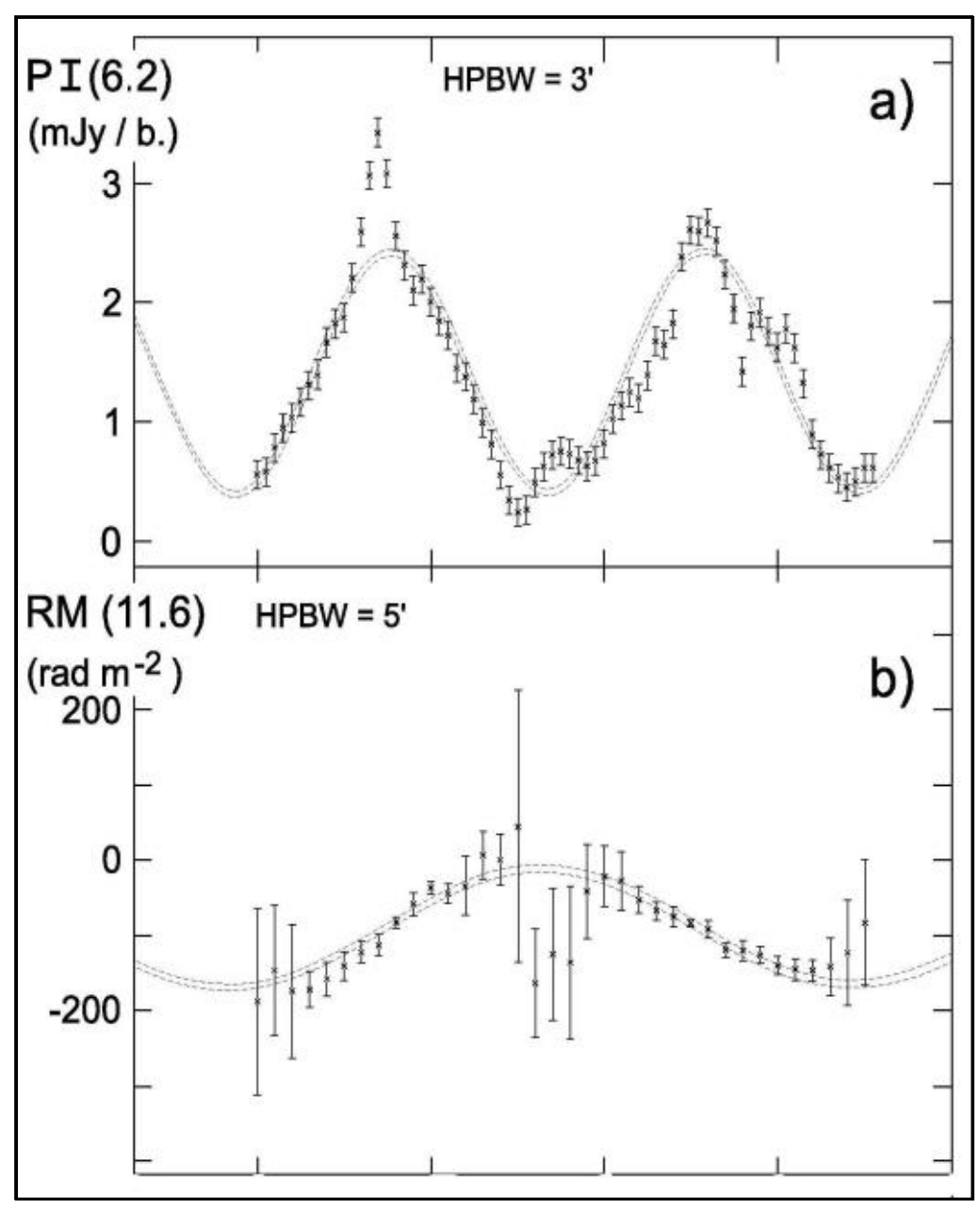

*Fig. 30: Spiral galaxy M31. (a) Polarized intensity and (b) Faraday rotation measures between 4.75 GHz (6.3 cm) and 2.7 GHz (11.1 cm) along the azimuthal angle in the plane of the galaxy, counted counter-clockwise from the northeastern major axis (the left side of Fig. 29) (Berkhuijsen et al. 2003).*

By measuring the signs of the RM distribution and the velocity field on both sides of a galaxy's major axis, the *inward* and *outward* directions of the radial component of the ASS field can be distinguished (Krause & Beck 1998). Dynamo models predict that both signs should have about similar probabilities, which is confirmed by the observations (Table 4). Three galaxies show a large-scale magnetic field in their central regions directed opposite (outward) to that in the disk (inward): M31, IC342, and NGC2997. Only one galaxy was found so far with a large-scale field reversal, from inward to outward, at a certain radius within the disk, namely NGC4666 (Fig. 31).

*Table 3: Decomposition of large-scale regular magnetic fields in galaxy disks into azimuthal modes of order m by mode fitting, using polarization angles at several frequencies averaged in the radial range given in column 2. Columns 3-5 give the amplitudes relative to the strongest mode. "?" indicates a mode that was not sought. "0" indicates that no signature of this mode was found. Modes of order >2 cannot be detected with the resolution of present-day telescopes. "a" indicates a result derived from Fourier decomposition of the azimuthal RM variation, not by mode fitting (Beck et al. 2019).*

| Galaxy | Radial Range [kpc] | $m = 0$ | $m = 1$ | $m = 2$ |
|---|---|---|---|---|
| M 31 | 6.8–9.0 | 1 | $\approx 0$ | $0.54 \pm 0.13$ |
| | 9.0–15.8 | 1 | $\approx 0$ | $\approx 0$ |
| M 31 | 9.0–11.0 | 1 [a] | $0.18 \pm 0.06$ [a] | ? |
| M 33 | 1–5 | 1 | $\approx 0$ | $0.62 \pm 0.07$ |
| M 51 (disc) | 2.4–3.6 | 1 | $\approx 0$ | $0.72 \pm 0.06$ |
| | 3.6–7.2 | 1 | $\approx 0$ | $0.52 \pm 0.07$ |
| M 51 (halo) | 2.4–3.6 | $0.30 \pm 0.09$ | 1 | $\approx 0$ |
| | 3.6–7.2 | $\approx 0$ | 1 | $\approx 0$ |
| M 81 | 9–12 | $< 0.5$ | 1 | ? |
| M 83 | 4–12 | $0.4 \pm 0.3$ [a] | 1 [a] | ? |
| NGC 253 | 1.4–6.7 | 1 | ? | ? |
| NGC 1097 | 3.75–5.0 | 1 | $0.33 \pm 0.05$ | $0.48 \pm 0.05$ |
| NGC 1365 | 2.625–14.875 | 1 | $0.9 \pm 0.6$ | $0.9 \pm 0.3$ |
| NGC 4254 | 4.8–7.2 | 1 | $0.58 \pm 0.07$ | ? |
| NGC 4414 | $\approx$2–7 | 1 | $\approx 0.6$ | $\approx 0.4$ |
| NGC 4449 | 1–3 | 1 | ? | ? |
| NGC 6946 | 0–18 | $\approx 1$ [a] | ? | $\approx 1$ [a] |
| IC 342 | 5.5–17.5 | 1 | ? | ? |
| LMC | 0–3 | 1 [a] | ? | ? |

*Table 4: Directions of the radial component of spiral fields in the disks of 15 spiral galaxies and one irregular galaxy (NGC4449) with a dominating axisymmetric mode (Beck et al. 2019).*

| Direction | Galaxies |
|---|---|
| Inward (disc) | M 33, NGC 253, NGC 6946 |
| Outward (central region) & inward (disc) | M 31, NGC 2997, IC 342 |
| Outward (disc) | M 51, M 83, NGC 891, NGC 4013, NGC 4254, NGC 4414, NGC 4449 NGC 4736, NGC 5775 |
| Inward (central region) & outward (disc) | – |
| Inward (inner disc) & outward (outer disc) | NGC 4666 |
| Outward (inner disc) & inward (outer disc) | Milky Way |

M81, M83, and a galaxy at a redshift of 0.4 in front of a radio jet (Kronberg et al. 1992) are the only candidates so far for a dominating bisymmetric spiral (BSS) field (m = 1), characterized by a double-periodic RM variation, but the data quality is limited in all these cases. Thus dominating BSS fields are rare, as predicted by dynamo models. It was proposed that tidal interaction can excite the BSS mode, but no preference for BSS was found even in the most heavily interacting galaxies in the Virgo cluster (section 4.8). The idea that galactic fields are wound-up primordial intergalactic fields and hence should be of BSS type (section 2.7) can also be excluded from the existing observations.

Faraday rotation in NGC6946 (Fig. 32) and in other similar galaxies with two magnetic arms can be described by a superposition of two azimuthal dynamo modes (m = 0 and m = 2) with about equal amplitudes, where the quadrisymmetric spiral (QSS) m = 2 mode is phase shifted with respect to the density wave (Beck 2007). This model is based on the RM pattern of NGC6946 that shows different field directions in the northern and southern magnetic arm (Fig. 32) and is supported by the detection of four field reversals along azimuthal angle (Kurahara & Nakanishi 2019). A weaker QSS mode superimposed onto the dominating ASS mode is indicated in the disk of M51 (Fletcher et al. 2011) and in the inner part of the ring of M31 (Fletcher et al. 2004). A superposition of ASS and BSS modes can describe the fields of M31, M33, and NGC4254, while three modes (ASS+BSS+QSS) are needed for several other galaxies (Table 3 and Tables A4-A6 in the Appendix).

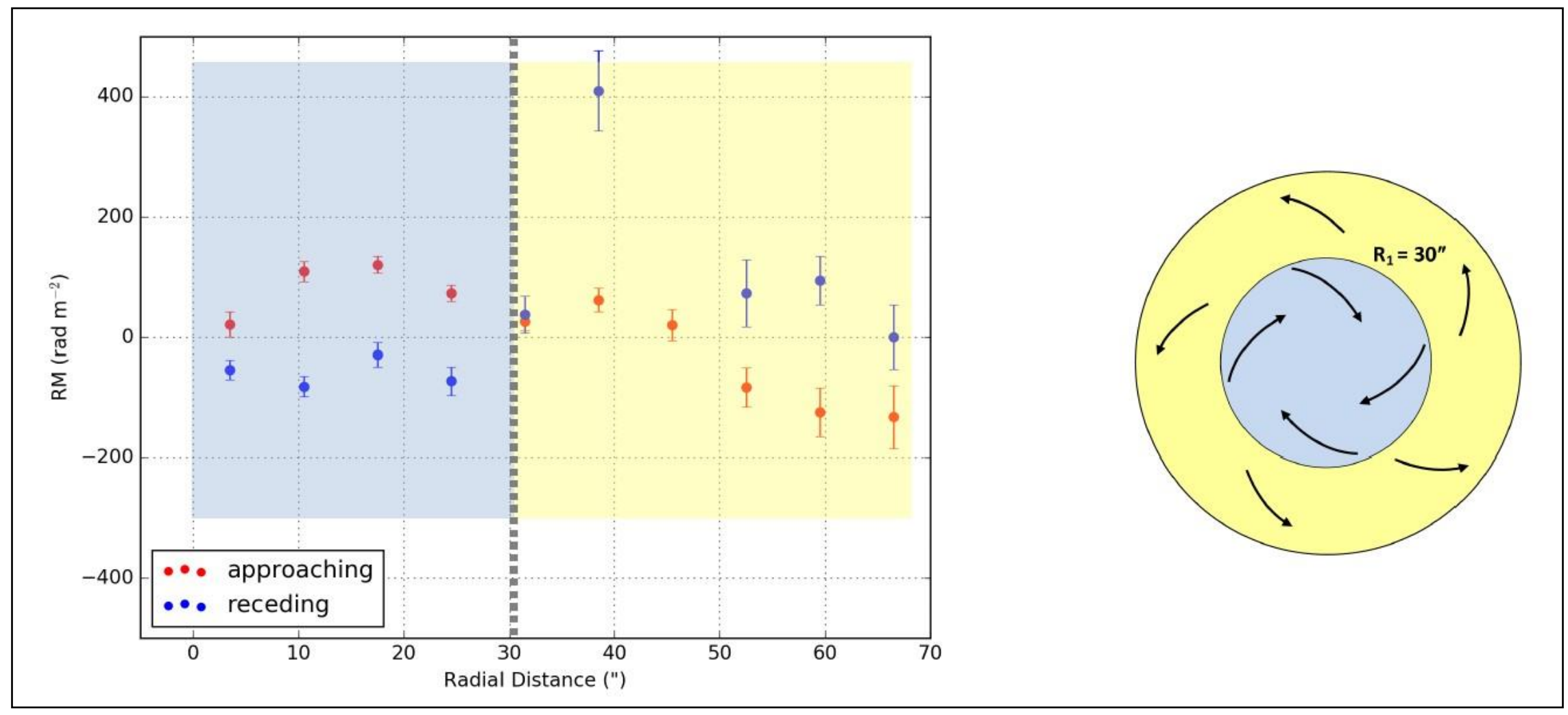

*Fig. 31: Edge-on galaxy NGC4666. Left: RMs measured from C-band observations, averaged in boxes along the galaxy plane with increasing distance from the center. RMs on the dynamically approaching side are plotted red, those on the receding side in blue. A reversal of the large-scale field is indicated at about 30" (4 kpc) radius. Right: simplified model for the ASS disk field (Stein et al. 2019a).*

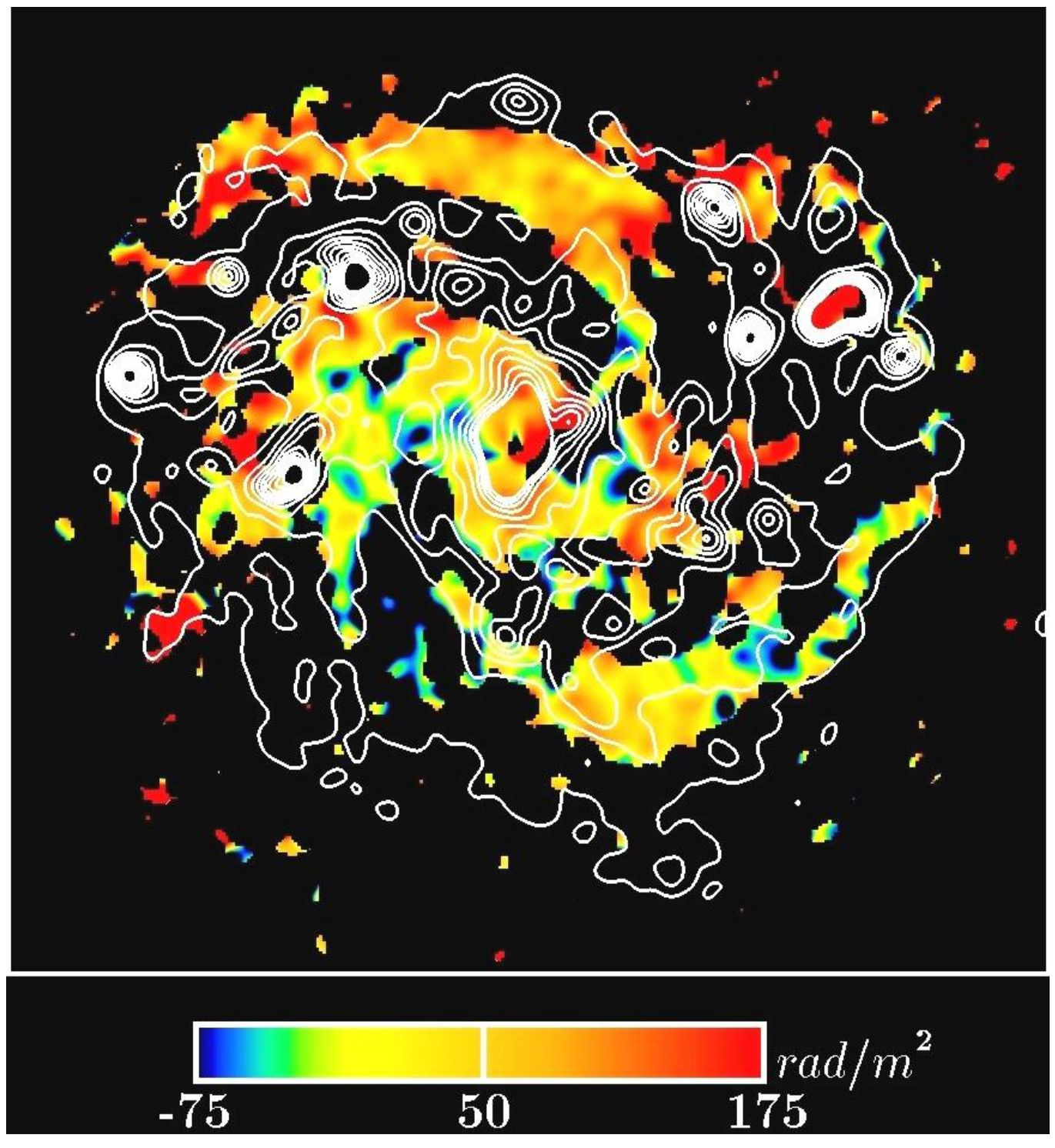

*Fig. 32: Spiral galaxy NGC6946. Total radio intensity at 4.86 GHz (6.2 cm) (contours) and Faraday rotation measures between 8.46 GHz (3.5 cm) and 4.86 GHz (colours), derived from combined observations with the VLA and Effelsberg telescopes (Beck 2007). The average rotation measure of about +50 rad $m^{-2}$ originates in the foreground medium of the Milky Way.*

In most star-forming galaxies observed so far, a spiral polarization pattern was found, often with a large-scale RM pattern as a signature of large-scale regular fields. Often the available polarization data is insufficient to derive reliable RMs. Even if the data quality is high, large-scale RM patterns are not always visible. In density-wave galaxies, strong compression and shearing flows can generate *anisotropic turbulent* fields or tangle the regular field (causing frequent reversals, but no observable RM). The anisotropic turbulent field is generally stronger than the underlying large-scale regular field (Table 2), sometimes much stronger, as in M51 (Fletcher et al. 2011; Kierdorf et al. 2020) and in IC342 (Beck 2015). Gravitational interactions with other galaxies may have disturbed the regular field. Another reason for a weak regular field could be that the timescale for the generation of large-scale modes is longer than the galaxy's lifetime, so that the regular field is not yet fully organized and restricted to limited regions (Moss et al. 2012).

Large-scale *field reversals* along the radial direction were discovered from pulsar RMs in the Milky Way (section 3.5), but only in one spiral galaxy so far (NGC4666, Fig. 31), although high-resolution RM maps of Faraday rotation are available for many spiral galaxies. A satisfying explanation is still lacking (section 3.5). The disk fields of several galaxies may be composed of a mixture of modes, where reversals emerge in a limited radial and azimuthal range of the disk, like in NGC4414 (Soida et al. 2002). Reversals on smaller scales are probably frequent, e.g. tangled regular fields, but difficult to observe in external galaxies with the resolution of present-day telescopes. In the barred galaxy NGC7479, where a jet serves as a bright polarized background (Fig. 50), several reversals on 1 – 2 kpc scale were detected in the foreground disk of the galaxy (Laine & Beck 2008). The complex RM pattern observed in NGC6946 (Fig. 32) and in M51 (Mao et al. 2015; Kierdorf et al. 2020) indicates reversals on scales of a few kpc, due to tangled regular fields in the disk and/or due to vertical regular fields in the halo (Parker instability). The similarity to the synthetic RM patterns obtained from numerical MHD simulations of the Parker instability (Rodrigues et al. 2016) is striking.

The central regions of M31 (Fig. 26), IC342 (Fig. 27), and NGC2997 host regular spiral fields that appear disconnected from the disk fields (Gießübel & Beck 2014; Beck 2015; Han et al. 1999). As the direction of the radial field component points outwards, *opposite* to that of the disk field, two separate dynamo regimes seem to be active in these galaxies.

While the azimuthal symmetry of dynamo modes is known for many galaxies, the vertical symmetry (*even* or *odd*) is much harder to determine. The toroidal field of odd modes reverses its sign above and below the galactic plane (Fig. 6 top). The RM patterns of even and odd modes observed in face-on or mildly inclined galaxies are similar. When RMs are derived from diffuse internal emission, the average RM through an odd field is about half of that in an even field, which cannot be distinguished because the ionized gas density is not known with sufficient accuracy. If RMs from background sources are used, significant RMs can be observed for even fields, while RMs from odd fields cancel. In the LMC and in M51, RMs of background sources (corrected for foreground RM) are significant, so that the large-scale fields must be of even parity (Mao et al. 2012b; 2015). The symmetry type becomes directly visible in strongly inclined galaxies as a sign reversal of RMs above and below the plane. Only even-symmetry fields were found so far, namely in M31, NGC253, NGC891, NGC4013, NGC4666, and NGC5775 (section 4.7), in agreement with the prediction of dynamo models.

In summary, magnetic field structures in spiral galaxies are as complex as in the Milky Way. The observations can best be explained by a superposition of dynamo-generated modes of regular fields coupled to the diffuse warm gas, plus anisotropic turbulent fields by shearing and compressing flows, plus isotropic turbulent fields or tangled regular fields, coupled to the cold gas in spiral arms. The magnetic fields in barred galaxies behave similarly (section 4.5). For a more detailed model of the physics of the field-gas interaction, spectro-polarimetric data (RM Synthesis, section 2.4), and high resolution with future telescopes are required (section 5).

### 4.5 Magnetic fields in barred galaxies

Gas and stars in the gravitational potential of strongly barred galaxies move in noncircular orbits. Numerical models show that gas streamlines are deflected in the bar region along shock fronts, behind which the cold gas is compressed in a fast shearing flow (Athanassoula 1992). Compression regions traced by dust lanes develop along the edge of the bar that is leading with respect to the galaxy's rotation because the gas rotates faster than the bar pattern. The warm, diffuse gas has a higher sound speed and is not compressed. According to simulations, the shearing flows around a bar amplify magnetic fields and generate complex field patterns (Otmianowska-Mazur et al. 2002). Field reversals may appear between the bar and the spiral arms (Kulpa-Dybeł et al. 2011). The asymmetric gas flow can enhance large-scale dynamo action and excite the QSS (m = 2) mode (Moss et al. 2001).

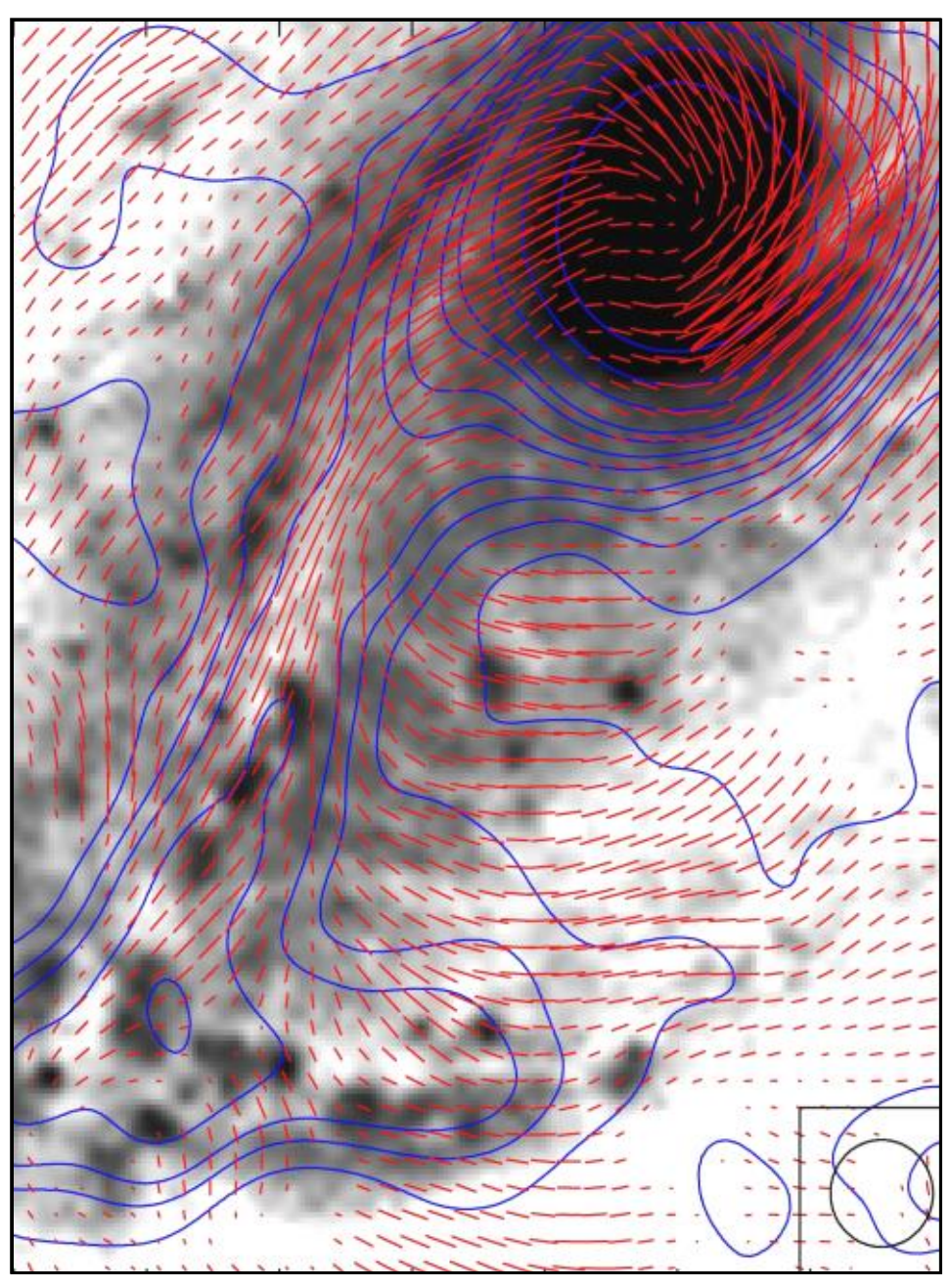

*Fig. 33: Southern half of the barred galaxy NCC1097. Total radio intensity (contours) and magnetic field orientations at 8.46 GHz (3.5 cm), observed with the VLA (Beck et al. 2005a). The lengths of the lines are proportional to the polarized intensity. The background optical image is from Halton Arp (Cerro Tololo Observatory).*

20 galaxies with large bars were observed with the Very Large Array (VLA) and with the Australia Telescope Compact Array (ATCA) (Beck et al. 2002; 2005a). The total radio luminosity (measure of the total magnetic field strength) is strongest in galaxies with high far-infrared luminosity (high star-formation activity), a result similar to non-barred galaxies. Average radio intensity, radio luminosity, and star-formation activity all correlate with the relative bar length. The pattern of the ordered field in the galaxies with long bars (NGC1097, 1365, 1559, 1672, 2442, and 7552) is significantly different from that in non-barred galaxies: Field enhancements occur outside of the bar (upstream) and the field lines are oriented at large angles with respect to the bar.

NGC1097 (Fig. 33) is one of the nearest barred galaxies and hosts a huge bar of about 16 kpc length. The total radio intensity (not shown in the figure) and the polarized intensity are strongest in the downstream region of the dust lanes (southeast of the center). This can be explained by a compression of isotropic turbulent fields in the bar's shock, leading to strong and anisotropic turbulent fields in the downstream region. The surprising result is that the polarized intensity is also strong in the upstream region (south of the center in Fig. 33), where RM data indicate that the field is regular. The pattern of field lines in NGC1097 is similar to that of the gas streamlines as obtained in numerical simulations (Athanassoula 1992). This suggests that the ordered (partly regular) magnetic field is aligned with the flow and amplified by strong shear. Remarkably, the optical image of NGC1097 shows dust filaments in the upstream region, oriented almost perpendicular to the bar and aligned with the ordered field. Between the region upstream of the southern bar and the downstream region the field lines smoothly change their orientation by almost 90°. The ordered field is probably coupled to the diffuse gas (having a larger sound speed) and thus avoids being shocked in the bar. The magnetic energy density in the upstream region is sufficiently high to affect the flow of the diffuse gas.

NGC1365 (Fig. 34) is similar to NGC1097 in its overall properties, but the polarization data indicate that shear due to the bar is weaker. The ordered field bends more smoothly from the upstream region into the bar, again with no indication of a shock. M83 is the nearest barred galaxy, but with a short bar; it shows compressed ordered fields at the leading edges of the bar on both sides of the nucleus and some polarization in the upstream regions (Frick et al. 2016). In all other barred galaxies observed so far (section A.2) the resolution is insufficient to separate the bar and upstream regions.

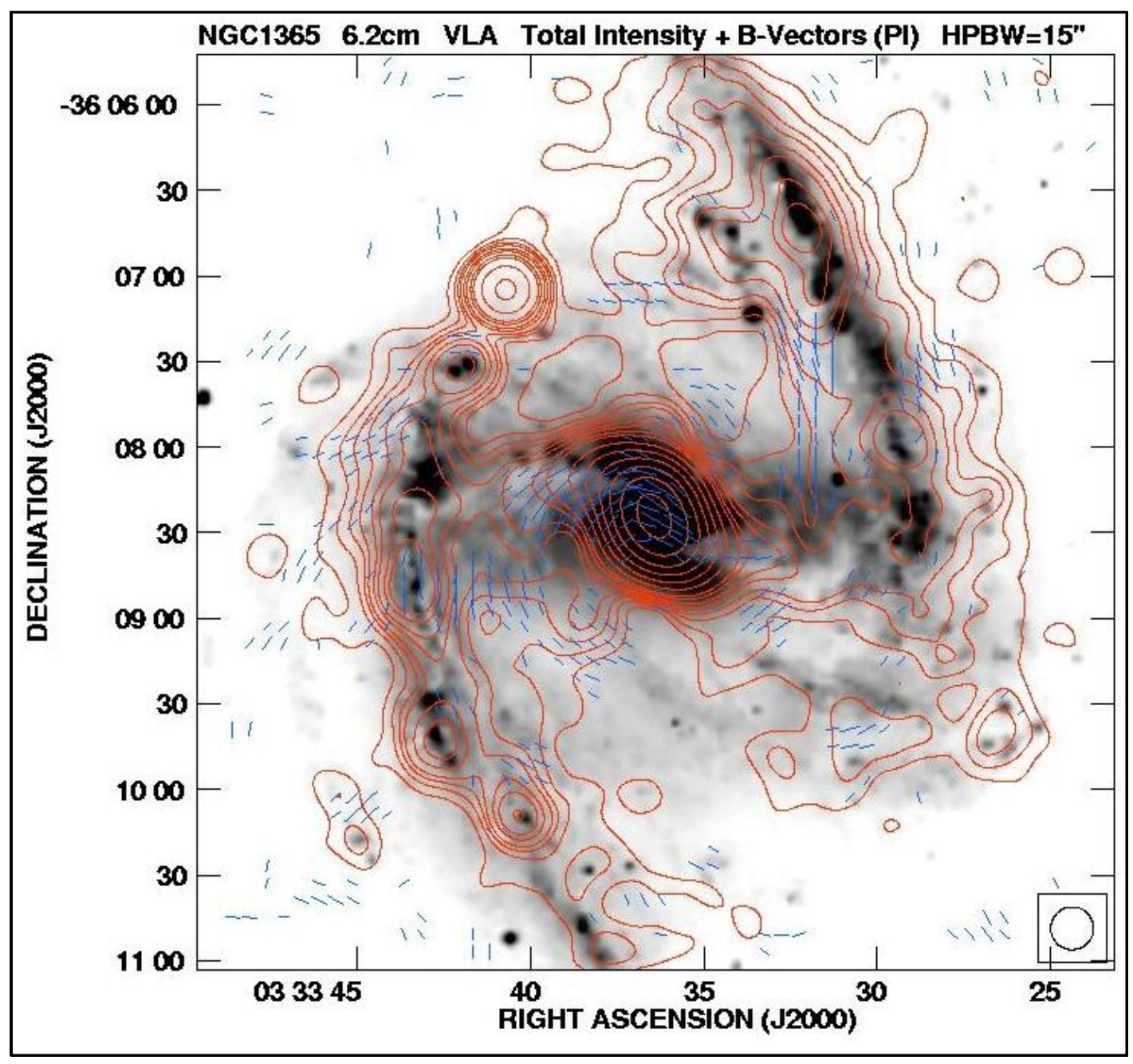

*Fig. 34: Barred galaxy NGC1365. Total radio intensity (contours) and magnetic field orientations at 4.86 GHz (6.2 cm), observed with the VLA (Beck et al. 2005a). The lengths of the lines are proportional to the polarized intensity. The background optical image is from Per Olof Lindblad (ESO).*

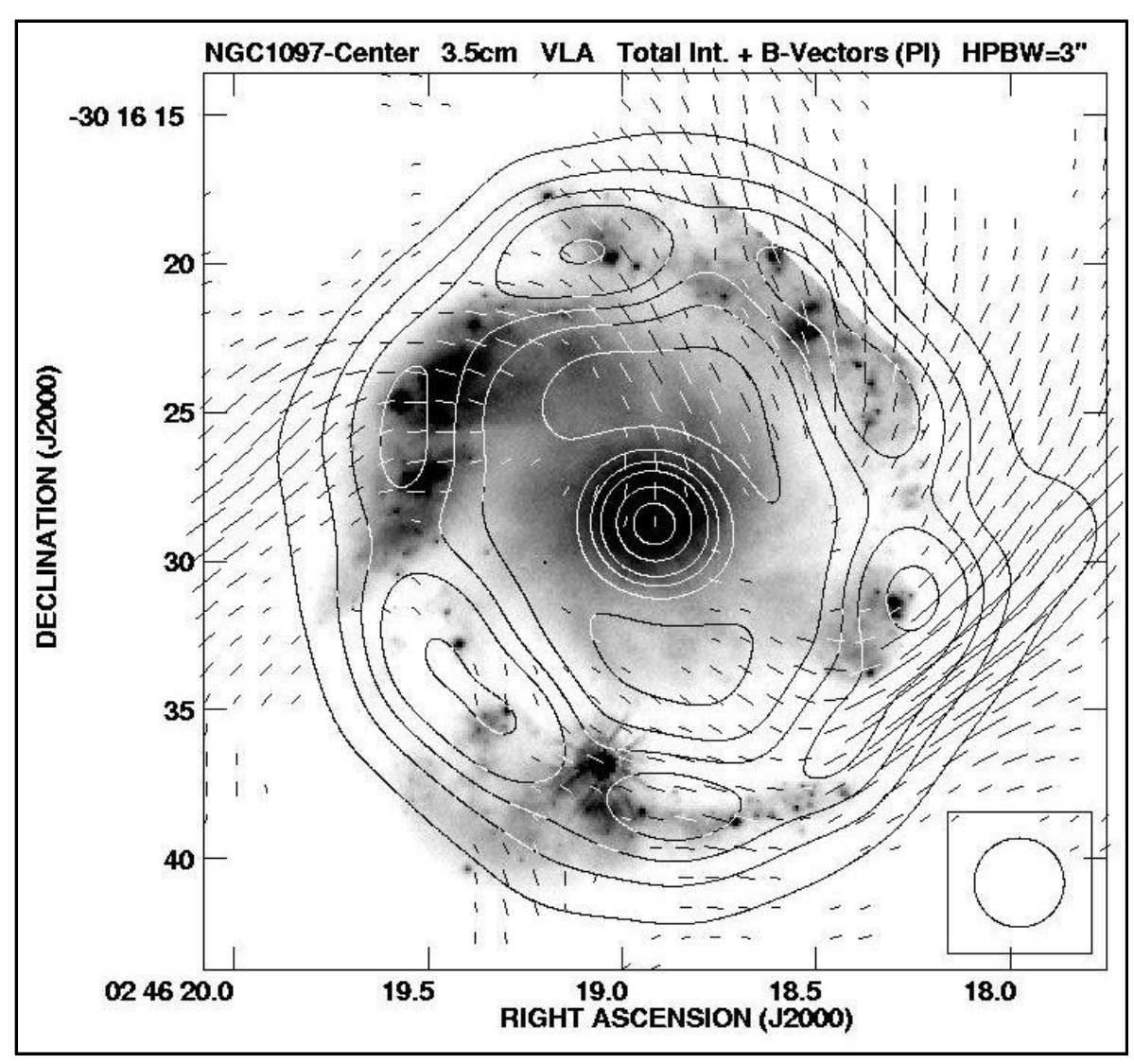

*Fig. 35: Central star-forming ring of the barred galaxy NGC1097. Total radio intensity (contours) and magnetic field orientations at 8.46 GHz (3.5 cm), observed with the VLA (Beck et al. 2005a). The lengths of the lines are proportional to the polarized intensity. The background optical image is from the Hubble Space Telescope.*

The central regions of barred galaxies are often sites of ongoing intense star formation and strong magnetic fields that can affect the gas flow. Radio emission from ring-like regions was found in NGC1097, NGC1672, and NGC7552 (Beck et al. 2005b). NGC1097 hosts a bright ring with about 1.5 kpc diameter and an active nucleus in its center (Fig. 35). Star-formation activity is high, but is slowed down by the strong magnetic field (Tabatabaei et al. 2018). The ordered field in the ring seen in radio (Beck et al. 2005a) and from CO velocity gradients (Hu et al. 2022) has a spiral pattern and extends towards the nucleus, while the field seen in FIR is located at the contact regions between the ring and the inner bar (Lopez-Rodriguez el al. 2021b). The orientation of the innermost spiral field agrees with that of the spiral dust filaments visible on optical images. Magnetic stress in the circum-nuclear ring can drive mass inflow at a rate of $dM/dt = -h/\Omega\ (\langle b_r\, b\rangle + B_r\, B_\Phi)$, where h is the scale height of the gas, $\Omega$ its angular rotation velocity, b the strength of the turbulent field and B that of the ordered field, and r and $\Phi$ denote the radial and azimuthal field components (Balbus & Hawley 1998). For NGC1097, h ≈ 100 pc, v ≈ 450 km/s at 1 kpc radius, $b_r \approx b_\Phi \approx 50$ µG gives an inflow rate of several solar masses per year, which is sufficient to fuel the activity of the nucleus (Beck et al. 2005a). A similar inflow rate was determined for the central region of the starburst galaxy M82 (Adebahr et al. 2017).

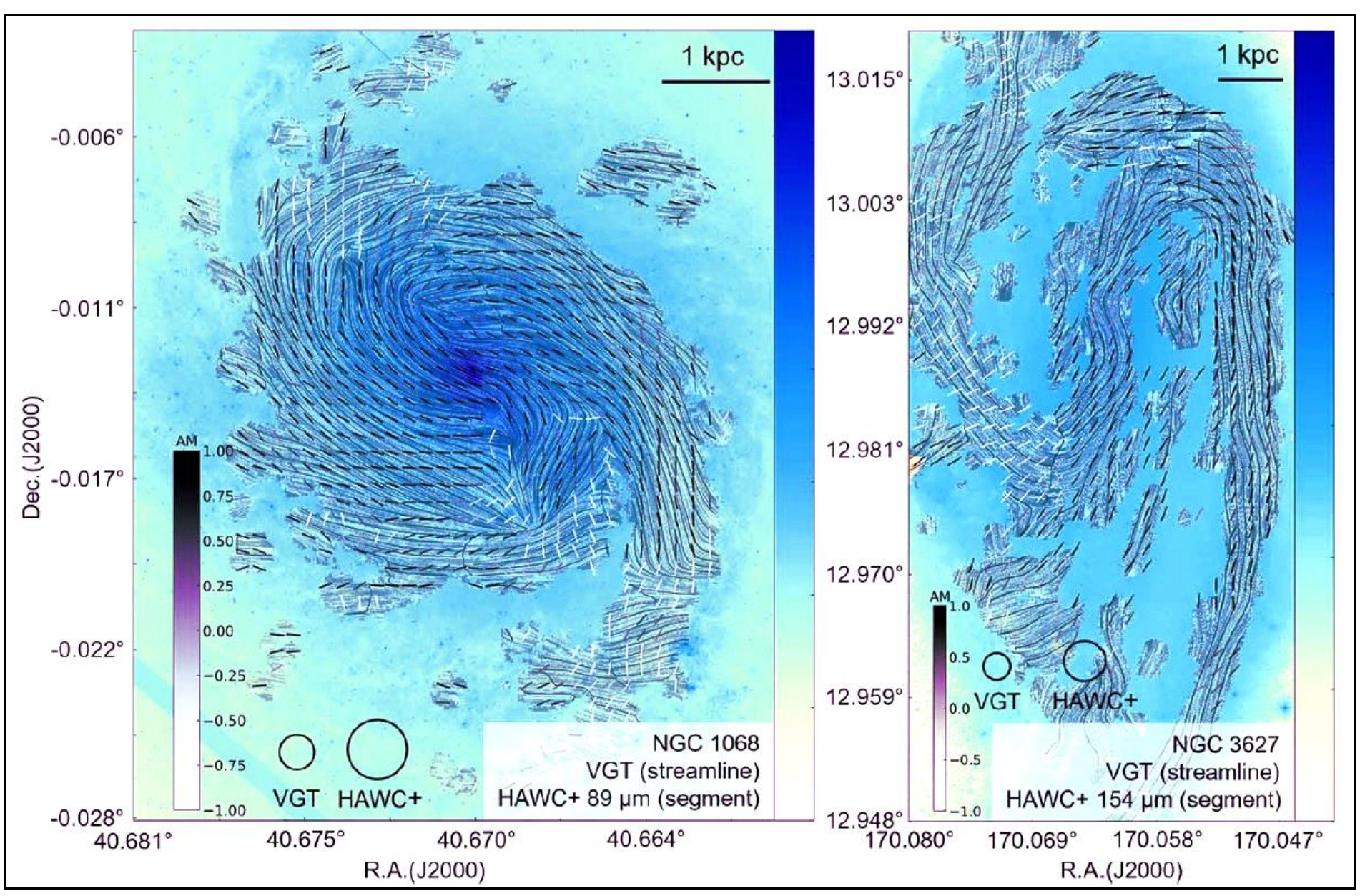

*Fig. 36: Barred galaxies NGC1068 and NGC3627. Magnetic field orientations from SOFIA/HAWC+ far-infrared polarization (Lopez-Rodriguez et al. 2022b) are shown as short lines with constant length, superimposed onto coloured streamlines of field orientations derived from CO velocity gradients (Hu et al. 2022). Black lines indicate agreement between the two methods, white lines major discrepancies.*

Field orientations were also determined from far-infrared polarization and CO velocity gradients (Fig. 36). These two methods reveal similar spiral patterns and are globally compatible, except in the interface regions between the bar and the spiral arms in NGC1068 and in the southern bar of NGC3627. The probable reason are distortions of the molecular gas disk by interactions, inflows, or outflows (Hu et al. 2022).

In summary, the isotropic turbulent field in galaxies with massive bars is coupled to the cold gas and compressed in the bar's shock. The ordered field outside the bar region follows the general flow of the cold and warm gas, possibly due to shear, but decouples from the cold gas in front of the shock and goes with the diffuse warm gas. The polarization pattern in barred galaxies can be used a tracer of the flow of diffuse gas in the sky plane and hence complements spectroscopic measurements of radial velocities. Detailed comparisons between polarimetric and spectroscopic data are required, as well as MHD models including the back-reaction of the magnetic fields onto the gas flow.

Radio polarization data have revealed differences, but also similarities between the behaviours of ordered magnetic fields in barred and non-barred galaxies. In galaxies without bars and without strong density waves the field lines have a spiral shape, they do not follow the gas flow and are probably amplified by dynamo action. In galaxies with massive bars or strong density waves the field lines mostly follow the flow of the diffuse warm gas. Near the shock fronts galaxies with strong bars and with strong density waves (section 4.4) reveal a similar behaviour: Isotropic turbulent fields, coupled to the cold gas, are shocked and become anisotropic turbulent, while ordered fields are coupled to the warm diffuse gas and hence avoid shock compression.

### 4.6 Flocculent and irregular galaxies

*Flocculent* galaxies have disks, but no prominent spiral arms. Nevertheless, spiral magnetic patterns are observed in all flocculent galaxies, indicative that the $\alpha$-$\Omega$ dynamo works independently of density waves. The multi-wavelength data of M33 and NGC4414 call for a mixture of dynamo modes or an even more complicated field structure (Table A4 in the Appendix). Ordered magnetic fields with strengths similar to those in grand-design spiral galaxies have been detected in the flocculent galaxies M33 (Fig. 37), NGC3521, NGC5055, and NGC4414, and also the mean degree of polarization is similar between grand-design and flocculent galaxies (Knapik et al. 2000).

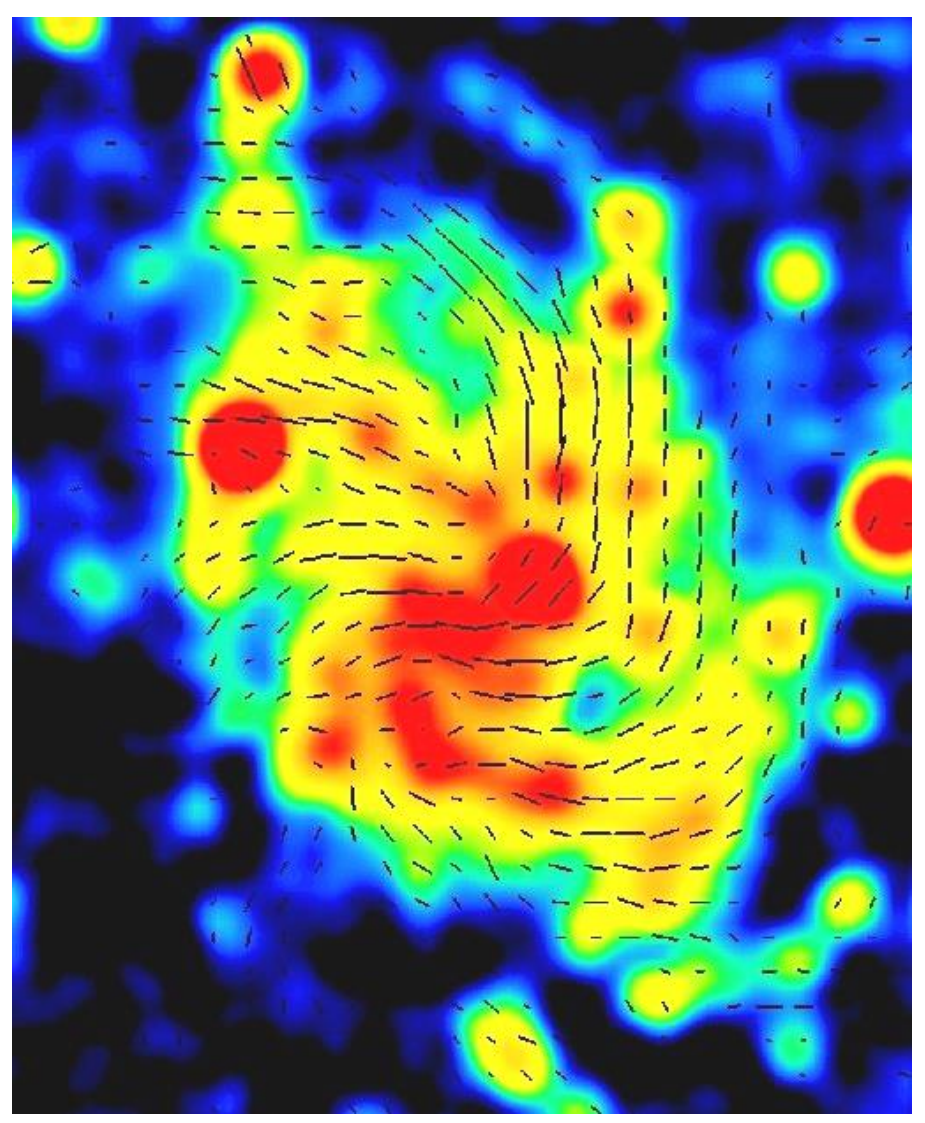

*Fig. 37: Flocculent galaxy M33. Total radio intensity (colours) and magnetic field orientations at 8.35 GHz (3.6 cm), observed with the Effelsberg telescope (Tabatabaei et al. 2008). The lengths of the lines are proportional to the polarized intensity.*

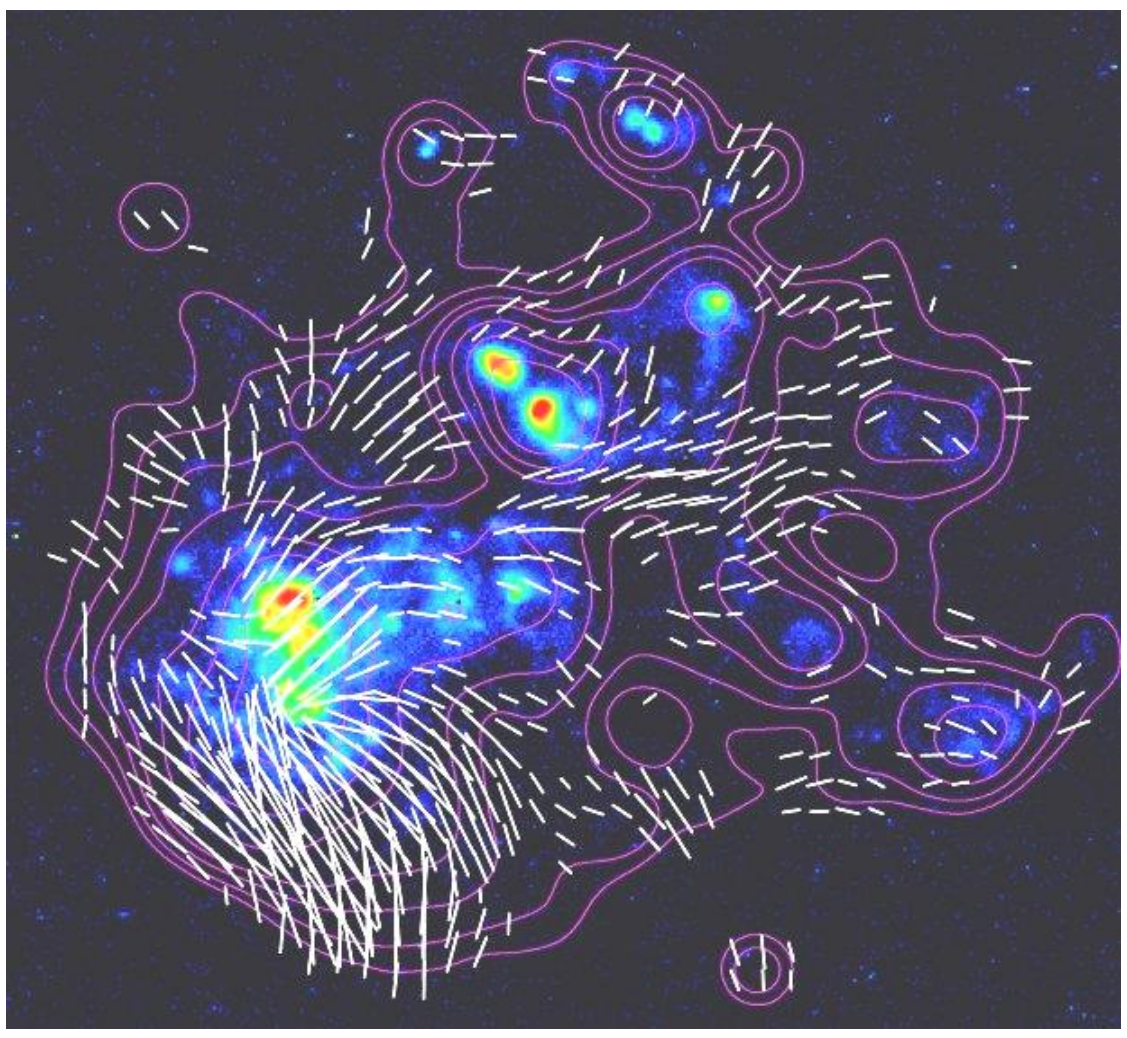

*Fig. 38: Irregular galaxy IC10. Total radio intensity (contours) and magnetic field orientations at 4.86 GHz (6.2 cm), observed with the VLA (from Chris Chyży, Kraków University). The lengths of the lines are proportional to the polarized intensity. The background H$\alpha$ image is from Dominik Bomans.*

Radio continuum maps of irregular, slowly rotating galaxies may reveal strong total equipartition magnetic fields, e.g. in the Magellanic-type galaxy NGC4449, IC10 (Fig. 38), and NGC2976. IC10 and NGC2976 are able to drive galactic winds and inflate a huge radio halo with vertical magnetic fields (Chyży et al. 2016; Drzazga et al. 2016). In NGC4449 some fraction of the field is ordered with about 7 μG strength and a spiral pattern. Faraday rotation shows that this ordered field is partly regular and the α-Ω dynamo is operating in this galaxy. The total field is of comparable strength (10 − 15 μG) in starburst dwarfs like NGC1569 (Kepley et al. 2010), where star formation activity is sufficiently high for the operation of the small-scale dynamo (section 2.7). In these galaxies the energy density of the magnetic fields is only slightly smaller than that of the (chaotic) rotation of the gas and thus may affect the evolution of the whole system. The starburst dwarf galaxy NGC1569 shows polarized emission, but no large-scale regular field. In dwarf galaxies with very weak star-forming activity, no polarized emission is detected and the isotropic turbulent field strength is several times smaller than in spiral galaxies (Chyży et al. 2011), sometimes less than 5 μG (Chyży et al. 2003). The latter value may indicate a sensitivity limit of present-day observations or a threshold for small-scale dynamo action.

The Magellanic Clouds are the closest irregular galaxies and deserve special attention. Polarization surveys with the Parkes single-dish telescope at several wavelengths had low angular resolution and revealed weak polarized emission. Two magnetic filaments were found in the LMC south of the 30 Dor star-formation complex (Klein et al. 1993). ATCA surveys of an RM grid towards background sources show that the LMC probably contains a large-scale, even-symmetry regular field similar to large spirals (Gaensler et al. 2005; Mao et al. 2012b), while that the field of the SMC is weak and uniformly directed away from us, possibly part of a pan-Magellanic field joining the two galaxies (Mao et al. 2008) with a strength of -0.3 ± 0.1 μG, compared to a ≈ 5 μG random field (Livingston et al. 2022).

### 4.7 Radio halos

Radio halos are observed around the disks of most edge-on galaxies, but their radio intensity varies significantly. The halo luminosity in the radio range correlates with those in Hα and X-rays (Tüllmann et al. 2006), although the halo shapes differ strongly between the different spectral ranges. The data suggests that star formation in the disk is the energy source for halo formation.

In spite of the different intensities and vertical extents of radio halos, their exponential scale heights at ≈ 5 GHz were found to be similar (Dumke & Krause 1998; Heesen 2009a). The scale heights of the weakest halos, like NGC4565, and of the brightest ones known, NGC253 (Fig. 39) and NGC891 (Fig. 41), are similar. The average scale height of a sample of 13 galaxies edge-on galaxies from the CHANG-ES sample (Wiegert et al. 2015) is 1.1 ± 0.3 kpc at 6 GHz and 1.4 ± 0.7 kpc at 1.5 GHz (Krause et al. 2018). The halo scale heights increase about linearly with the optical and radio diameters, while no significant relations were found with the star-formation rate nor with the magnetic field strength. Heesen et al. (2018) modelled the vertical profiles of the synchrotron emission and the spectral indices of 12 nearby edge-on galaxies to derive the exponential scale heights of the total magnetic field; they also found no correlations with the total star-formation rate nor with the star-formation rate per surface area.

In case of equipartition between the energy densities of magnetic field and cosmic rays, the exponential scale height of the total field $h_B$ is at least (3 – α) times larger than the synchrotron scale height (where α ≈ -1 is the synchrotron spectral index in the halo). The values from Krause et al. (2018) yield $h_B \geq 5$ kpc. The real value depends on the energy losses of the cosmic-ray electrons in the halo (section 2.2). With such scale heights, the magnetic energy density in halos is much higher than that of the thermal gas, while still lower than that of the kinetic energy of the gas outflow.

Radio halos grow in size with decreasing observation frequency. The scale heights of the halo of NGC891, NGC4157, and NGC4631 are about twice higher in the LOFAR image at about 150 MHz compared to the VLA image at 1.5 GHz, while two edge-on galaxies (NGC3432 and NGC4013) show similar scale heights (Mulcahy et al. 2018; Stein et al. 2023). The halo extent is limited by energy losses of the cosmic-ray electrons (CREs), i.e. synchrotron, inverse Compton, or adiabatic losses, or by escape from the halo (Heesen 2021; Krause et al. 2018). The stronger magnetic field in the central region causes stronger synchrotron loss, leading to the “dumbbell” shape of many radio halos, e.g. around NGC253 (Fig. 39). From the radio scale heights of NGC253 at three frequencies and the electron lifetimes (due to synchrotron, inverse Compton, and adiabatic losses) an outflow bulk speed of about 300 km/s was measured (Fig. 40), which is larger than the escape velocity. The scale heights of the halo of NGC891, measured at 150 MHz and 1.5 GHz in different regions along the galaxy’s major axis, decrease with increasing magnetic field strength, which is a signature of dominating synchrotron losses of CREs (Mulcahy et al. 2018).

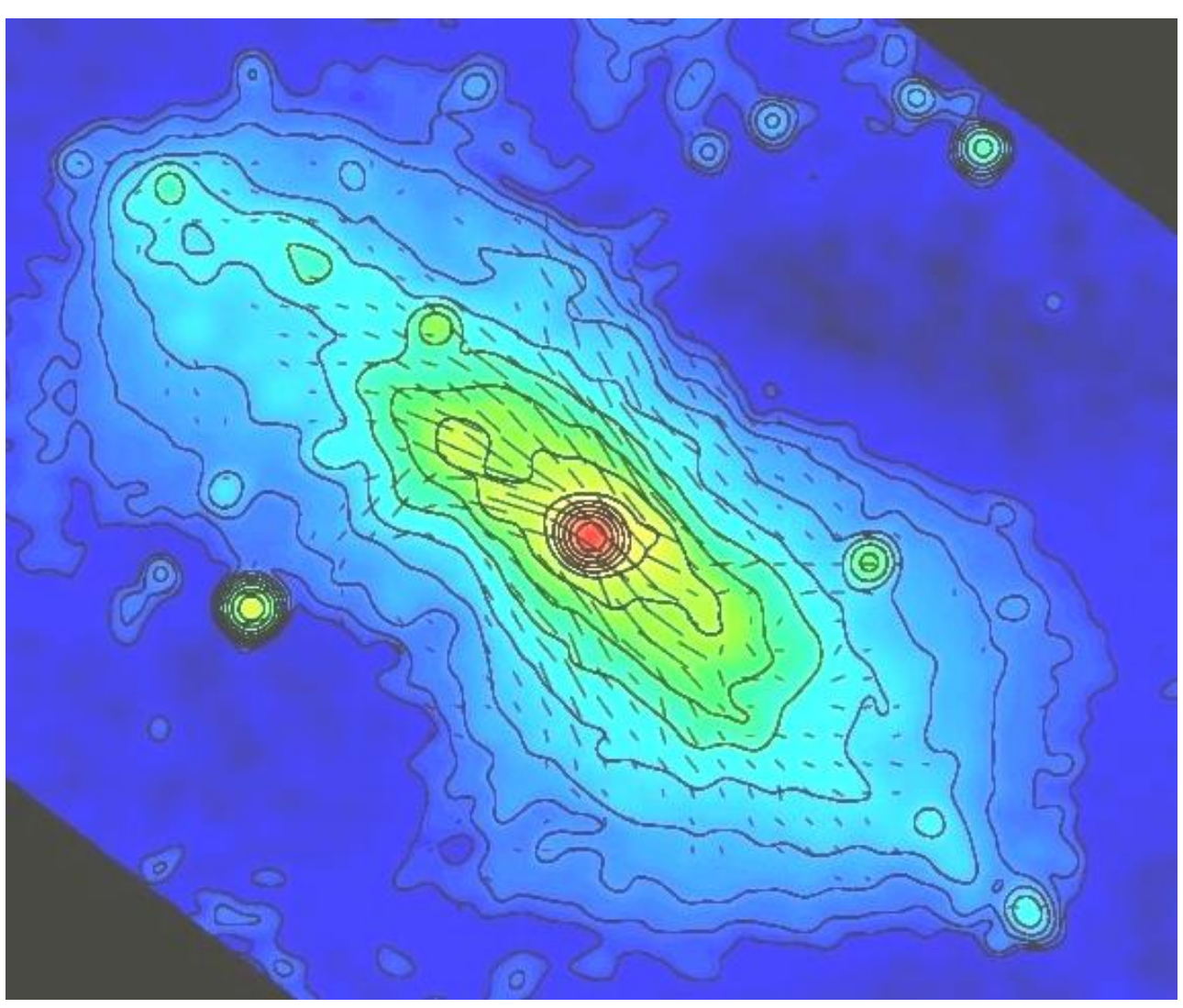

*Fig. 39: Almost edge-on spiral galaxy NGC253. Total radio intensity (contours) and magnetic field orientations at 4.86 GHz (6.2 cm), combined from observations with the VLA and the Effelsberg telescope (Heesen et al. 2009b). The lengths of the lines are proportional to the polarized intensity.*

Escaping flows are called *galactic winds*, while outflows returning to the disk are called *fountain flows.* Indications for galactic winds were found in many edge-on galaxies (Heesen et al. 2018; Stein et al. 2023). The modelled outflow velocities range from about 100 km/s to about 700 km/s, similar to the escape velocities, and display a significant correlation with the star-formation rate, as expected from numerical MHD simulations (Vijayan et al. 2020). The halo scale height reveals a tight and almost linear anti-correlation with the mass surface density that is a measure of the gravitational potential of a galaxy, indication for a gravitational deceleration of the outflow (Krause et al. 2018).

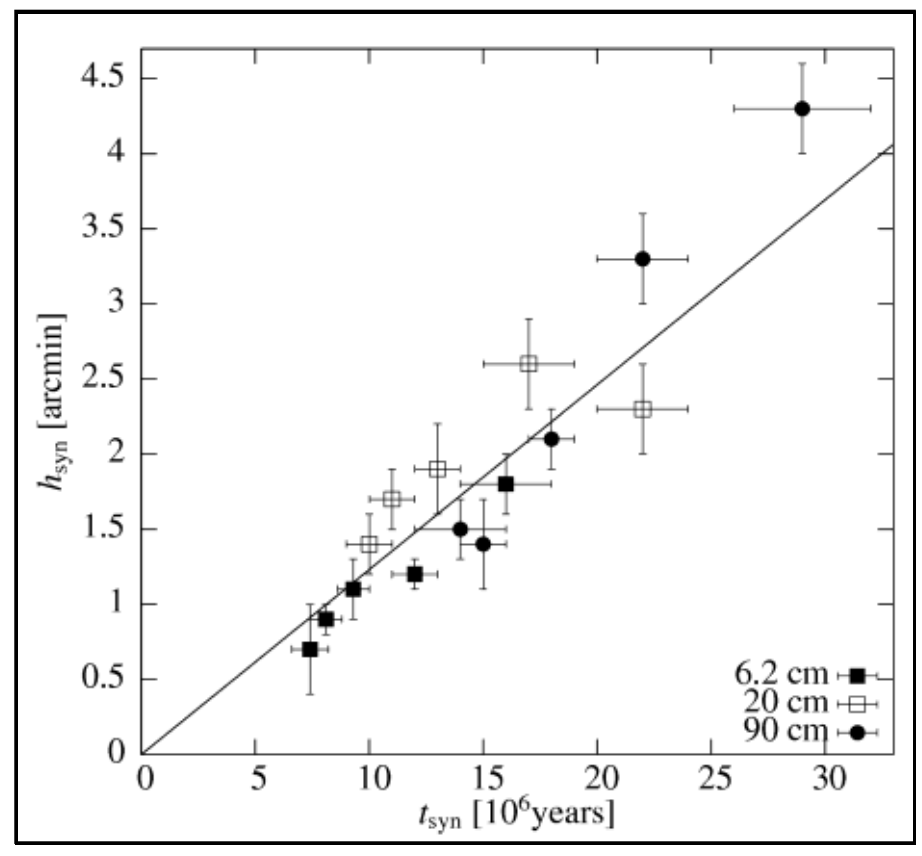

*Fig. 40: Synchrotron scale heights of the northern radio halo of NGC253 at different distances from the center and at different wavelengths, as a function of synchrotron lifetime of cosmic-ray electrons. The slope of the linear fit corresponds to a bulk outflow speed of about 300 km/s (Heesen et al. 2009a).*

Radio polarization observations of nearby galaxies seen edge-on at high frequencies and high resolution (VLA C-band) generally show a disk-parallel field near the disk plane (Krause et al. 2020). FIR polarimetry of NGC891 confirms the plane-parallel field in the inner disk (Jones et al. 2020). High-sensitivity observations of several edge-on galaxies like NGC253 (Fig. 39), NGC891 (Fig. 41), NGC5775 (Tüllmann et al. 2000; Soida et al. 2011), M104 (Krause et al. 2006), and many galaxies of the CHANG-ES sample (Krause et al. 2020) revealed vertical field components that become stronger with increasing height above and below the galactic plane and also with increasing radius, the so-called *X-shaped* halo fields. The X-pattern is even seen in NGC4565 with its low star-formation rate and a radio-faint halo (Krause 2009) and is a general phenomenon.

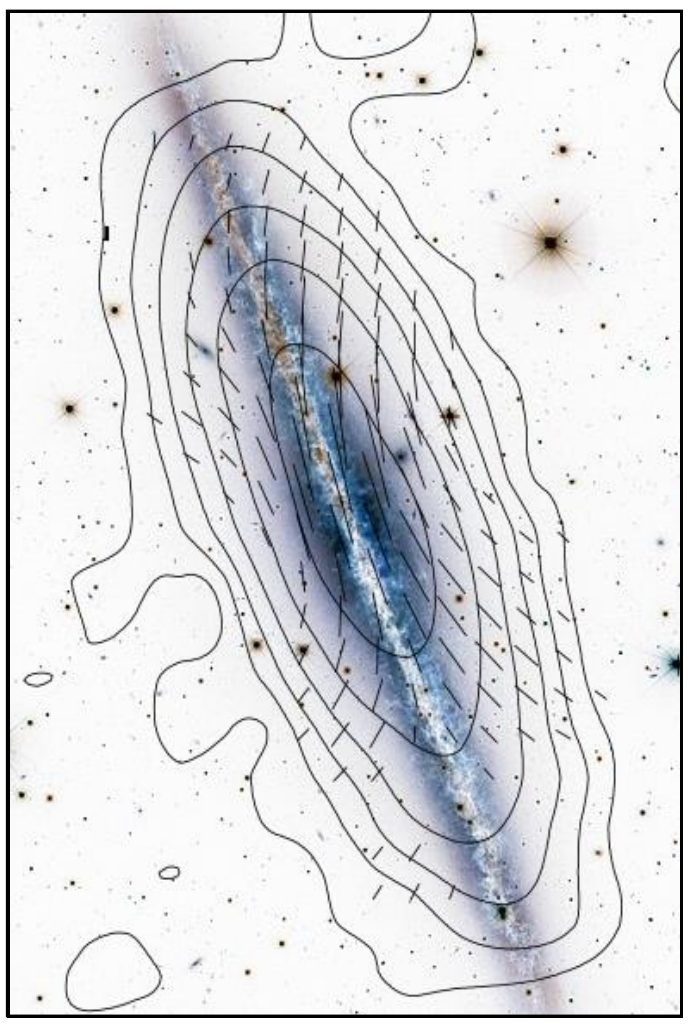

*Fig. 41: Edge-on spiral galaxy NGC891. Total radio intensity (contours) and magnetic field orientations at 8.35 GHz (3.6 cm), observed with the Effelsberg telescope (Krause 2009). The lengths of the lines are proportional to the polarized intensity. The background optical image is from the CFHT.*

Polarization orientations do not distinguish between halo fields that are sheared into elongated loops with reversals or regular dynamo-type fields. Faraday rotation measure (RM) is a signature of regular fields (section 2.4). RM patterns are hard to measure in halos of edge-on galaxies because the field components along the line of sight are expected to be small. A detailed analysis of the multi-frequency observations of the highly inclined galaxy NGC253 (Fig. 39) allowed the identification of an axisymmetric disk field with even symmetry and an X-shaped halo field, also of even symmetry (Fig. 43). The combined analysis of RMs of the diffuse emission and extragalactic sources revealed an even-symmetry halo field also in the LMC (Mao et al. 2012b). On the other hand, the analysis of CHANG-ES polarization data between 5 cm (VLA C-band) and 20 cm (VLA L-band) suggests odd-symmetry patterns fields in the central regions of several galaxies (Myserlis & Contopoulos 2021). However, the 20 cm data may be severely affected by Faraday depolarization (section 2.4), which calls for further investigation.

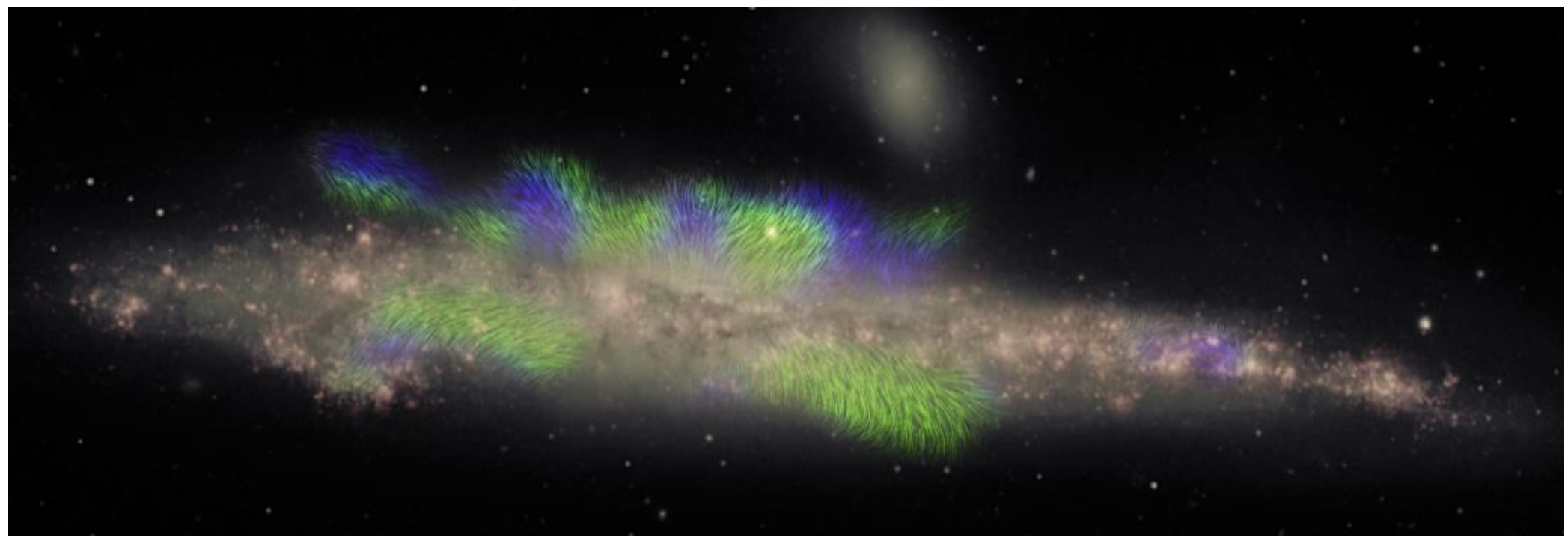

*Fig. 42: Edge-on irregular galaxy NGC4631. Magnetic field orientations at 6.0 GHz (5.0 cm), observed with the VLA (Mora-Partiarroyo et al. 2019). In the green areas the field points towards us, in the blue areas away from us. The background optical image is from the Mayall 4-m telescope (Patterson & Walterbos, New Mexico State Univ., USA). Composite image courtesy: Jayanne English (University of Manitoba, Canada).*

The magnetic field in the radio halo of the irregular galaxy NGC4631 is not X-shaped, but mostly vertically oriented in the inner part, while parallel to the plane in the outer regions (Krause 2009; Mora-Partiarroyo et al. 2013; 2019). RMs in the northern halo show a quasi-periodical variation (Fig. 42) with field reversals separated by about 2 kpc, called “giant magnetic ropes”, with a strength of the regular halo field of about 4 μG. Parker instabilities were proposed as an explanation (Tharakkal et al. 2023).

The observation of field patterns is of fundamental importance to understand the field origin in halos. $\alpha$-$\Omega$ dynamo models for thin galaxy disks predict fields of even symmetry, in the simplest case a poloidal component of quadrupolar shape (section 2.7). The field strength of a purely quadrupole-type field is expected to decrease rapidly with 1/r, where r is the distance from the center along the galaxy plane (e.g. Prouza & Šmída 2003), whereas the observed radial profiles of polarized emission show a slow exponential decrease. The vertical component of a quadrupolar field is largest near the rotation axis and decreases with distance from the rotation axis. Such an effect is seen only in two galaxies in the sample of Krause et al. (2020), namely NGC3044, NGC3448, while in several other edge-on galaxies like NGC891 (Fig. 41) the vertical field component increases with increasing distance from the rotation axis, giving rise to an X-shape. In the north-western halo of the edge-on galaxy NGC4217, an outflow with a helical field extends to a height of nearly 7 kpc (Stein et al. 2020). The field structure of NGC4631 is even more complex (Fig. 42). Hence, the field structures in galaxy halos cannot be simply quadrupolar.

The starburst galaxy M82 has a large radio halo with strong vertical field components (Reuter et al. 1992; Adebahr et al. 2017). FIR polarimetry of M82 also shows a mostly vertical field (Jones et al. 2019; Lopez-Rodriguez et al. 2021a). This indicates that the wind transport is more efficient here than in other spiral galaxies. A fast galactic outflow could be driven by regions of star formation activity. A weak gravitational potential or external forces by neighbouring galaxies may also be responsible for high outflow velocities. The action of the $\alpha$-$\Omega$ dynamo can be suppressed under such conditions (Chamandy et al. 2015).

Analytical descriptions of divergence-free, X-shaped, helical magnetic fields in galaxy halos were presented by Ferrière & Terral (2014) and Farrar (2016) (Fig. 17), possibly generated by dynamos including outflows with moderate velocities (section 2.7). An alternative analytical model invokes a rotational velocity lag in the halo and predicts an outwards-winding helical field (Henriksen & Irwin 2016; Henriksen et al. 2018; Henriksen 2022). Numerical MHD models also predict outflows and strong vertical fields (e.g. Pakmor et al. 2017; 2018; Steinwandel et al. 2020; Pfrommer et al. 2022).

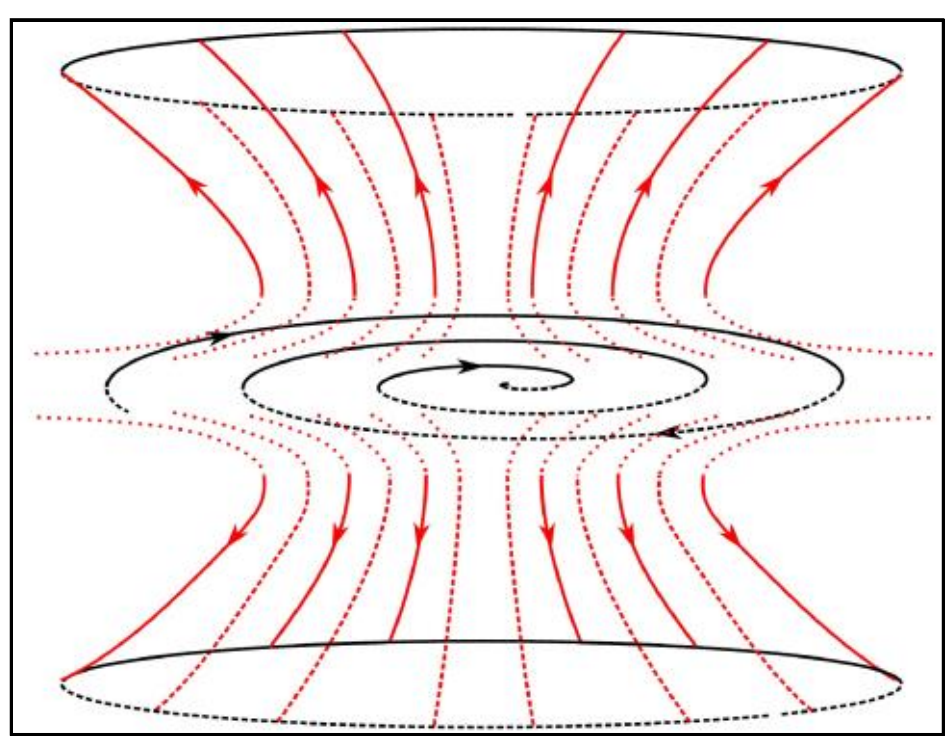

*Fig. 43: NGC253. Model of an even-symmetry (quadrupolar) halo field (red) that is outwards-directed. The spiral disk field (black) is also symmetric with respect to the plane (from Heesen et al. 2009b).*

Outflows may also magnetize the circumgalactic medium. High-precision RMs from the LOFAR RM catalogue at 144 MHz indicate an excess towards nearby galaxies for impact parameters of < 100 kpc and for sources within 45° of the galaxies' minor axis (Heesen et al. 2023b), suggesting regular fields along the line of sight of around 0.5 $\mu$G on average.

In summary, the detection of X-shaped fields in all galaxies observed so far can be explained by dynamo action and/or outflows and/or rotational velocity lag. If outflows are a general phenomenon in galaxies, they can magnetize the intergalactic medium (IGM). Starburst galaxies in the early Universe were especially efficient in magnetizing the IGM. The extent of magnetic fields into the IGM is not yet visible. Energy losses of CREs prevent the emission of radio waves beyond some height while magnetic fields may still exist much further outwards. Low-energy CREs live longer, can propagate further into the IGM and emit synchrotron emission at low frequencies (section 2.2). Indeed, observations with the Low Frequency Array (LOFAR) reveal larger radio halos than at higher frequencies (Mulcahy et al. 2018; Stein et al. 2019b; Stein et al. 2023).

#### 4.8 Interacting galaxies

Gravitational interaction between galaxies leads to asymmetric gas flows, compression, shear, enhanced turbulence and outflows. Compression and shear of gas flows can also modify the structure of galactic and intergalactic magnetic fields. In particular, fields can become aligned along the compression front or perpendicular to the velocity gradients. Such gas flows make turbulent fields highly anisotropic.

The classical interacting galaxy pair is NGC4038/39, the "Antennae" (Fig. 44). It shows bright, extended radio emission filling the volume of the whole system, with no dominant nuclear sources. In the interaction region between the galaxies, where star formation did not yet start, and at the north-eastern edge of the system, the magnetic field seen in radio and FIR is partly ordered, probably the result of compression and shearing motions along the tidal tail, respectively (Basu et al. 2017; Lopez-Rodriguez et al. 2023). The regular field in the tidal tail is about 8 μG strong and extends to about 20 kpc distance from the galaxy (Basu et al. 2017). Particularly strong, almost unpolarized emission comes from a region of violent star formation, hidden in dust, at the southern end of a dense cloud complex extending between the galaxies. In this region, highly turbulent magnetic fields reach strengths of ≈ 20 μG. The mean total magnetic field is stronger than in normal spirals, but the mean degree of polarization is unusually low, implying that the ordered field, generated by compression, has become tangled in the region with violent star formation. After an interaction, the magnetic field strength in a galaxy is expected to decrease again to its normal value (Drzazga et al. 2011).

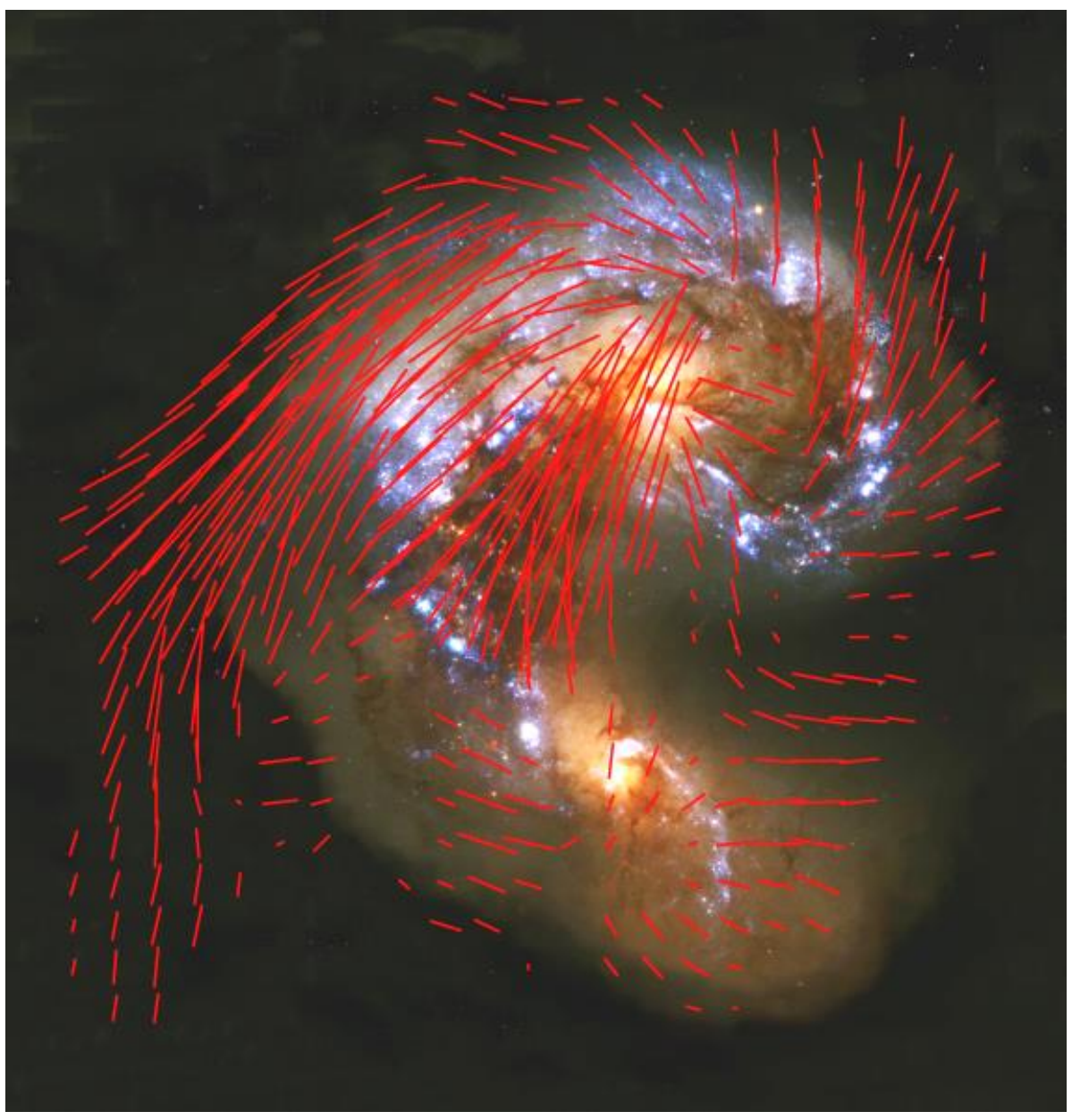

*Fig. 44: "Antennae" galaxy pair NGC4038/39. Magnetic field orientations at 4.86 GHz (6.2 cm), observed with the VLA (Chyży & Beck 2004). The lengths of the lines are proportional to the polarized intensity. The background optical image is from the Hubble Space Telescope.*

Interaction with a dense intergalactic medium also imprints unique signatures onto magnetic fields and thus the radio emission. The Virgo cluster is a location of especially strong interaction effects, and almost all cluster galaxies observed so far show asymmetries of their polarized emission (Table A6 in the Appendix). In NGC4254, NGC4522, and NGC4535 (Fig. 45), the polarized emission on one side of the galaxy is shifted towards the edge of the spiral arm, an indication for shear by tidal tails or ram pressure by the intracluster medium. The heavily disrupted galaxy NGC4438 (Vollmer et al. 2007) has almost its whole radio emission (total power and polarized) displaced towards the giant elliptical M86, to which it is also connected by a chain of Hα-emitting filaments.

Interaction may also induce violent star-formation activity in the nuclear region or in the disk, which may produce huge radio lobes due to outflowing gas and magnetic field. The lobes of the Virgo spiral NGC4569 reach out to at least 24 kpc from the disk and are highly polarized (Fig. 46). However, there is neither an active nucleus nor a recent starburst in the disk, so that the radio lobes are probably the result of nuclear activity in the past.

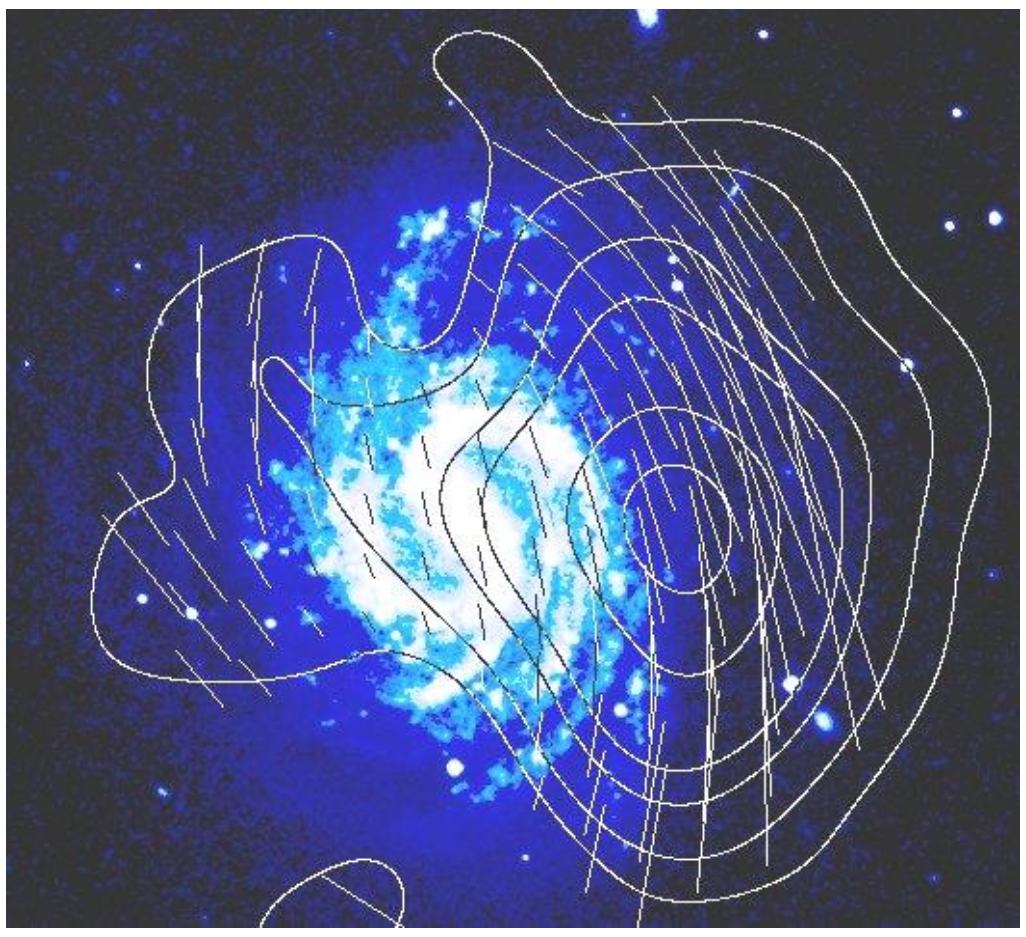

*Fig. 45: Spiral galaxy NGC4535 in the Virgo cluster. Polarized radio intensity (contours) and magnetic field orientations at 4.75 GHz (6.3 cm), observed with the Effelsberg telescope (Weżgowiec et al. 2007). The lengths of the lines are proportional to the polarized intensity. The background optical image is from the Digital Sky Survey.*

Tidal interaction is also the probable cause of the asymmetric appearance of the magnetic field in NGC3627 within the Leo Triplet (Fig. 47; Liu et al. 2023). While the ordered field in the western half is strong and follows the dust lanes, a bright magnetic arm in the eastern half crosses the optical arm and its massive dust lane at a large angle. No counterpart of this feature was detected in any other spectral range. Either the optical arm was recently deformed due to interaction or ram pressure, or the magnetic arm is an out-of-plane feature generated by interaction.

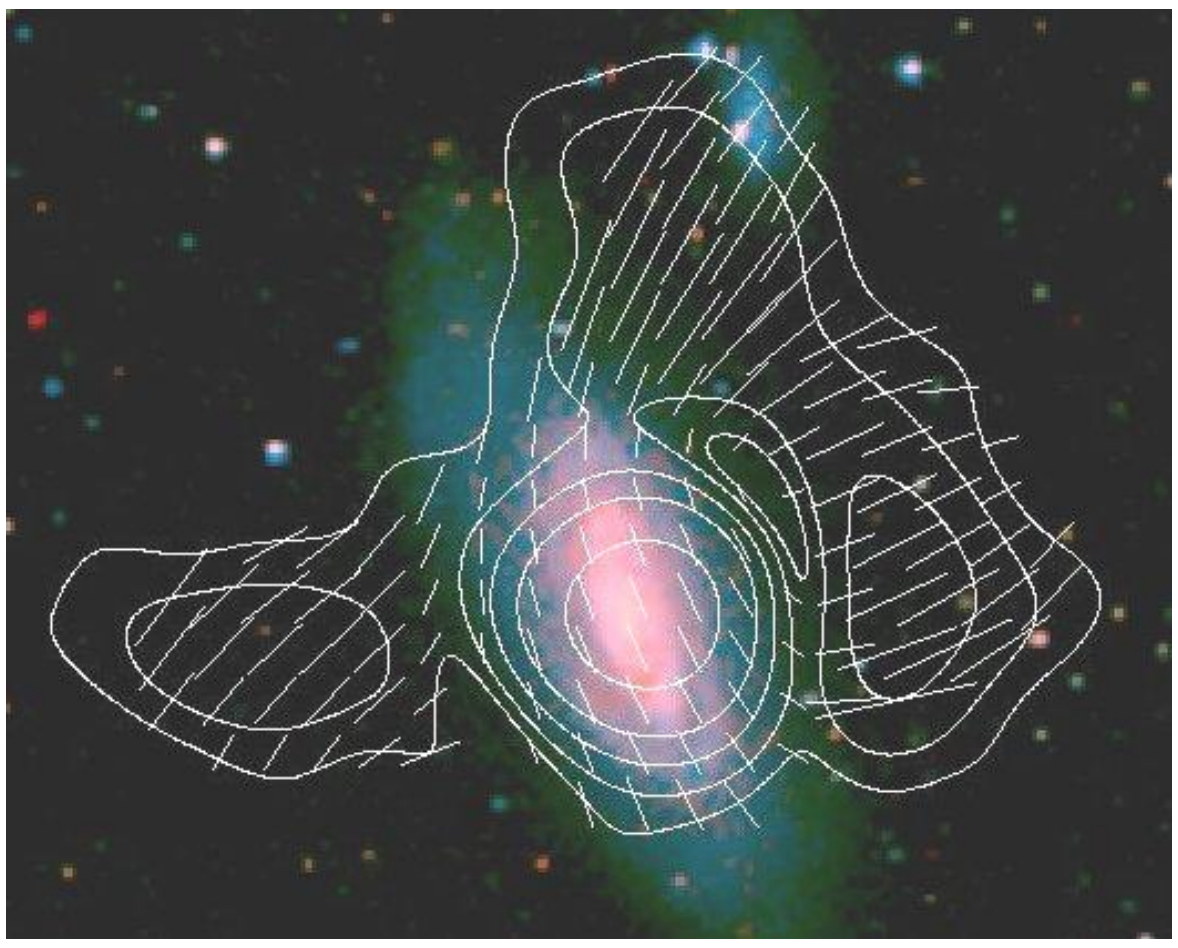

*Fig. 46: Spiral galaxy NGC4569 in the Virgo cluster. Polarized radio intensity (contours) and magnetic field orientations at 4.75 GHz (6.3 cm), observed with the Effelsberg telescope (Chyży et al. 2006). The lengths of the lines are proportional to the polarized intensity. The background optical image is from the Digital Sky Survey.*

In a few cases a gaseous bridge between colliding galaxies is associated with radio emission that is due to relativistic electrons pulled out from the disks together with gas and magnetic fields. This phenomenon (called “taffy galaxies”) seems to be rare. Only two objects, UGC12914/5 and UGC813/6, were found so far (Condon et al. 2002; Drzazga et al. 2011). This may be due to the steep spectrum of the bridges, making them hard to detect at centimeter radio wavelengths.

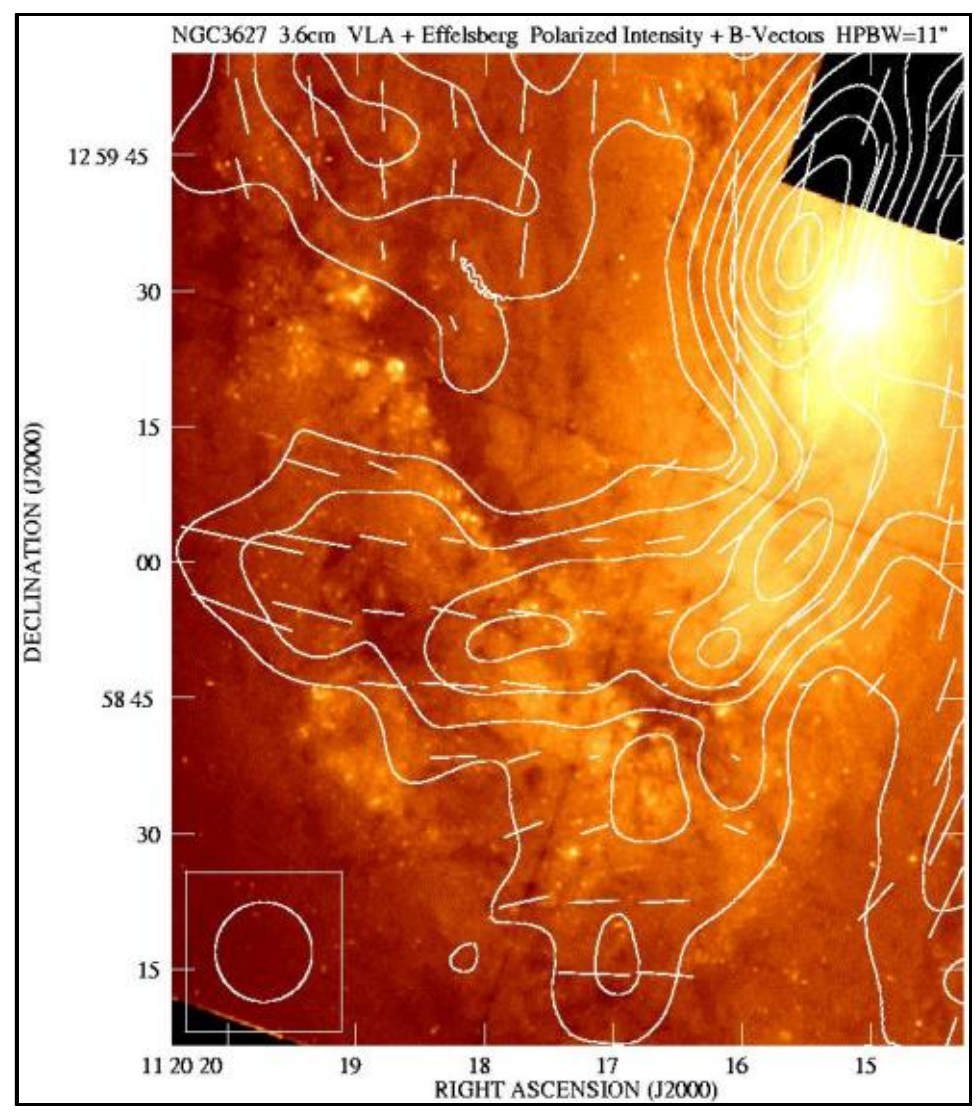

*Fig. 47: Interacting spiral galaxy NGC3627. Polarized radio intensity (contours) and magnetic field orientations at 8.46 GHz (3.5 cm), combined from observations with the VLA and Effelsberg telescopes (Soida et al. 2001). The lengths of the lines are proportional to the degree of polarization. The background optical image is from the Hubble Space Telescope.*

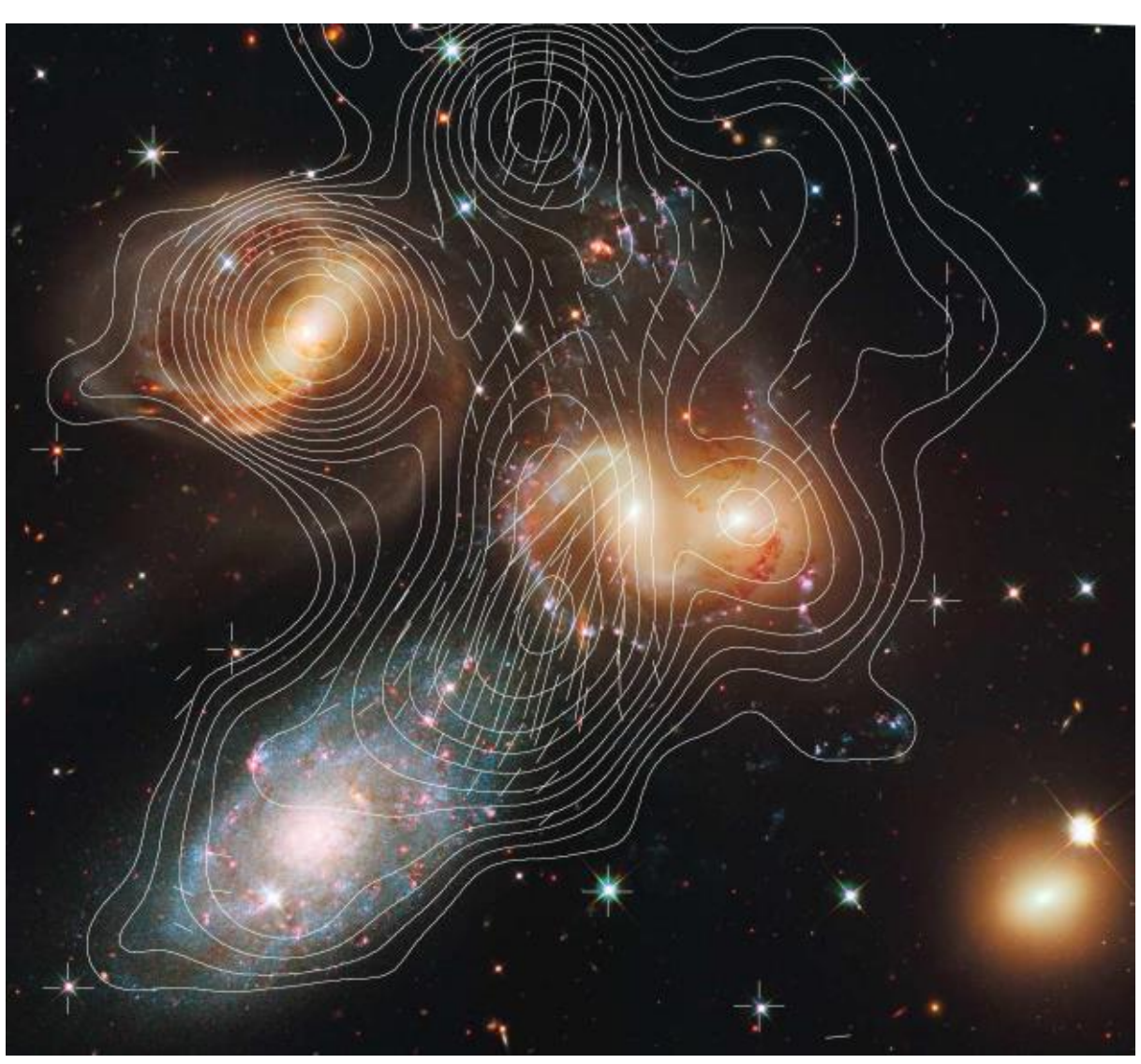

*Fig. 48: Stephan's Quintet of interacting galaxies. Total radio intensity (contours) and magnetic field orientations at 4.86 GHz (6.2 cm), observed with the VLA by Marian Soida (Kraków University). The lengths of the lines are proportional to the polarized intensity. The background optical image is from the Hubble Space Telescope.*

In compact galaxy groups tidal interactions may trigger rapid star formation in one or more member galaxies, causing supersonic outflows of hot gas. Some compact groups have long HI tails, indicating strong, tidally-driven outflows of the neutral gas from the system. If the expelled gas was magnetized it might provide the supply of magnetic fields into the intergalactic space. Starburst galaxies (either dwarf and massive) constitute the basic source responsible for the enrichment of the intra-group medium with relativistic particles and magnetic fields. There are grounds to expect that the compact galaxy groups show diffuse radio emission, with a spectrum rapidly steepening away from the cosmic-ray sources in galactic disks.

The best studied example of a compact group is Stephan's Quintet (at a distance of about 85 Mpc), with its pool of hot gas extending between the galaxies (Nikiel-Wroczyński et al. 2013b). It shows a huge filament visible in radio continuum. Strong polarization of this intra-group emission (Fig. 48) indicates a substantial content of ordered (probably shock-compressed) magnetic fields. Intergalactic fields were also found in the compact groups HCG15, HCG60, and possibly in HCG44 (Nikiel-Wroczyński et al. 2017).

In summary, polarized radio emission is an excellent tracer of tidal effects between galaxies and of ram pressure in the intracluster medium. As the decompression and diffusion timescales of the field are very long, it keeps memory of events in the past, up to the lifetime of the illuminating cosmic-ray electrons. Low-frequency radio observations will trace interactions that occurred many Gyr ago and are no longer visible in other spectral ranges. Tidal tails from interacting galaxies may also constitute a significant source of magnetic fields in the intracluster and intergalactic media.

### 4.9 Galaxies with jets

Nuclear jets are observed in several spiral galaxies. These jets are weak and small compared to those of radio galaxies and quasars. Detection is further hampered by the fact that they emerge at some angle with respect to the disk, so that little interaction with the ISM occurs. If the nuclear disk is oriented almost perpendicular to the disk, the jet hits a significant amount of ISM matter, cosmic-ray electrons are accelerated in shocks, and the jet becomes radio-bright. This geometry was found for NGC4258 by observations of the water maser emission from the nuclear disk that has an inner radius of 0.13 pc and is seen almost edge-on (Greenhill et al. 1995). This is probably why NGC4258 is one of the rare cases, where a large and bright radio jet of at least 15 kpc length is observed (van Albada & van der Hulst 1982; Krause & Löhr 2004). The total intensity map of NGC4258 (Fig. 49) reveals that the jets emerge from the Galactic center perpendicular to the nuclear disk that is oriented in east-west orientation, and bend out to become the ”anomalous radio arms”, visible out to the boundaries of the spiral galaxy. The magnetic field orientation is mainly along the jet. The equipartition field strength is about 300 µG (at the resolution of about 100 pc), which is a lower limit due to energy losses of the cosmic-ray electrons and the limited resolution.

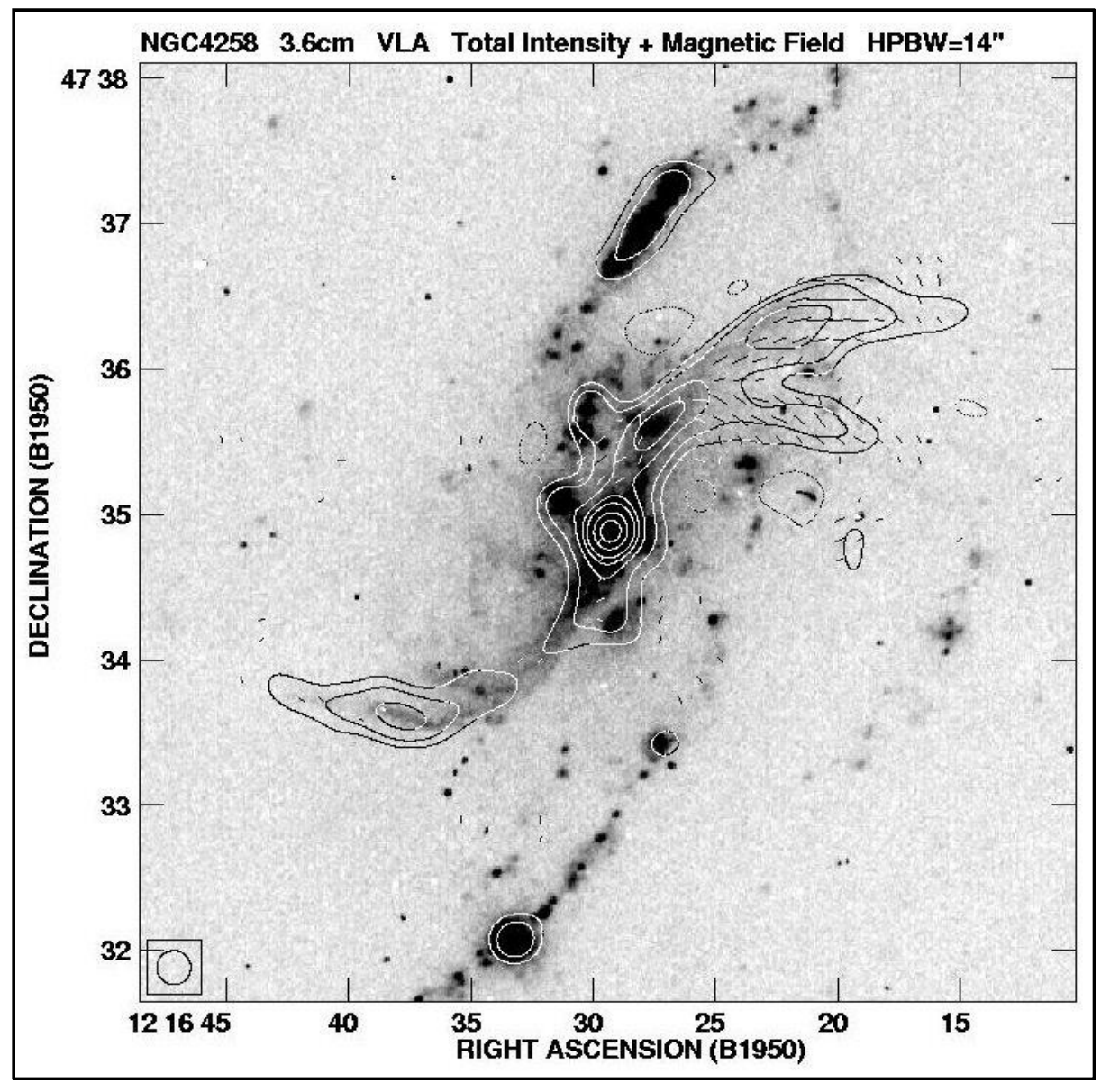

*Fig. 49: Spiral galaxy NGC4258 with two jets. Total radio intensity (contours) and magnetic field orientations at 8.46 GHz (3.5 cm), observed with the VLA (Krause & Löhr 2004). The lengths of the lines are proportional to the polarized intensity. The background $H\alpha$ image is from the Hoher List Observatory of the University of Bonn.*

The barred galaxy NGC7479 also shows remarkable jet-like radio continuum features: bright, narrow, 12 kpc long in projection, and containing an aligned magnetic field (Fig. 50). The lack of any optical or near-infrared emission associated with the jets suggests that at least the outer parts of the jets are extraplanar features, although close to the disk plane. The equipartition strength is 35 – 40 µG for the total and about 10 µG for the ordered magnetic field in the jets. According to Faraday rotation data, the large-scale regular magnetic field along the bar points towards the nucleus on both sides. Multiple reversals on scales of 1 – 2 kpc are detected, probably occurring in the galaxy disk in front of the eastern jet by anisotropic turbulent fields in the shearing gas flow in the bar potential.

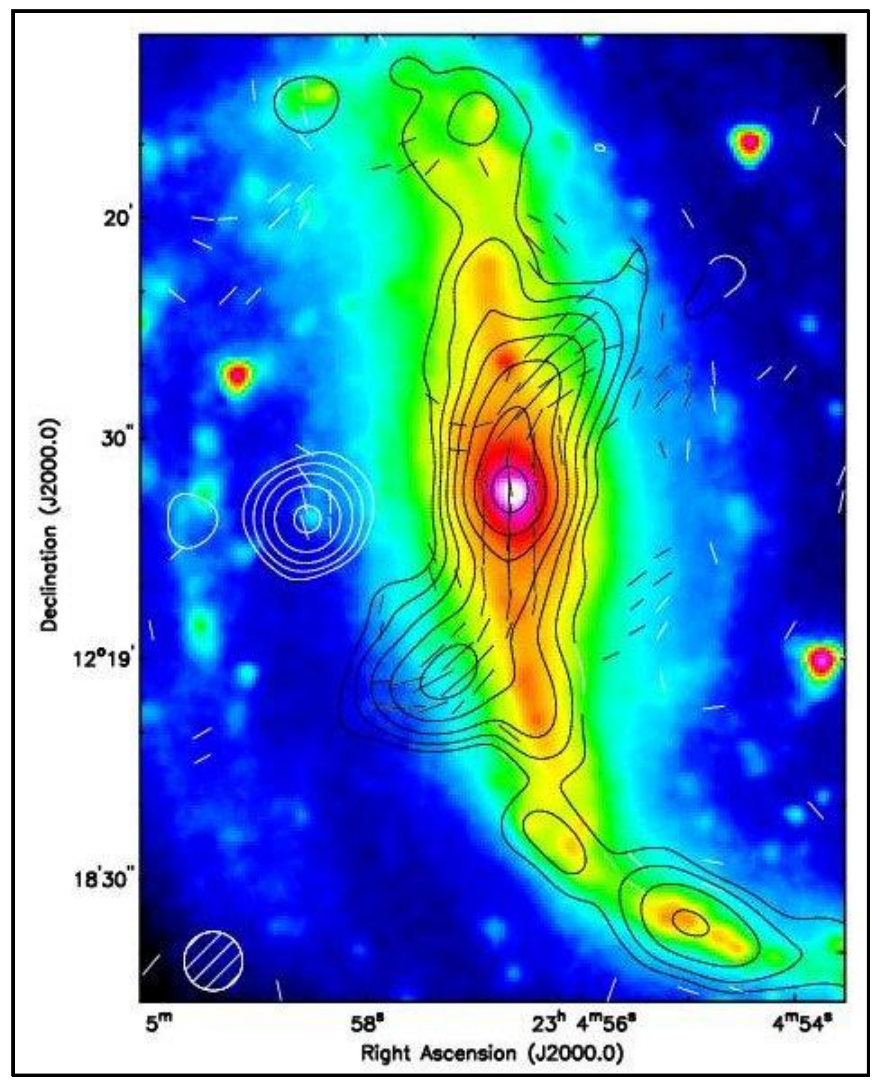

*Fig. 50: Barred spiral NGC7479 with two jets. Total radio intensity (contours) and magnetic field orientations at 8.46 GHz (3.5 cm), observed with the VLA (Laine & Beck 2008). The lengths of the lines are proportional to the polarized intensity. The background shows a Spitzer/IRAC infrared image at 3.6 µm (NASA/JPL-Caltech/Seppo Laine).*

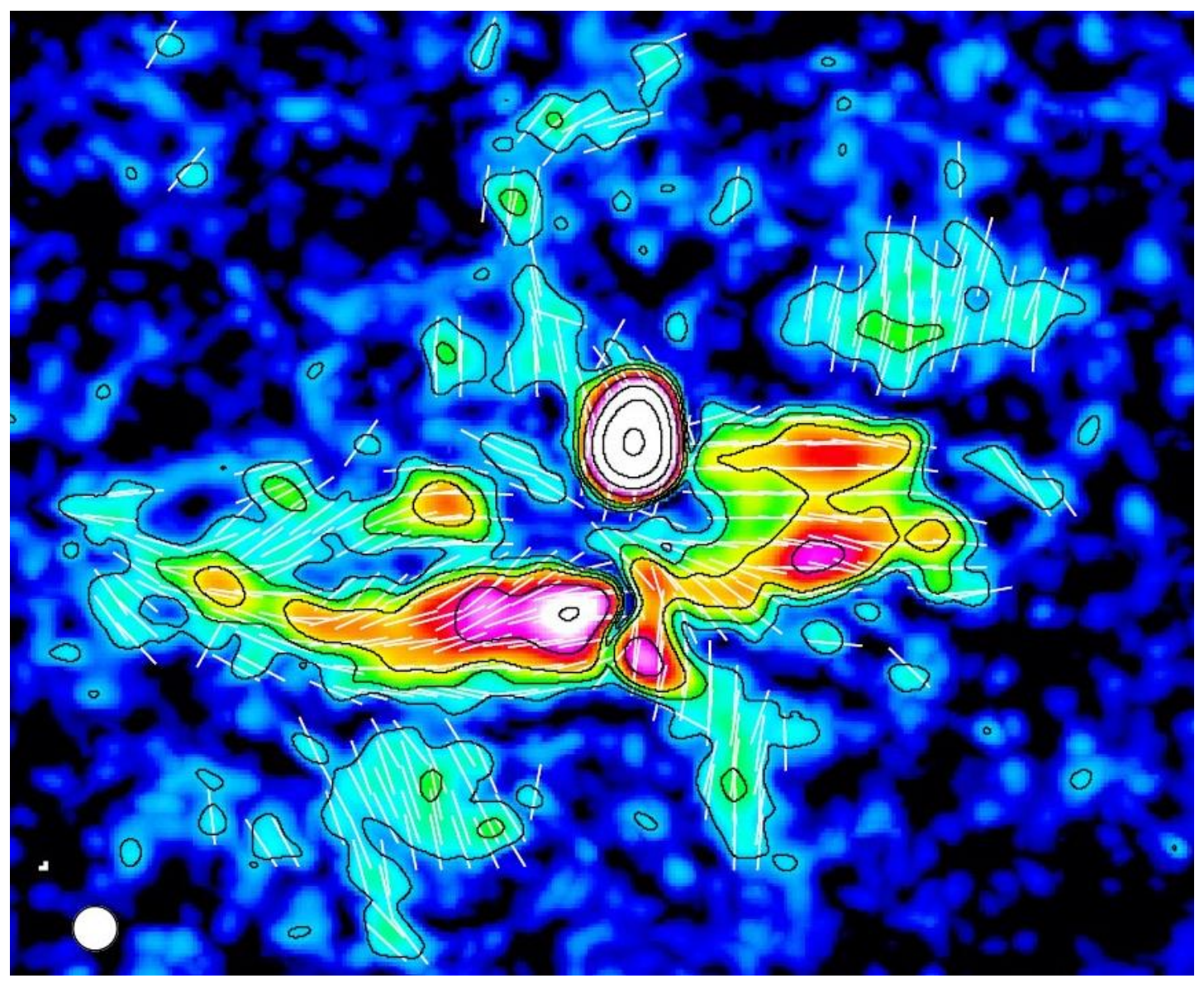

*Fig. 51: Edge-on spiral galaxy NGC4388 in the Virgo cluster. Polarized radio intensity (contours) and magnetic field orientations at 6.0 GHz (5.0 cm), observed with the VLA (Damas-Segovia et al. 2016). The lines are plotted with constant length.*

NGC4388 in the Virgo cluster has a nuclear outflow emerging from the active Seyfert–type nucleus extending to about 5 kpc from the center (Fig. 51). The radio emission is highly polarized, with RMs indicating a helical field. Further blobs of polarized emission above the plane in the north-west and south-east with vertical fields probably arise from a galactic wind. Highly polarized radio emission from kpc-sized jets has also been detected e.g. in NGC3079 (Cecil et al. 2001) and in the outflow lobes of the Circinus Galaxy (Elmouttie et al. 1995).

Jets in spiral galaxies may be more frequent than the available radio observations suggest. Future low-frequency observations may help, because they may show weak synchrotron emission from interface regions between the jets and the low-density halo gas.

### 4.10 Early-type and dwarf spheroidal galaxies

Elliptical galaxies with active nuclei are among the brightest known radio sources. Their jets and radio lobes are generated by magneto-hydrodynamic processes that are discussed elsewhere. Radio emission from quiet elliptical and S0 galaxies, if any, is also associated with their nuclei (Fabbiano et al. 1987). While elliptical galaxies without an active nucleus and galaxies of type Sa and S0 have little overall star formation activity, a significant fraction of fast-rotating early-type galaxies is still forming stars in the inner regions and emits synchrotron emission (Nyland et al. 2017). Their radio luminosity follows a similar radio – IR correlation as that for spiral galaxies (section 4.3), indicating amplification of turbulent magnetic fields. Polarized emission has not yet been detected, except for the Sa galaxy M104 with its prominent dust ring, where Krause et al. (2006) found weak, ordered magnetic fields. Slowly rotating quiescent elliptical galaxies were not detected or very weak in radio continuum.

The existence of magnetic fields in the halos of ellipticals without active nucleus and without any star formation is a matter of speculation. Regular fields are not expected in ellipticals because the lack of ordered rotation prevents the action of the $\alpha$-$\Omega$ dynamo. Dwarf spheroidal galaxies have some ordered rotation, but lack turbulent gas. Models of turbulence in the hot gas of large quiescent ellipticals indicate that a small-scale dynamo generates turbulent fields with less than 1 μG strength in the central regions and with turbulent scales of a few 100 pc (Seta et al. 2021). However, it is unclear whether the shock fronts of Type Ia supernovae can generate a significant number of cosmic-ray electrons that would emit synchrotron emission. Future detection of turbulent magnetic fields is more promising via the dispersion of Faraday rotation measures towards polarized background sources. However, most large ellipticals are located in galaxy clusters, where Faraday rotation is dominated by the turbulent fields of the intracluster gas. For small ellipticals, the number of observable polarized background sources will be sufficiently high only with very sensitive radio telescopes like the SKA.

Dwarf spheroidal galaxies are of interest to search for synchrotron emission from secondary electrons and positrons generated by the decay of dark-matter by WIMP annihilations, e.g. neutralinos (Colafrancesco et al. 2007). These galaxies do not generate thermal emission or primary electrons from star formation. Detection of diffuse radio synchrotron emission would be of high importance, but all attempts failed so far (Spekkens et al. 2013, Regis et al. 2014, Basu et al. 2021; Gajović et al. 2023). The main uncertainty is the origin of magnetic fields in such systems (see above). If the field strength is a few μG, detection of synchrotron emission from dark-matter decay may be possible, e.g. with the Square Kilometre Array (SKA) (Colafrancesco et al. 2015).

## 5. Outlook

Thanks to radio polarization observations, the global properties of interstellar magnetic fields in external galaxies and the field structures on pc and sub-pc sizes in the Milky Way are reasonably well known, while the processes connecting the features at large and small scales are not understood because the angular resolution in external galaxies is too low with present-day radio telescopes. Most of the existing polarization data was observed in single frequency bands and may suffer from depolarization by gradients of Faraday rotation or by different Faraday rotation components within the beam or along the line of sight. Modern radio telescopes are equipped with broadband receivers and multichannel polarimeters, allowing application of RM Synthesis (section 2.4), so that components in the Faraday spectrum can be resolved, which will help to identify field patterns (Omae et al. 2023). This method is presently revolutionizing radio polarization observations.

New and planned telescopes will widen the range of observable magnetic phenomena. The importance of polarimetry for the planned giant optical telescopes still needs to be established. The PLANCK satellite and several balloon instruments (PILOT, BLASTPol) has improved the sensitivity of polarimetry in the FIR and sub-mm range at arcminute resolution. SOFIA/HAWC+ enabled excellent FIR polarimetry at even higher resolution, but does not operate anymore. The Atacama Large Millimetre Array (ALMA) provides greatly improved sensitivity at arcsecond resolution for detailed imaging diffuse polarized emission from dust grains and for detection of the Zeeman effect in molecular clouds in the Milky Way and in external galaxies. The detection of a large-scale magnetic field in a lensed galaxy at z = 2.6 by observing red-shifted FIR polarization with ALMA (Geach et al. 2023) opens new perspectives to observe magnetic fields in distant galaxies.

High-resolution, deep observations at high radio frequencies (≥ 2 GHz), where Faraday effects are small, require a major increase in sensitivity for continuum observations that can be achieved with MeerKAT and the Square Kilometre Array (SKA) under construction. The detailed structure of the magnetic fields in the ISM of nearby galaxies and in galaxy halos will be observed with the SKA, giving direct insight into the interaction between magnetic fields and the various gas components (Beck et al. 2015). High angular resolution is also needed to distinguish between regular, tangled regular, and anisotropic turbulent fields and to test various models of the interaction between spiral shocks and magnetic fields. The power spectra of turbulent magnetic fields can be measured down to small scales. The SKA will also allow to measure the Zeeman effect in much weaker magnetic fields in the Milky Way and also in nearby galaxies.

The SKA will detect synchrotron emission from Milky Way-type galaxies at redshifts of z ≤ 1.5 (Fig. 52) and their polarized emission to z ≤ 0.5 (assuming 10% polarization) and lensed galaxies at even higher redshift. Bright starburst galaxies can be observed at larger redshifts, but are not expected to host ordered or regular fields. Total synchrotron emission, signature of total magnetic fields, can be detected in starburst galaxies with the SKA out to large redshifts, depending on luminosity and magnetic field strength (Fig. 52). However, for fields weaker than 3.25 μG $(1+z)^2$, energy losses of cosmic-ray electrons are dominated by the inverse Compton effect with photons of the cosmic microwave background, so that the particle energy is transferred mostly to X-rays and not to the radio domain. Furthermore, in the strong fields of young galaxies the energy range of electrons emitting in the GHz range is shifted to low energies, where ionization and bremsstrahlung losses are strong. The expected evolution of luminosities and synchrotron spectra of galaxies between redshifts of 5 and 0.1 were computed by Schober et al. (2016). The maximum redshift, until which synchrotron emission can be detected, can constrain models of the evolution of magnetic fields in young galaxies (Schleicher & Beck 2013).

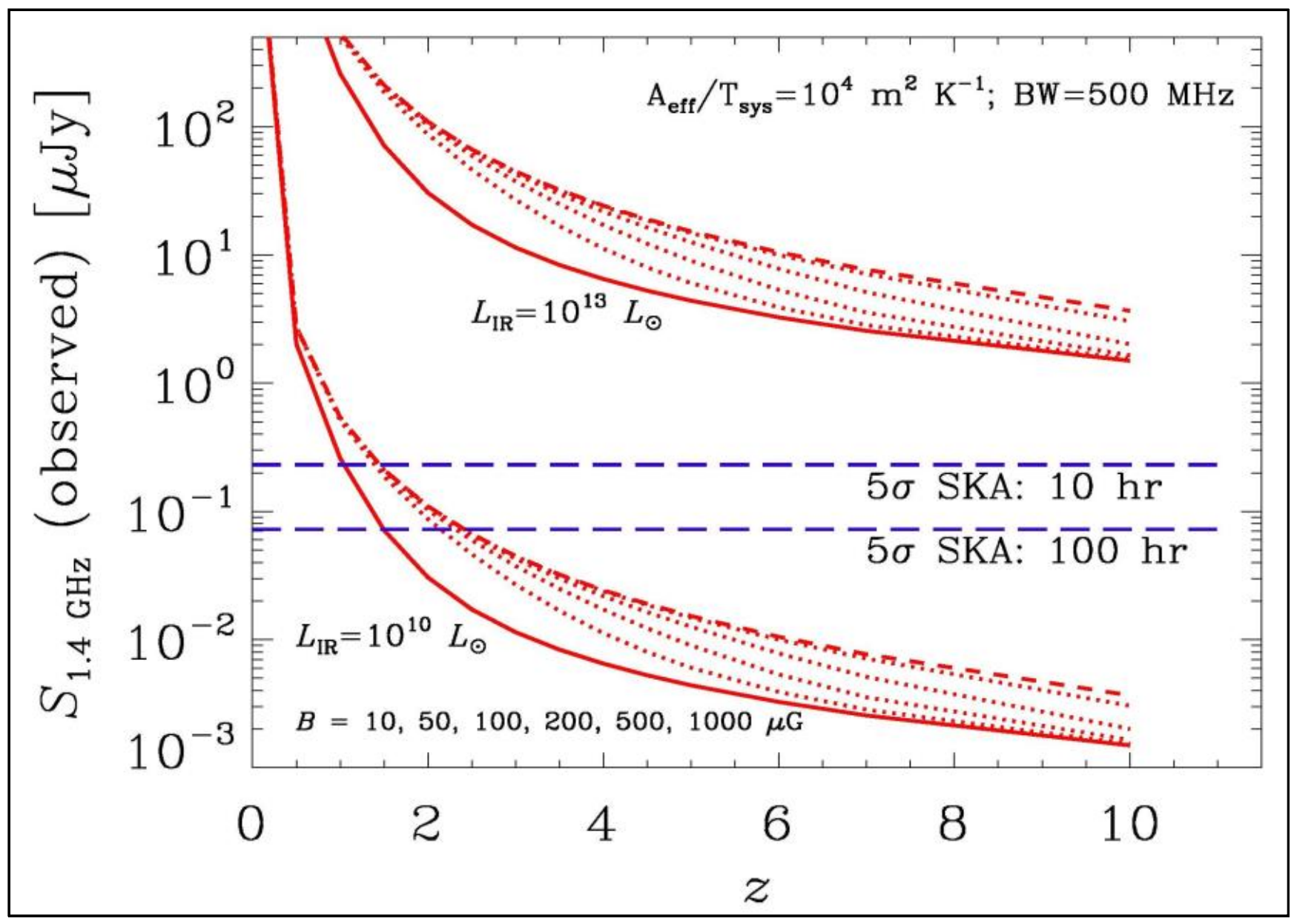

*Fig. 52: Total synchrotron emission at 1.4 GHz as a function of redshift z, total magnetic field strength B and total infrared luminosity $L_{IR}$. The 5σ detection limits for 10 h and 100 h integration time with the SKA are also shown (Murphy 2009).*

Ordered fields of nearby galaxies seen edge-on near the disk plane are preferably oriented parallel to the plane (section 4.7). As a result, polarized emission can be detected from distant, *unresolved* galaxies if they are symmetric (not distorted by interaction) and their inclination is larger than about 20° (Stil et al. 2009). This opens another method to search for ordered fields in distant galaxies. As the plane of polarization is almost independent of wavelength, distant spiral galaxies with known orientation of their major axis can also serve as background polarized sources to search for Faraday rotation by intergalactic fields in the foreground.

If polarized emission of galaxies is too weak to be detected, the method of *RM grids* towards polarized background QSOs can still be applied to measure the strength and structure of regular fields. Faraday rotation in an intervening galaxy (identified via absorption lines in its optical spectrum) occurs if its field is regular on spatial scales larger than that corresponding to the angular size of the background source. Circumgalactic fields in the outer halos of nearby galaxies were found with help of high-precision RMs from LOFAR (Heesen et al. 2023b). Regular fields of several μG strengths were discovered in galaxies up to redshifts of about 2 (Bernet et al. 2008; 2013; Kronberg et al. 2008; Farnes et al. 2014b; 2017). The distance limit of this method is determined by the polarized flux of the background QSO that can be much larger than that of the intervening galaxy, so that it can be applied to very large distances, even near to those of young QSOs ($z \geq 5$). The *gravitational lens effect* allows to measure several lines of sight through the lensing galaxy and hence to unambiguously measure its RM. The first application of this method led to the discovery of a regular field of a few μG strength in a galaxy at $z = 0.439$ (Mao et al. 2017). This opens the prospects to investigate magnetic fields in young galaxies and to search for their first fields with future radio telescopes like the SKA.

A reliable model for the structure of the magnetic field of an intervening galaxy needs many RM values, hence a sufficiently large number density of polarized background sources. If the background QSO has a polarized jet, some information about the large-scale field pattern in the intervening galaxy can be obtained (Kronberg et al. 1992). At least 10 randomly distributed background sources behind a galaxy disk are needed to recognize simple magnetic patterns, and several 1000 sources for a full reconstruction (Stepanov et al. 2008). The RM values measured today have been reduced by the redshift dilution factor of $(1+z)^{-2}$, so that high RM accuracy is needed. Present-day observations are not sensitive enough, and one has to wait for the SKA and its precursor telescopes.

Detection of regular fields in young galaxies is a critical test of $\alpha$-$\Omega$ dynamo models. Dynamo theory predicts timescales of amplification and coherent ordering of magnetic fields in galaxies (section 2.7). Based on models describing the formation and evolution of dwarf and disk galaxies, the probable evolution of turbulent and regular magnetic fields can be tested observationally (Arshakian et al. 2009; Rodrigues et al. 2015; 2019):

- Strong isotropic turbulent fields (in equipartition with turbulent gas motions) and hence unpolarized synchrotron emission are expected in galaxies at $z < 10$.
- Strong regular fields (that are coherent over a scale of about 1 kpc) and hence polarized synchrotron emission and fluctuating RMs are expected in galaxies at $z \leq 3$.
- Fully coherent regular fields and hence polarized synchrotron emission together with large-scale RM patterns are expected in dwarf and Milky-Way type galaxies at $z \approx 1 - 1.5$.
- More massive galaxies should build up regular fields earlier than less massive ones.
- Generating fully coherent regular fields takes more time in large galaxies (disk radius > 15 kpc), so that giant galaxies may not yet host fully coherent regular fields at $z = 0$.
- Major mergers enhance turbulent fields, but tangle or even destroy regular fields, and hence delay the formation of fully coherent fields. The lack of regular fields in nearby galaxies can be a signature of major mergers in the past.

The detections of total synchrotron emission in distant starburst galaxies at $z \leq 4$, far-infrared/submm polarized emission in a lensed galaxy at $z = 2.6$, and of RMs from intervening galaxies at $z \leq 2$ (see above) are consistent with dynamo theory. The overall field patterns observed in nearby galaxies are in general agreement with the predictions of the $\alpha$-$\Omega$ dynamo (sections 2.7, 4.4, and 4.7). Crucial tests of dynamo action will be possible in distant galaxies. Detection of regular fields at $z > 3$ would call for a faster type of dynamo or a different process. On the other hand, the failure to detect globally coherent patterns of regular fields in galaxies $z \geq 1$ would indicate that the time needed for field ordering is longer than the current $\alpha$-$\Omega$ dynamo models predict. However, if the bisymmetric spiral (BSS) magnetic pattern would turn out to dominate in distant galaxies, in contrast to that in nearby galaxies, this would challenge dynamo theory and indicate that the protogalactic fields could be of primordial origin (Sofue et al. 1986; Kulsrud et al. 1997).

The SKA "Cosmic Magnetism" Key Science Project plans to observe a polarization survey over the entire accessible sky with the SKA-MID Band 2 at around 1.4 GHz (Johnston-Hollitt et al. 2015; Heald et al. 2020). Within 1 h integration per field this will allow detection of about 10 million discrete extragalactic sources and measurement of their RMs, about 300 RMs per square degree, about 10000 RMs from pulsars in the Milky Way (Fig. 53) and several 100 extragalactic pulsars. About 1000 extragalactic sources are expected in the area around M31 (Fig. 54). This fundamental survey will be used to model the structure and strength of the magnetic fields in nearby and intervening galaxies (Beck et al. 2015), in galaxy clusters, and in the intergalactic medium. The SKA will unravel the 3D structure and configurations of magnetic fields in the Milky Way on sub-parsec to galaxy scales, including the field structure in the Galactic Center (Haverkorn et al. 2015; Han et al. 2015). The global configuration of the Milky Way's magnetic field in the disk, probed through pulsar RMs, will resolve the controversy about reversals within the Galactic plane. Characteristics of magnetic interstellar turbulence can also be determined from the RM grid. A pilot all-sky survey called POSSUM with the Australian SKA Precursor (ASKAP) is ongoing. MeerKAT, the South African SKA precursor, and APERTIF, the Dutch SKA pathfinder telescope, have higher sensitivity, but a smaller field of view than ASKAP and will concentrate on measuring RM grids centered on individual objects.

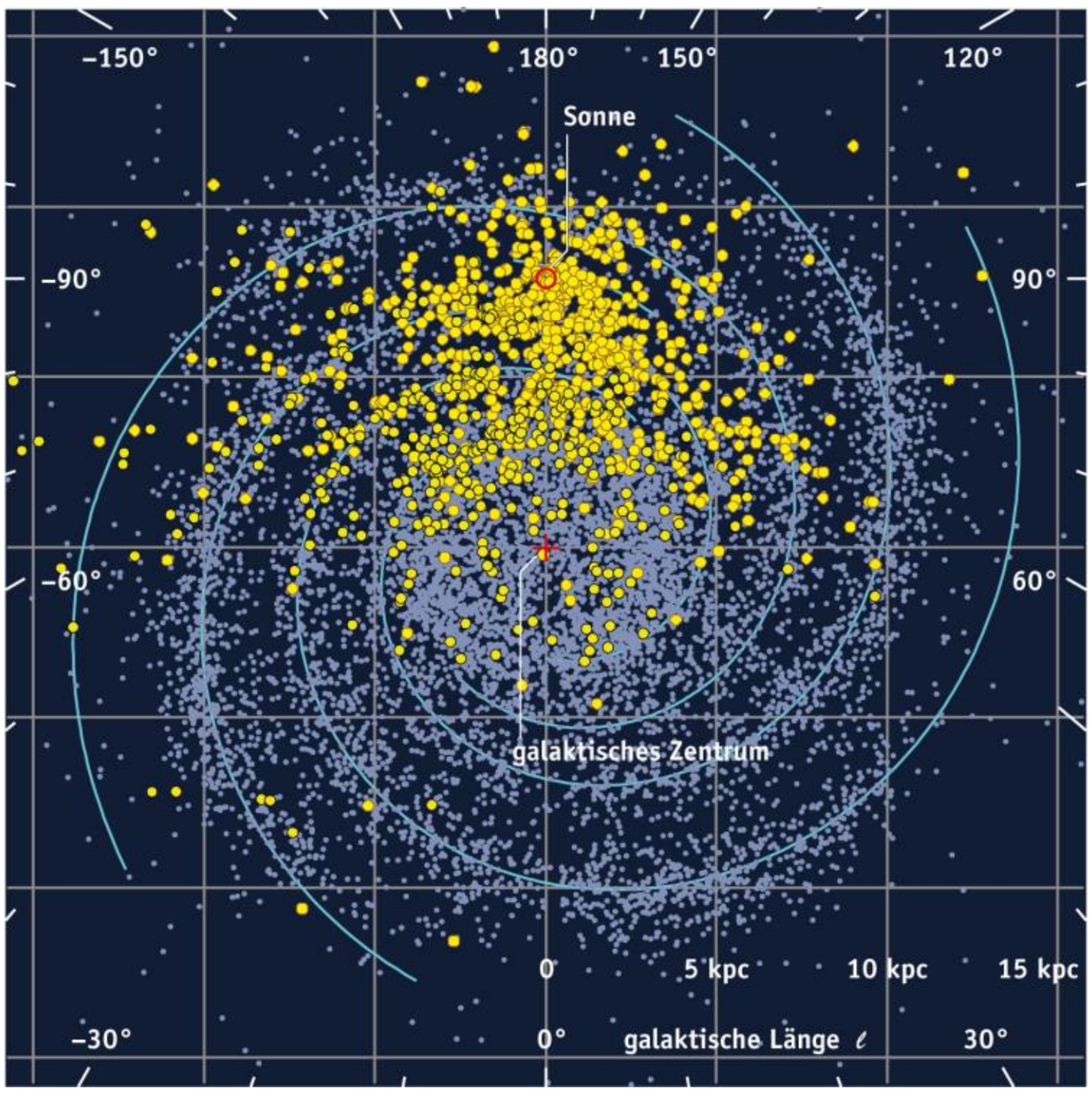

*Fig. 53: Simulation of pulsars in the Milky Way that can be detected with the SKA (blue), compared to about 2000 pulsars known today (yellow) (from Jim Cordes, Cornell University). Graphics: Sterne und Weltraum.*

Progress is also expected at low radio frequencies. Present-day measurements of galactic magnetic fields are limited by the lifetime and diffusion length of the cosmic-ray electrons that illuminate the fields. With typical diffusion lengths of only 1 kpc away from the acceleration sites in star-forming regions, the size of galaxies at centimeter wavelengths is not much larger than that in the optical or infrared spectral ranges. There is indication that magnetic fields probably extend much further into the intergalactic space (section 4.7). The Low Frequency Array (LOFAR), the Murchison Widefield Array (MWA), and the low-frequency part of the planned SKA are suitable instruments to search for extended synchrotron radiation at the lowest possible levels in outer galaxy disks and halos and investigate the transition to intergalactic space. While most of the disk is depolarized at low frequencies (Mulcahy et al. 2014), polarization may still be detectable from the outer regions.

The filaments of the local Cosmic Web may contain *intergalactic magnetic fields*, possibly enhanced by IGM shocks, and this field may be detectable by direct observation of total synchrotron emission or by Faraday rotation towards background sources (Akahori et al. 2014). For fields of $10^{-8}$ – $10^{-7}$ G with 1 Mpc coherence length and $10^{-5}$ $cm^{-3}$ electron density, |RM| of 0.1 – 1 rad $m^{-2}$ is expected. An overall intergalactic field is much weaker and may become evident as increased |RM| towards QSOs at redshifts of $z > 3$ by averaging over a large number of sources (Vacca et al. 2016). As the Faraday rotation angle increases with $\lambda^2$, searches for low |RM| should preferably be done at low frequencies. RM data from LOFAR indicates a typical field strength in a filament of ≈ 10 - 20 nG, scaled to z = 0 (Carretti et al. 2023).

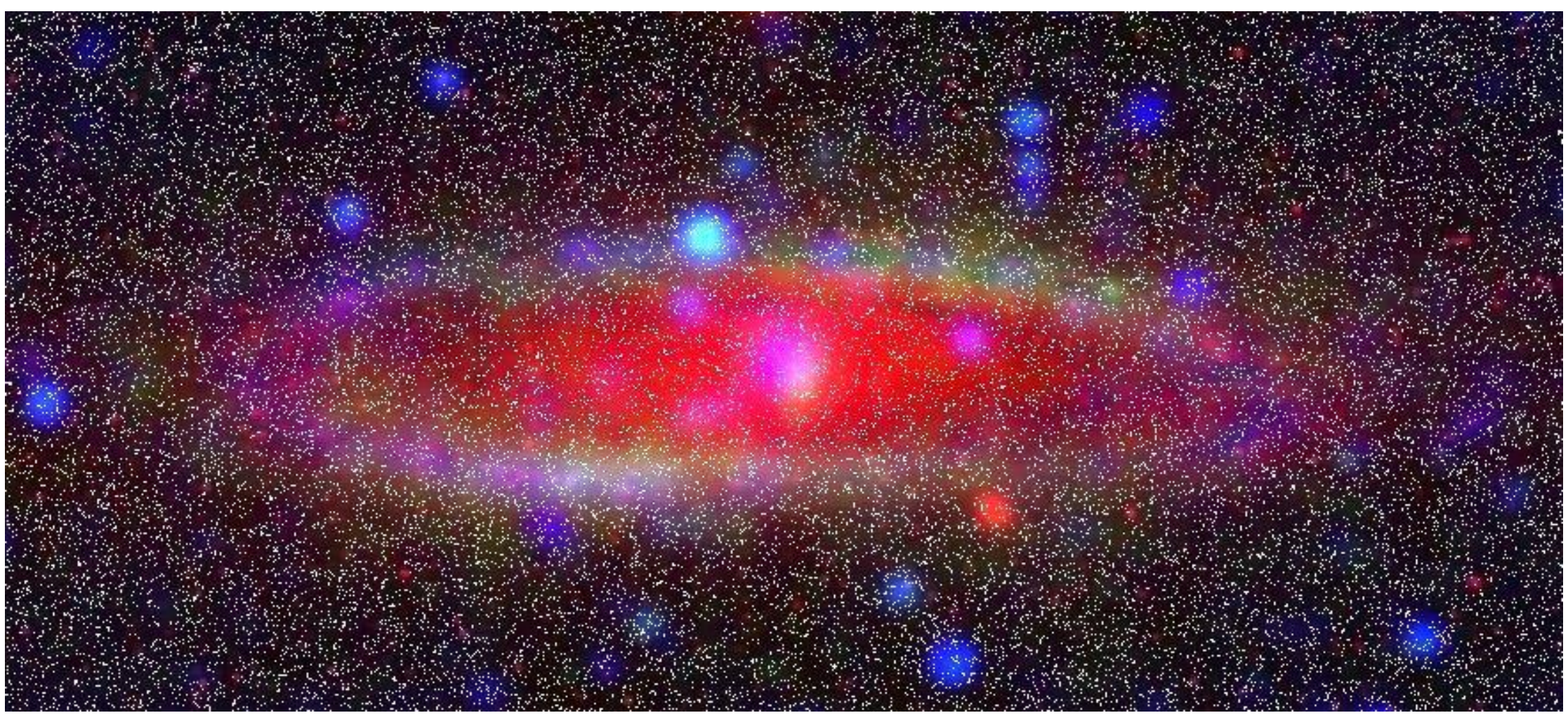

*Fig. 54: Simulation of RMs towards background sources (white points) in the region of M31 observable with the SKA within 1 h integration time. Optical emission from M31 is shown in red, diffuse total radio continuum intensity in blue and diffuse polarized intensity in green. Courtesy: Bryan Gaensler (University of Toronto).*

In summary, the SKA and its pathfinders (VLA, LOFAR, MWA, APERTIF) and precursors (ASKAP, MeerKAT) will measure the structure and strength of the magnetic fields in the Milky Way, in intervening galaxies, and possibly in the intergalactic medium. Hopefully, a new large FIR polarimetric telescope will follow. Looking back into time, the future telescopes could shed light on the origin and evolution of cosmic magnetic fields. The observational methods are:

- RM grids of extragalactic sources and pulsars to map the detailed 3D structure of the Milky Way's magnetic field (0.2–1 GHz)
- High-resolution mapping of total and polarized synchrotron or FIR emission and of RMs from galaxy disks and halos of nearby galaxies at high frequencies (≥ 5 GHz)
- Mapping of the total and polarized synchrotron emission from the outer disks and halos of nearby galaxies and galaxy groups at low frequencies (≤ 1 GHz)
- Reconstruction of 3D field patterns in nearby galaxies by RM Synthesis of diffuse polarized emission observed with broadband receivers (≥ 2 GHz)
- Reconstruction of 3D field patterns in nearby galaxies from RM grids towards polarized background sources (≥ 1 GHz)
- Recognition of patterns of regular fields in galaxies from RM grids towards polarized background sources (at $z \leq 0.02$)
- Search for polarized synchrotron emission from distant galaxies (at $z \leq 0.5$)
- Search for polarized synchrotron or FIR/submm emission from distant lensed galaxies (at $z \leq 3$)
- Search for RM patterns in distant lensed galaxies (≤ 1 GHz)
- Search for total synchrotron emission from distant starburst galaxies (at $z \leq 3$)
- Search for regular fields in very distant intervening galaxies towards QSOs (at $z \leq 5$).

Fundamental questions are waiting to be answered:

- When were the first magnetic fields generated and amplified: in young galaxies, in protogalactic clouds, or are they relics from the early Universe, before galaxies were formed?
- How and how fast were magnetic fields amplified and ordered in the interstellar medium?
- How did magnetic fields affect the evolution of galaxies?
- How important are magnetic fields for the physics of galaxies, like the efficiency to form stars from gas, the formation of spiral arms, and the generation of outflows?
- How strong and how ordered are magnetic fields in intergalactic space?
- Can outflows from galaxies magnetize the intergalactic space?
- What is the detailed structure of the magnetic fields in the Milky Way and in external galaxies, in particular the number and amplitudes of azimuthal modes, the frequency of field reversals on small and large scales, and the importance of tangled regular fields?
- How strongly are extragalactic ultrahigh-energy cosmic rays deflected by magnetic fields in the Milky Way and in intergalactic space?

## Acknowledgements

The authors would like to thank many of our colleagues, who have pursued the studies of magnetic fields in the Milky Way and in galaxies for the past 45 years, especially Elly M. Berkhuijsen, Marita Krause, Sui Ann Mao, Patricia Reich, and Wolfgang Reich, all at MPIfR Bonn. Many excellent cooperation projects in this field were performed with groups in Bochum and Potsdam (Germany), Kraków (Poland), DRAO Penticton (Canada), NAOC Beijing (China), Moscow and Perm (Russia), and Newcastle (UK). We thank Chris Chyży, JinLin Han, Marian Soida, and Marek Weżgowiec for providing figures. Elly M. Berkhuijsen, Katia Ferrière, Andrew Fletcher, Marijke Haverkorn, Cathy Horellou, Marita Krause, and Anvar Shukurov are acknowledged for carefully reading the manuscript and for many valuable comments. – This work was supported by the DFG Research Unit FOR1254.

# Appendix

## A.1 Catalogue of radio surveys of the Milky Way

*Table A1: All-sky or all-hemisphere radio total intensity surveys*

| Frequency | Beam size | Reference |
|---|---|---|
| 45 MHz | ≈4° | Guzmán et al. 2011 |
| 150 MHz | 3°.6 | Landecker & Wielebinski 1970 |
| 408 MHz | 2° | Haslam et al. 1982 |
| 1.4 GHz | 0°.6 | Reich 1982; Reich & Reich 1986; Reich et al. 2001 |
| 2.3 GHz | 0°.33 | Jonas et al. 1998 (southern hemisphere) |
| 2.7 GHz | 0°.33 | Reif et al. 1987 (northern hemisphere) |
| 23 – 94 GHz | 0.°8 – 0°.2 | Hinshaw et al. 2009 |
| 20 – 100 GHz | 1° | Ade et al. (Planck Collaboration) 2016b |

*Table A2: All-sky or all-hemisphere total intensity and linear polarization surveys*

| Frequency | Beam size | Reference |
|---|---|---|
| 120-168 MHz | 5” – 4’ | Shimwell et al. 2017; Van Eck et al. 2019; Erceg et al. 2022 (LOTSS, northern hemisphere) |
| 189 MHz | 15.6’ | Bernardi et al. 2013 (southern hemisphere, 2400 $deg^2$) |
| 408 MHz | 7.5° | Wielebinski et al. 1962 (northern hemisphere) |
| | 2° | Berkhuijsen & Brouw 1963 (northern hemisphere) |
| | 7.5° | Wielebinski & Shakeshaft 1964 (northern hemisphere) |
| | ≈1° | Mathewson & Milne 1965 (southern hemisphere) |
| 0.3 – 1.8 GHz | 30’ – 81’ | Wolleben et al. 2009; 2019 (GMIMS; all-sky) |
| 1.28 – 1.75 GHz | 40’ | Wolleben et al. 2010a; 2010b; 2021; Sun et al. 2015; Hill et al. 2017 |
| 1.4 GHz | 36’ | Wolleben et al. 2006; Testori et al. 2008 (all-sky) |
| 1.4 GHz | ≈13’ | Rudnick & Brown 2009 (northern hemisphere) |
| 2.3 GHz | 8.9’ | Carretti et al. 2019 (S-PASS; southern hemisphere) |
| 2 – 4 GHz | 2.5” | Lacy et al. 2020 (VLASS; all sky visible to the VLA) |
| 4.8 GHz | 45’ | Dickinson et al. 2019 (C-BASS; all sky) |
| 23 – 94 GHz | 0.°8 – 0°.2 | Kogut et al. 2007; Hinshaw et al. 2009 (all-sky) |
| 28.4 GHz | 1° | Ade et al. (Planck Collaboration) 2016b (all-sky) |
| 353 GHz | 1° | Ade et al. (Planck Collaboration) 2015 (all-sky) |

*Table A3: Radio surveys of the Galactic plane*

| Frequency | Beam | Area (Galactic coordinates) | | Reference |
|---|---|---|---|---|
| 325 MHz | ≈4’ | selected areas | | Wieringa et al. 1993 |
| 350 MHz | ≈5’ | selected areas | | Haverkorn et al. 2003; 2004 |
| 408 MHz | ≈3’ | $147.3^\circ > l > 74.2^\circ$ | $-7.7^\circ < b < 8.7^\circ$ | Taylor et al. 2003 |
| 408 MHz | ≈3’ | $193^\circ > l > 52^\circ$ | $-6.5^\circ < b < 8.5^\circ$ | Tung et al. 2017 |
| 1.4 GHZ | ≈1’ | $147.3^\circ > l > 74.2^\circ$ | $-3.6^\circ < b < 5.6^\circ$ | Taylor et al. 2003 |
| 1.4 GHz | ≈1’ | $67^\circ > l > 18^\circ$ | $b \pm 1^\circ.5$ | Stil et al. 2006 |
| 1.4 GHz (PI) | ≈1’ | $358^\circ > l > 253^\circ$ | $b \pm 1^\circ.5$ | Haverkorn et al. 2006 |
| 1.4 GHz (PI) | ≈1’ | $175^\circ > l > 66^\circ$ | $-3^\circ < b < 5^\circ$ | Landecker et al. 2010 (CGPS) |
| 1.4 GHz | 9’ | $162^\circ > l > 93^\circ$ | $b \pm 4^\circ$ | Kallas & Reich 1980 |
| 1.4 GHz | 9.4’ | $240^\circ > l > 95.5^\circ$ | $-4^\circ < b < 5^\circ$ | Reich et al. 1990a; 1997 |
| 1.4 GHz (PI) | 9.4’ | $230^\circ > l > 25^\circ$ | $-20^\circ < b < 20^\circ$ | Uyanıker et al. 1999; Reich et al. 2004 (EMLS) |
| 1.4 GHz (PI) | ≈1’ | $332.5^\circ > l > 325.5^\circ$ | $-0.5^\circ < b < 3.5^\circ$ | Gaensler et al. 2001 |
| 1 – 2 GHz (PI) | 15” | $67.25^\circ > l > 14.5^\circ$ | $b \pm 1.25^\circ$ | Shanahan et al. 2019 (THOR) |
| 2.3 GHz (PI) | 10.75’ | $34^\circ > l > 10^\circ$ | $b \pm 5^\circ$ | Sun et al. 2014 |
| 2.4 GHz | 10.4’ | $238^\circ > l > 365^\circ$ | $b \pm 5^\circ$ | Duncan et al. 1995 |
| 2.4 GHz (PI) | 10.4’ | $238^\circ > l > 5^\circ$ | $b \pm 5^\circ$ | Duncan et al. 1997 |
| 2.7 GHz | 4.3’ | $357.4^\circ < l < 76^\circ$ | $b \pm 1.5^\circ$ | Reich et al. 1984 |
| 2.7 GHz (PI) | 6’ | $74^\circ > l > 4^\circ.9$ | $b \pm 1.5^\circ$ | Junkes et al. 1987 |
| 2.7 GHz | 4.4’ | $76^\circ > l > 358^\circ$ | $b \pm 5^\circ$ | Reich et al. 1990b |
| 2.7 GHz | 4.4’ | $240^\circ > l > 76^\circ$ | $b \pm 5^\circ$ | Fürst et al. 1990 |
| 2.7 GHz (PI) | 4.3’ | $74^\circ > l > 4^\circ.9$ | $b \pm 5^\circ$ | Duncan et al. 1999 |
| 5 GHz (PI) | 9’ | $129^\circ > l > 122^\circ$ | $b \pm 5^\circ$ | Sun et al. 2007 |
| 5 GHz (PI) | 9’ | $230^\circ > l > 129^\circ$ | $b \pm 5^\circ$ | Gao et al. 2010 |
| 5 GHz (PI) | 9’ | $122^\circ > l > 60^\circ$ | $b \pm 5^\circ$ | Xiao et al. 2011 |
| 5 GHz (PI) | 9’ | $60^\circ > l > 10^\circ$ | $b \pm 5^\circ$ | Sun et al. 2011; 2014 |

(PI): with linear polarization data

### A.2 Catalogue of radio polarization observations of nearby galaxies

In radio continuum the typical degrees of polarization are much higher than those in the other spectral ranges and further benefit comes from the development of large instruments and sensitive receivers. This is why almost all of our knowledge on interstellar magnetic fields in galaxies is based on their polarized radio emission and Faraday rotation.

Lists of spiral, barred, irregular and dwarf galaxies detected in radio polarization until the year 2020 are given in Tables A4-A6. Most detections were made in the wavelength range 3 – 13 cm, where Faraday depolarization is small. At $\lambda \approx 20$ cm, the polarized intensity is generally smaller by a factor of several (Fig. 3). At wavelengths longer than 20 cm, only one detection of polarized emission from spiral galaxies has been reported so far (M31). Results on FIR polarization are compiled in Table A7.

Radio images in colour and download links are compiled on: http://www.mpifr-bonn.mpg.de/atlasmag.

*Table A4: Radio polarization observations and magnetic field structures of galaxies with **low or moderate inclination***

| Galaxy | Telescope & wavelength (cm) | Structure | Reference |
|---|---|---|---|
| M33 | E21,18,11,6,3 | BSS | Buczilowski & Beck 1991 |
| | E6,4, V21 | ASS+BSS+QSS | Tabatabaei et al. 2008 |
| M51 | W21,6 | Spiral | Segalovitz et al. 1976 |
| | V21,18 | Spiral | Horellou et al. 1992 |
| | E6,3, V21,6 | ASS+BSS (disk) | Neininger 1992 |
| | | +ASS (halo) | Berkhuijsen et al. 1997 |
| | W21 | BSS (halo) | Heald et al. 2009 |
| | E6,4, V21,6,4 | ASS+QSS (disk), +BSS (halo) | Fletcher et al. 2011 |
| | V20L | BSS+vertical (halo) | Mao et al. 2015 |
| | L200 | *Not detected* | Mulcahy et al. 2014 |
| | V10S | Tangled or vertical fields | Kierdorf et al. 2020 |
| M74 =NGC0628 | W21 | Incomplete spiral | Heald et al. 2009 |
| | E11,4, V10S | Magnetic arms | Mulcahy et al. 2017 |
| M81 | E6, V21 | BSS? | Krause et al. 1989b |
| | E6, V21 | BSS? (+ASS) | Sokoloff et al. 1992 |
| | V21,6 | Interarm fields | Schoofs 1992 |
| M83 | V21 | Spiral | Sukumar & Allen 1989 |
| | P6 | Spiral | Harnett et al. 1990 |
| | E6,3 | BSS? | Neininger et al. 1991; 1993 |
| | A13 | Magnetic arms | Ehle 1995; Frick et al. 2016 |
| | V6 | Magnetic arms +‖bar | Frick et al. 2016 |
| | K21 | Reversal between disk and halo? | Heald et al. 2016 |
| M101 | E11,6,3 | Spiral | Gräve et al. 1990 |
| | E11,6 | Spiral | Berkhuijsen et al. 2016 |
| NGC876/7 | V6 | Spiral | Drzazga et al. 2011 |
| NGC1097 | V21,18,6,4 | ASS+BSS+QSS +‖bar +nuclear spiral | Beck et al. 2002; 2005a |
| NGC1156 | V21,6,4, E6,4 | Patches of ordered field | Kepley et al., in prep. |
| NGC1365 | V21,18,6,4 | ASS+BSS+QSS + ‖bar +nuclear spiral | Beck et al. 2005a |
| NGC1559 | A13,6 | ‖bar | Beck et al. 2002 |
| NGC1566 | A21,13,6 | Spiral, interarm | Ehle et al. 1996 |
| NGC1569 | W21, V6,4 | Spiral, bubbles, loops | Kepley et al. 2010 |
| NGC1672 | A13,6 | Spiral, interarm | Beck et al. 2002 |
| NGC2207 | V6 | Spiral +radial streamers | Drzazga et al. 2011 |
| NGC2276 | V21,6 | BSS? | Hummel & Beck 1995 |
| NGC2403 | E11,6 | Spiral | Beck, unpubl. |
| | W21 | Diffuse | Heald et al. 2009 |
| NGC2442 | A13,6 | Spiral +‖bar | Harnett et al. 2004 |
| NGC2841 | W21 | Two arcs | Heald et al. 2009 |
| NGC2903 | E6,3, V21 | Spiral | Beck, unpubl. |
| | W21 | Spiral | Heald et al. 2009 |
| NGC2976 | V21, W21, E6,4 | Distorted disk+halo field | Drzazga et al. 2016 |
| NGC2997 | V21,6,4, A13 | Spiral +inner ASS? | Han et al. 1999 |
| NGC3521 | E3 | Spiral, compressed | Knapik et al. 2000 |
| NGC3627 | E3 | Spiral +‖dust lane | Soida et al. 1999 |
| | V6,4 | Anomalous arm | Soida et al. 2001 |
| | V6,4 | BSS | Kurahara et al. 2021 |
| | W21 | Spiral | Heald et al. 2009 |
| | E11 | Spiral | Nikiel-Wroczyński et al. 2013a |
| NGC3938 | W21 | Spiral | Heald et al. 2009 |
| NGC4038/9 | V21,6,4 | ‖tidal arm | Chyży & Beck 2004 |
| | V10S | Regular field ‖tidal arm | Basu et al. 2017 |
| NGC4214 | V6 | *No ordered field* | Kepley et al. 2011 |
| | E6 | Fragment of a spiral | Drzazga 2008 |
| NGC4258 | W21, V21 | In anomalous arms | van Albada & van der Hulst 1982 |
| | V21,6 | ‖anomalous arms | Hummel et al. 1989 |
| | V4, E3 | ‖nuclear jet + an. arms | Krause & Löhr 2004 |
| NGC4414 | V6,4 | ASS+BSS+QSS | Soida et al. 2002 |
| NGC4449 | E6,3 | ‖opt. filaments | Klein et al. 1996 |
| | V6,4 | Spiral+radial field | Chyży et al. 2000 |
| NGC4490/85 | E6,4 | Radial halo field | Knapik et al., in prep. |
| NGC4736 | V6,4 | Spiral, ASS? | Chyży & Buta 2008 |
| | W21 | Outer lobe | Heald et al. 2009 |
| NGC5033 | W21 | Inner disk | Heald et al. 2009 |
| NGC5055 | E3 | Spiral | Knapik et al. 2000 |
| | W21 | Spiral | Heald et al. 2009 |
| NGC5426/7 | V6 | Spiral +spiral | Drzazga et al. 2011 |
| NGC6822 | E11,6,3 | *No ordered field* | Chyży et al. 2003; 2011 |
| NGC6907/8 | V6 | Spiral | Drzazga et al. 2011 |
| NGC6946 | E21,11,6,3 | ASS? | Ehle & Beck 1993; Beck 2007 |
| | V21,18,6,4 | ASS+QSS | Beck 1991; 2007; Kurahara & Nakanishi 2019 |
| | W21 | ASS (halo) | Heald et al. 2009 |
| | W21,13 | ASS (disk+halo) | Williams et al. 2021 |
| NGC7479 | V21,6,4 | ‖spiral jet | Beck et al. 2002; Laine & Beck 2008 |
| NGC7552 | A6 | Spiral +‖bar | Beck et al. 2002 |
| | G49 | Nucleus | Farnes et al. 2014a |
| IC10 | E11,6,3 | ‖Hα filament | Chyży et al. 2003; 2011 |
| | V6 | Filaments | Heesen et al. 2011a |
| | V6,4 | Filaments, X-shaped halo | Chyży et al. 2016 |
| IC342 | E11,6 | ASS | Gräve & Beck 1988 |
| | E6, V21 | ASS | Krause et al. 1989a |
| | E6, V21 | ASS | Sokoloff et al. 1992 |
| | E21,11,6,3 | ASS | Beck 2015a |
| | V21,6,4 | Magnetic arm, helically twisted field | Beck 2015a |
| | L200 | *Not detected* | Van Eck et al. 2017 |
| IC1613 | E11,6 | *No ordered field* | Chyży et al. 2011 |
| UGC813/6 | V6 | ⊥bridge | Drzazga et al. 2011 |
| UGC12914/5 | V6 | ‖bridge | Drzazga et al. 2011 |
| Holmberg II | E11,6 | *No ordered field* | Chyży et al., unpubl. |
| SMC | P21,13 | ‖main ridge | Haynes et al. 1986 |
| | A21 | Pan-Magellanic? | Mao et al. 2008 |
| | A10S+20L | Coherent field | Livingston et al. 2022 |
| LMC | P21,13,6 | Magn. loop near 30 Dor | Haynes et al. 1991; Klein et al. 1993 |
| | A21 | ASS | Gaensler et al. 2005 |
| | P21, A21 | ‖filaments, even-symmetry halo field | Mao et al. 2012b |
| | A10S+20L | | Livingston et al., in prep. |
| Leo Triplet | E11 | *No intergalactic field* | Nikiel-Wroczyński et al. 2013a |
| Stephan's Quintet | E4, V21,6 | Intergalactic field | Nikiel-Wroczyński et al. 2013b |
| HCG15 | E6, V21 | Intergalactic field | Nikiel-Wroczyński et al. 2017 |
| HCG44 | E6, V21 | Intergalactic field | Nikiel-Wroczyński et al. 2017 |
| HCG60 | E6, V21 | Intergalactic field | Nikiel-Wroczyński et al. 2017 |
| PKS1229-021 | V21,6,2 | BSS? | Kronberg et al. 1992 |
| CLASS B1152+199 | V20L,10S,5C | ASS? | Mao et al. 2017 |

*Table A5: Radio polarization observations and magnetic field structures of galaxies with* ***high inclination***

| Galaxy | Telescope & λ(cm) | Structure | Reference |
|---|---|---|---|
| M31 | E21,11,6 | Even-symmetry ASS | Beck 1982; Beck et al. 1989 |
| | V21 | Ring-like, rudimentary | Beck et al. 1998 |
| | V6,4, E6,4 | Reversed ASS (center) | Gießübel & Beck 2014 |
| | E11,6, V21 | Even ASS (+QSS) | Berkhuijsen et al. 2003; Fletcher et al. 2004 |
| | E4 | Ring-like | Gießübel 2012 |
| | W90 | Ring-like, strongly depol. | Gießübel et al. 2013 |
| | E11,6,4 | ASS +BSS | Beck et al. 2020 |
| M82 | V6,4 | Radial halo field | Reuter et al. 1994 |
| | E1 | ‖disk +vertical halo field | Wielebinski 2006 |
| | V6,3, W21 | Helical field in outflow | Adebahr et al. 2017 |
| M104 | V21,6 | ‖disk | Bajaja et al. 1988; |
| | | +X-shaped halo field | Krause et al. 2006 |
| NGC253 | P6,4 | ‖plane | Harnett et al. 1990 |
| | V21,6 | ‖plane | Carilli et al. 1992 |
| | E6,3 | ‖plane, halo spurs | Beck et al. 1994 |
| | E6,4, V21,6 | Even ASS disk field | |
| | | +even halo field | Heesen et al. 2009a; 2009b |
| | V21,6,4 | Helical field in | Heesen et al. 2011b |
| | | outflow cone | |
| NGC660 | V6 | ‖polar ring +X-shaped | Drzazga et al. 2011 |
| | V5C | ‖plane +X-shaped | Wiegert et al. 2015; Krause et al. 2020 |
| NGC891 | V6 | ‖plane +halo spurs | Sukumar & Allen 1991 |
| | V21 | ‖plane +halo spurs | Hummel et al. 1991 |
| | E3 | ‖plane +tilted | Dumke et al. 1995 |
| | E4 | Even ASS disk field | |
| | | +X-shaped halo field | Krause 2009 |
| | V20L,5C | X-shaped | Wiegert et al. 2015; Schmidt et al. 2019; |
| | | | Krause et al. 2020 |
| NGC1808 | V21,6 | Halo spurs | Dahlem et al. 1990 |
| NGC2613 | V5C | ‖plane | Krause et al. 2020 |
| NGC2683 | V5C | ‖plane | Wiegert et al. 2015 |
| NGC2820 | V5C | ‖plane | Krause et al. 2020 |
| NGC3044 | V20L,5C | X-shaped | Wiegert et al. 2015; Krause et al. 2020 |
| NGC3079 | V6 | Extraplanar outflow | Duric & Seaquist 1988 |
| | V5C | ‖plane +extraplanar | Krause et al. 2020 |
| NGC3432 | E6 | Vertical, weak | Drzazga 2008 |
| NGC3448 | V5C | X-shaped | Wiegert et al. 2015; Krause et al. 2020 |
| NGC3556 | V20L,5C | ‖plane +vertical filaments | Wiegert et al. 2015; Krause et al. 2020 |
| NGC3628 | V21 | Fragments of ord. field | Reuter et al. 1991 |
| | E3 | ‖plane | Dumke et al. 1995 |
| | E4 | ‖plane | |
| | | +X-shaped halo field | Krause, unpubl. |
| | E11 | ‖plane | Nikiel-Wroczyński et al. 2013a |
| | V20L,5C | X-shaped | Wiegert et al. 2015; Krause et al. 2020 |
| NGC3735 | V20L,5C | X-shaped | Wiegert et al. 2015; Krause et al. 2020 |
| NGC4013 | V5C | ‖plane | Wiegert et al. 2015; Krause et al. 2020 |
| | V5C | ‖plane +vertical filaments | Stein et al. 2019b |
| NGC4157 | V20L,5C | ‖plane +X-shaped | Wiegert et al. 2015; Krause et al. 2020 |
| NGC4192 | V5C | ‖plane | Krause et al. 2020 |
| NGC4217 | V6 | X-shaped | Soida 2005 |
| | V20L,5C | X-shaped | Wiegert et al. 2015; Krause et al. 2020 |
| | V20L,5C | X-shaped +helical outflow | Stein et al. 2020 |
| NGC4236 | E6 | *No ordered field* | Chyży et al. 2007 |
| NGC4302 | V5C | ‖plane +vertical filaments | Krause et al. 2020 |
| NGC4565 | V21 | ‖plane | Sukumar & Allen 1991 |
| | E3 | ‖plane | Dumke et al. 1995 |
| | E6,4, V6 | ‖plane | |
| | | +X-shaped halo field | Krause 2009 |
| | V20L,5C | ‖plane | Wiegert et al. 2015; Krause et al. 2020 |
| NGC4631 | V21 | ⊥plane | Hummel et al. 1991 |
| | V6,4 | ⊥plane, spurs | Golla & Hummel 1994 |
| | V21,18 | X-shaped halo field | Beck 2009 |
| | E4 | ‖plane +vertical central field +X-shaped halo field | Krause 2009; Mora-Partiarroyo et al. 2013 |
| | E6 | X-shaped | Mora et al. 2013 |
| | V6 | ‖plane +vertical field | Irwin et al. 2012; Mora-Partiarroyo et al. 2013 |
| | W21 | X-shaped | Heald et al. 2009 |
| | V20L,5C | X-shaped | Wiegert et al. 2015; Krause et al. 2020 |
| | V5C | Magnetic ropes | Mora-Partiarroyo et al. 2019 |
| NGC4656 | E6 | No ordered field | Chyży et al. 2007 |
| NGC4666 | V21,6 | X-shaped | Dahlem et al. 1997 |
| | V6 | X-shaped | Soida 2005 |
| | V20L,5C | X-shaped | Wiegert et al. 2015; Krause et al. 2020 |
| | V20L,5C | ASS +reversal | Stein et al. 2019a |
| NGC4945 | P6,4 | Halo spurs | Harnett et al. 1989; 1990 |
| NGC5775 | V21,6 | X-shaped | Tüllmann et al. 2000 |
| | V4 | Even ASS disk field +X-shaped halo field | Soida et al. 2011 |
| | V20L,5C | X-shaped | Wiegert et al. 2015; Krause et al. 2020 |
| NGC5907 | E6,3, V21 | ‖plane +X-shaped | Dumke 1997; Dumke et al. 2000 |
| | E4 | ‖plane +X-shaped | Krause, unpubl. |
| | V5C | ‖plane | Krause et al. 2020 |
| NGC7090 | A21,6 | Vertical field | Heesen et al. 2016 |
| NGC7331 | E3 | ‖plane | Dumke et al. 1995 |
| | W21 | X-shaped | Heald et al. 2009 |
| NGC7462 | A21,6 | Vertical field | Heesen et al. 2016 |
| NGC7582 | G49 | Nucleus | Farnes et al. 2014a |
| Circinus | A21, 13 | ‖radio lobes | Elmouttie et al. 1995 |
| IC2574 | E6 | *No ordered field* | Chyży et al. 2007 |
| UGC10288 | V20L,5C | Ordered field in back-ground radio galaxy | Irwin et al. 2013 |

*Table A6: Radio polarization observations and magnetic field structures of galaxies in the* ***Virgo cluster***

| Galaxy | Telescope & $\lambda$(cm) | Structure | Reference |
|---|---|---|---|
| NGC4192 | E6,4 | ASS? +halo field | Weżgowiec et al. 2012 |
| | V21,6 | ‖disk +inclined | Vollmer et al. 2013 |
| | V20L,5C | ‖plane | Wiegert et al. 2015 |
| NGC4254 | E6,3 | Spiral | Soida et al. 1996 |
| | V21,6,4, E6,3 | ASS (+BSS), tidally stretched | Chyży 2008 |
| | W21 | Spiral | Heald et al. 2009 |
| NGC4294 | V21,6 | Halo field, inclined to disk | Vollmer et al. 2013 |
| NGC4299 | V21,6 | Fragments of a spiral | Vollmer et al. 2013 |
| NGC4298/ NGC4302 | E6,4 | ‖disk+intergal. bridge, locally vertical | Weżgowiec et al. 2012 |
| NGC4302 | V21,6 | Mostly ‖disk | Vollmer et al. 2013 |
| NGC4303 | E6,4 | ASS? | Weżgowiec et al. 2012 |
| | V21,6 | Spiral | Vollmer et al. 2013 |
| NGC4321 | V21,6 | Spiral | Vollmer et al. 2007; 2010 |
| | W21 | Spiral | Heald et al. 2009 |
| | E6,4 | BSS? +‖bar | Weżgowiec et al. 2012 |
| NGC4330 | V21,6 | Mostly ‖disk | Vollmer et al. 2013 |
| NGC4388 | E6,4 | Inclined to disk | Weżgowiec et al. 2012 |
| | V21,6 | ‖disk +incl. in halo | Vollmer et al. 2007; 2010 |
| | V5C | Helical field in nuclear outflow, ⊥plane in wind | Damas-Segovia et al. 2016; Krause et al. 2020 |
| NGC4396 | V21,6 | ‖disk +‖NW tail | Vollmer et al. 2007; 2010 |
| NGC4402 | V21,6 | ‖disk in southern halo, incl. in northern halo | Vollmer et al. 2007; 2010 |
| NGC4419 | V21,6 | ‖disk +X-shaped | Vollmer et al. 2013 |
| NGC4424 | V6 | X-shaped | Vollmer et al. 2013 |
| NGC4438 | E6 | ‖disk, ⊥outflow | Weżgowiec et al. 2007 |
| | V6 | ‖disk, displaced from disk in the east | Vollmer et al. 2007; 2010 |
| NGC4457 | V21,6 | Spiral | Vollmer et al. 2013 |
| NGC4501 | E6,4 | ‖disk, asymmetric | Weżgowiec et al. 2007 |
| | V21,6 | Compressed along SW disk edge | Vollmer et al. 2007; 2010 |
| NGC4522 | V21,6 | ‖plane, compressed | Vollmer et al. 2004 |
| NGC4532 | V6 | Huge halo field, inclined +vertical, X-shaped | Vollmer et al. 2013 |
| NGC4535 | V21,6,4 | Spiral | Beck et al. 2002 |
| | E6,4 | Spiral, asymmetric | Weżgowiec et al. 2007; 2012 |
| | V21,6 | ‖spiral arm, asymmetric | Vollmer et al. 2007; 2010 |
| | E4 | ASS? | Weżgowiec et al. 2012 |
| NGC4548 | E6 | ⊥bar | Weżgowiec et al. 2007 |
| NGC4567/ NGC4568 | V21,6 | Intergal. bridge | Vollmer et al. 2013 |
| NGC4569 | E6,4 | ‖disk +‖outflow | Chyży et al. 2006 |
| | W21 | ‖disk +‖outflow | Heald et al. 2009 |
| | V21,6 | ‖disk +‖outflow | Chyży et al., in prep. |
| NGC4579 | V21,6 | ‖bar, spiral in outer disk | Vollmer et al. 2013 |
| NGC4654 | E6,4 | ‖SW arm +‖gas tail | Weżgowiec et al. 2007 |
| | V21,6 | ‖arms, bending out towards gas tail | Vollmer et al. 2007; 2010 |
| NGC4689 | V21,6 | Fragments of a spiral | Vollmer et al. 2013 |
| NGC4713 | V21,6 | Spiral | Vollmer et al. 2013 |
| NGC4808 | V21,6 | Vertical, asymmetric | Vollmer et al. 2013 |

**Radio telescopes:** A = Australia Telescope Compact Array, E = Effelsberg 100-m, G = Giant Metrewave Radio Telescope, K = KAT-7, L = Low Frequency Array (LOFAR), P = Parkes 64-m, V = Very Large Array, W = Westerbork Synthesis Radio Telescope

**Wavelength codes:** 200 = 167-250 cm, 90 = 80-95 cm, 49 = 48-50 cm, 21 = 20-22 cm, 20L = 15-30 cm (L-band), 18 = 18.0 cm, 13 = 12.5-13.4 cm, 11 = 11.1 cm, 10S = 7.5-15 cm (S-band), 6 = 5.8-6.3 cm, 5C = 4.3 – 6.0 cm (C-band), 4 = 3.6 cm, 3 = 2.8 cm, 2 = 2.0 cm, 1 = 9 mm

**Field structures:** ASS = axisymmetric spiral, BSS = bisymmetric spiral, QSS = quadrisymmetric spiral, MSS = multimode spiral

*Table A7: Far-infrared polarization observations of galaxies with* ***SOFIA/HAWC+***

| Galaxy | Wavelength (μm) | Structure | Reference |
|---|---|---|---|
| M51 | 154 | Spiral | Jones et al. 2020 |
| | 154 | Spiral, partly different from radio | Borlaff et al. 2021 |
| M82 | 53,154 | ⊥disk (nuclear region), ‖disk (outer regions) | Jones et al. 2019; |
| | 53, 89, 154, 214 | | Lopez-Rodriguez et al. 2021a |
| M83 | 154 | Spiral | Lopez-Rodriguez et al. 2022a; 2022b |
| NGC253 | 89 | ‖disk | Jones et al. 2019; |
| | 53, 89, 154, 214 | | Lopez-Rodriguez et al. 2022a; 2022b |
| NGC891 | 154 | ‖disk | Jones et al. 2020; |
| | 154 | ‖bar, some vertical field | Kim et al. 2023 |
| NGC1068 | 53, 89 | Spiral | Lopez-Rodriguez et al. 2020; Lopez-Rodriguez et al. 2022a; 2022b |
| NGC1097 | 53, 89, 154 | In bar/ring contact regions | Lopez-Rodriguez et al. 2021b; Lopez-Rodriguez et al. 2022a; 2022b |
| NGC2146 | 53, 89, 154, 214 | Vertical | Lopez-Rodriguez et al. 2022a; 2022b |
| NGC3627 | 154 | | Lopez-Rodriguez et al. 2022a; 2022b |
| NGC4038/9 | 154 | Spiral in relic arm, between galaxies | Lopez-Rodriguez et al. 2023 |
| NGC4736 | 154 | Spiral | Lopez-Rodriguez et al. 2022a; 2022b |
| NGC4826 | 89 | | Lopez-Rodriguez et al. 2022a; 2022b |
| NGC6946 | 154 | Complex | Lopez-Rodriguez et al. 2022a; 2022b |
| NGC7331 | 154 | | Lopez-Rodriguez et al. 2022a; 2022b |
| LMC (30 Dor) | 53, 89, 154, 214 | Complex | Gordon et al. 2018; Tram et al. 2023 |
| CenA | 89 | Along warped molecular disk | Lopez-Rodriguez 2021 |
| Circinus | 53, 89, 214 | | Lopez-Rodriguez et al. 2022a; 2022b |

### A.3 Links to the SKA radio telescope and its precursor and pathfinder telescopes

https://www.skao.int
http://www.scholarpedia.org/article/Square_kilometre_array
https://research.csiro.au/ska
https://www.sarao.ac.za
https://science.nrao.edu/facilities/vla
http://www.lofar.org
https://www.astron.nl/telescopes/wsrt-apertif
http://www.mwatelescope.org